\documentclass[preprint,showpacs,showkeys,preprintnumbers,amsmath,amssymb,aps]{revtex4}
\usepackage[dvips]{graphicx}
\usepackage{xspace}
\usepackage{dcolumn}
\usepackage{rotating}
 
\begin{document}

\preprint{Phys. Rev. C}
 
\title {Separated Structure Functions for the Exclusive Electroproduction of
$K^+\Lambda$ and $K^+\Sigma^0$ Final States}

\newcounter{univ_counter}
\setcounter{univ_counter} {0}

\addtocounter{univ_counter} {1} 
\edef\FIU{$^{\arabic{univ_counter}}$ } 

\addtocounter{univ_counter} {1} 
\edef\JLAB{$^{\arabic{univ_counter}}$ } 

\addtocounter{univ_counter} {1} 
\edef\OHIOU{$^{\arabic{univ_counter}}$ } 

\addtocounter{univ_counter} {1} 
\edef\CMU{$^{\arabic{univ_counter}}$ } 

\addtocounter{univ_counter} {1} 
\edef\ANL{$^{\arabic{univ_counter}}$ } 

\addtocounter{univ_counter} {1} 
\edef\ASU{$^{\arabic{univ_counter}}$ } 

\addtocounter{univ_counter} {1} 
\edef\SACLAY{$^{\arabic{univ_counter}}$ } 

\addtocounter{univ_counter} {1} 
\edef\UCLA{$^{\arabic{univ_counter}}$ } 

\addtocounter{univ_counter} {1} 
\edef\CSA{$^{\arabic{univ_counter}}$ } 

\addtocounter{univ_counter} {1} 
\edef\CUA{$^{\arabic{univ_counter}}$ } 

\addtocounter{univ_counter} {1} 
\edef\CNU{$^{\arabic{univ_counter}}$ } 

\addtocounter{univ_counter} {1} 
\edef\UCONN{$^{\arabic{univ_counter}}$ } 

\addtocounter{univ_counter} {1} 
\edef\EDINBURGH{$^{\arabic{univ_counter}}$ } 

\addtocounter{univ_counter} {1} 
\edef\FAIRFIELD{$^{\arabic{univ_counter}}$ } 

\addtocounter{univ_counter} {1} 
\edef\FSU{$^{\arabic{univ_counter}}$ } 

\addtocounter{univ_counter} {1} 
\edef\GIESSEN{$^{\arabic{univ_counter}}$ } 

\addtocounter{univ_counter} {1} 
\edef\GWU{$^{\arabic{univ_counter}}$ } 

\addtocounter{univ_counter} {1} 
\edef\GLASGOW{$^{\arabic{univ_counter}}$ } 

\addtocounter{univ_counter} {1} 
\edef\IDAHO{$^{\arabic{univ_counter}}$ } 

\addtocounter{univ_counter} {1} 
\edef\INDO{$^{\arabic{univ_counter}}$ } 

\addtocounter{univ_counter} {1} 
\edef\INFNFR{$^{\arabic{univ_counter}}$ } 

\addtocounter{univ_counter} {1} 
\edef\INFNGE{$^{\arabic{univ_counter}}$ } 

\addtocounter{univ_counter} {1} 
\edef\ORSAY{$^{\arabic{univ_counter}}$ } 

\addtocounter{univ_counter} {1} 
\edef\ITEP{$^{\arabic{univ_counter}}$ } 

\addtocounter{univ_counter} {1} 
\edef\JMU{$^{\arabic{univ_counter}}$ } 

\addtocounter{univ_counter} {1} 
\edef\KYUNGPOOK{$^{\arabic{univ_counter}}$ } 

\addtocounter{univ_counter} {1} 
\edef\MIT{$^{\arabic{univ_counter}}$ } 

\addtocounter{univ_counter} {1} 
\edef\UMASS{$^{\arabic{univ_counter}}$ } 

\addtocounter{univ_counter} {1} 
\edef\MSU{$^{\arabic{univ_counter}}$ } 

\addtocounter{univ_counter} {1} 
\edef\UNH{$^{\arabic{univ_counter}}$ } 

\addtocounter{univ_counter} {1} 
\edef\NSU{$^{\arabic{univ_counter}}$ } 

\addtocounter{univ_counter} {1} 
\edef\ODU{$^{\arabic{univ_counter}}$ } 

\addtocounter{univ_counter} {1} 
\edef\PITT{$^{\arabic{univ_counter}}$ } 

\addtocounter{univ_counter} {1} 
\edef\RPI{$^{\arabic{univ_counter}}$ } 

\addtocounter{univ_counter} {1} 
\edef\RICE{$^{\arabic{univ_counter}}$ } 

\addtocounter{univ_counter} {1} 
\edef\URICH{$^{\arabic{univ_counter}}$ } 

\addtocounter{univ_counter} {1} 
\edef\SCAROLINA{$^{\arabic{univ_counter}}$ } 

\addtocounter{univ_counter} {1} 
\edef\UNION{$^{\arabic{univ_counter}}$ } 

\addtocounter{univ_counter} {1} 
\edef\VIRGINIA{$^{\arabic{univ_counter}}$ } 

\addtocounter{univ_counter} {1} 
\edef\VT{$^{\arabic{univ_counter}}$ } 

\addtocounter{univ_counter} {1} 
\edef\WM{$^{\arabic{univ_counter}}$ } 

\addtocounter{univ_counter} {1} 
\edef\YEREVAN{$^{\arabic{univ_counter}}$ } 


\author{ 
P.~Ambrozewicz,\FIU\
D.S.~Carman,\JLAB$\!\!^,$\OHIOU\
R.J.~Feuerbach,\JLAB\
M.D.~Mestayer,\JLAB\
B.A.~Raue,\FIU$\!\!^,$\JLAB\
R.A.~Schumacher,\CMU\
A.~Tkabladze,\GWU\
M.J.~Amarian,\ODU\
M.~Anghinolfi,\INFNGE\
B.~Asavapibhop,\UMASS
% ...
\
\\
(CLAS Collaboration)
}

\affiliation{\FIU Florida International University, Miami, Florida 33199}
\affiliation{\JLAB Thomas Jefferson National Accelerator Laboratory, Newport News, Virginia 23606}
\affiliation{\OHIOU Ohio University, Athens, Ohio  45701}
\affiliation{\CMU Carnegie Mellon University, Pittsburgh, Pennsylvania 15213}
\affiliation{\ANL Argonne National Laboratory, Argonne, Illinois, 60439} 
\affiliation{\ASU Arizona State University, Tempe, Arizona 85287}
\affiliation{\SACLAY CEA-Saclay, DAPNIA-SPhN, F91191 Gif-sur-Yvette Cedex, France}
\affiliation{\UCLA University of California at Los Angeles, Los Angeles, California  90095}
\affiliation{\CSA California State University, Dominguez Hills, Carson, California
90747}
\affiliation{\CUA Catholic University of America, Washington, D.C. 20064}
\affiliation{\CNU Christopher Newport University, Newport News, Virginia 23606}
\affiliation{\UCONN University of Connecticut, Storrs, Connecticut 06269}
\affiliation{\EDINBURGH Edinburgh University, Edinburgh EH9 3JZ, United Kingdom}
\affiliation{\FAIRFIELD Fairfield Iniversity, Fairfield, Connecticut 06824}
\affiliation{\FSU Florida State University, Tallahassee, Florida 32306}
\affiliation{\GIESSEN Physikalisches Institut der Universitaet Giessen,
35392 Giessen, Germany}
\affiliation{\GWU The George Washington University, Washington, DC 20052}
\affiliation{\GLASGOW University of Glasgow, Glasgow G12 8QQ, United Kingdom}
\affiliation{\IDAHO Idaho State University, Pocatello, Idaho 83209}
\affiliation{\INFNFR INFN, Laboratori Nazionali di Frascati, P.O. 13,00044 Frascati, Italy}
\affiliation{\INFNGE INFN, Sezione di Genova and Dipartimento di Fisica, Universit\`a di Genova, 16146 Genova, Italy}
\affiliation{\ORSAY Institut de Physique Nucleaire d'ORSAY, IN2P3, BP1, 91406 Orsay, France}
\affiliation{\ITEP Institute of Theoretical and Experimental Physics, Moscow, 117259, Russia}
\affiliation{\JMU James Madison University, Harrisonburg, Virginia 22807}
\affiliation{\KYUNGPOOK Kyungpook National University, Daegu 702-701, South Korea}
\affiliation{\MIT Massachusetts Institute of Technology, Cambridge, Massachusetts  02139}
\affiliation{\UMASS University of Massachusetts, Amherst, Massachusetts  01003}
\affiliation{\MSU Moscow State University, 119899 Moscow, Russia}
\affiliation{\UNH University of New Hampshire, Durham, New Hampshire 03824}
\affiliation{\NSU Norfolk State University, Norfolk, Virginia 23504}
\affiliation{\ODU Old Dominion University, Norfolk, Virginia 23529}
\affiliation{\PITT University of Pittsburgh, Pittsburgh, Pennsylvania 15260}
\affiliation{\RPI Rensselaer Polytechnic Institute, Troy, New York 12180}
\affiliation{\RICE Rice University, Houston, Texas 77005}
\affiliation{\URICH University of Richmond, Richmond, Virginia 23173}
\affiliation{\SCAROLINA University of South Carolina, Columbia, South Carolina 29208}
\affiliation{\UNION Union College, Schenectady, NY 12308}
\affiliation{\VIRGINIA University of Virginia, Charlottesville, Virginia 22901}
\affiliation{\VT Virginia Polytechnic Institute and State University, Blacksburg, Virginia 24061}
\affiliation{\WM College of William and Mary, Williamsburg, Virginia 23187}
\affiliation{\YEREVAN Yerevan Physics Institute, 375036 Yerevan, Armenia}

\date{\today}

\begin{abstract}
We report measurements of the exclusive electroproduction of $K^+\Lambda$ 
and $K^+\Sigma^0$ final states from a proton target using the CLAS 
detector at the Thomas Jefferson National Accelerator Facility.  The 
separated structure functions $\sigma_T$, $\sigma_L$, $\sigma_{TT}$, and 
$\sigma_{LT}$ were extracted from the $\Phi$- and $\epsilon$-dependent 
differential cross sections taken with electron beam energies of 2.567, 
4.056, and 4.247~GeV.  This analysis represents the first 
$\sigma_L/\sigma_T$ separation with the CLAS detector, and the first
measurement of the kaon electroproduction structure functions away from
parallel kinematics.  The data span a broad range of momentum transfers 
from $0.5\leq Q^2\leq 2.8$~GeV$^2$ and invariant energy from 
$1.6\leq W\leq 2.4$~GeV, while spanning nearly the full center-of-mass 
angular range of the kaon.  The separated structure functions reveal 
clear differences between the production dynamics for the $\Lambda$ and 
$\Sigma^0$ hyperons.  These results provide an unprecedented data sample 
with which to constrain current and future models for the associated 
production of strangeness, which will allow for a better understanding of 
the underlying resonant and non-resonant contributions to hyperon production.
\end{abstract}


\maketitle

\newpage

\section{Introduction}
\label{intro}

A necessary step toward understanding the structure and dynamics of strongly 
interacting matter is to fully understand the spectrum of excited states of 
the nucleon.   This excitation spectrum is a direct reflection of its 
underlying substructure.  Understanding nucleon resonance excitation, and 
hadro-production in general, continues to provide a serious challenge to 
hadronic physics due to the non-perturbative nature of the theory of strong 
interactions, Quantum Chromodynamics (QCD), at these energies.  Because of 
this, a number of approximations to QCD have been developed to understand 
baryon resonance decays.  One such approach is a class of semi-relativized
symmetric quark models~\cite{isgur,capstick} that invoke massive constituent 
quarks.  These models typically predict many more nucleonic states 
than have been found experimentally. A possible explanation to this so-called 
``missing resonance'' problem is that these nucleon resonances may have a 
relatively weak coupling to the pion-nucleon states through which many 
searches have been performed, and may, in fact, couple to other final 
states such as multi-pion or strangeness channels.  In this work we provide 
an extensive set of data that may be used to search for these hidden states 
in strangeness electroproduction reactions.  These data then provide for a 
complementary way in which to view the baryon resonance spectrum, as some of 
the ``missing'' states might be only ``hidden'' when studied in particular 
reactions.  It could also be the case that some dynamical aspect of hadronic 
structure is acting to restrict the quark model spectrum of states to the more 
limited set established by existing data~\cite{klempt}.

Beyond different coupling constants relative to single-pion production 
(e.g. $g_{KYN}$ vs. $g_{\pi NN}$), the study of the exclusive production 
of $K^+\Lambda$ and $K^+\Sigma^0$ final states has other advantages in the 
search for missing resonances.  The higher masses of the kaon and hyperons,
compared to pionic final states, kinematically favor a two-body decay mode 
for resonances with masses near 2~GeV, a situation that is experimentally 
advantageous.  New information is also provided by comparing $K^+\Lambda$ 
to $K^+\Sigma^0$.  Note that although the two ground-state hyperons have 
the same valence quark structure ($uds$), they differ in isospin, such that 
intermediate $N^*$ resonances can decay strongly to $K^+\Lambda$ final states, 
but intermediate $\Delta^*$ states cannot.  Because $K^+\Sigma^0$ final 
states can have contributions from both $N^*$ and $\Delta^*$ states, the 
hyperon final state selection constitutes an isospin filter.  

Electro-excitation of the nucleon has served to quantify the structure of 
many excited states that decay to single pions.  These studies have been
used to test models of the internal structure of the excitations using
the interference structure functions and the behavior with $Q^2$ for
various regions of $W$.  This experiment is the first to allow analogous 
investigations to start when the intermediate electro-excited nucleon 
resonances couple to kaon-hyperon final states. 

The search for missing resonances requires more than identifying features 
in the relevant mass spectrum.  It also requires an iterative approach in 
which experimental measurements constrain the dynamics of various 
hadrodynamic models.  The tuned models can in turn be used to interpret 
$s$-, $t$- and $u$-channel spectra in terms of the underlying resonances. 
As emphasized by Lee and Sato~\cite{lee}, QCD cannot be directly tested with 
$N^*$ spectra without a model for the production dynamics.  The key to 
constraining models and unraveling the contributing resonant and non-resonant 
diagrams that contribute to the dynamics, is to measure as many observables 
over as wide a kinematic range as possible.  

In this paper, we present measurements of the separated structure functions
$\sigma_U$, $\sigma_{TT}$, and $\sigma_{LT}$ for exclusive electroproduction 
of $K^+\Lambda$ and $K^+\Sigma^0$ final states for a range of momentum 
transfer $Q^2$ from 0.5 to 2.8~GeV$^2$ and invariant energy $W$ from 1.6 to 
2.4~GeV, while spanning the full center-of-mass angular range of the kaon.  
Our center-of-mass angular coverage is unprecedented.  These are the
first data published on exclusive $KY$ electroproduction that extend beyond 
very forward kaon angles in the center of mass.  At one value of $Q^2$, 
$Q^2$= 1.0~GeV$^2$, we were also able to separate the unpolarized structure 
function, $\sigma_U$, into its components $\sigma_T$ and $\sigma_L$ using a 
traditional Rosenbluth separation and also an alternative $\epsilon-\Phi$ 
Rosenbluth technique (where $\epsilon$ is the transverse polarization of the 
virtual photon and $\Phi$ is the angle between the electron and hadron 
planes).  In this alternative method, we obtain the four structure functions 
in a single fit.  This extensive data set should provide substantial 
constraints on the various hadrodynamic models (discussed in 
Section~\ref{sec:theory}).  Due to the very large number of analysis bins 
encompassed by this work, only a portion of our available data is included 
here.  The full set of our data is available in Ref.~\cite{database}.  

After a brief review of the relevant formalism in Section~\ref{formalism}, an 
overview of previous experimental work in this area in Section~\ref{recent}, 
and a brief review of the current theoretical approaches in 
Section~\ref{sec:theory}, we present our measurements made using the CLAS 
detector in Hall B at Jefferson Laboratory (JLab) in Sections~\ref{analysis} 
through \ref{systematics}.  In Section~\ref{results}, the results are 
examined phenomenologically and compared to predictions from 
several models that have not been ``tuned'' to this data set.  
Finally, we present our conclusions regarding the potential impact 
of these data in Section~\ref{conclusions}.  Our conclusions 
regarding the $s$-channel baryon spectrum are, unfortunately, rather 
limited.  Real progress on identifying heretofore ``missing'' resonances
will only result from a judicious fitting of the theoretical and 
phenomenological models to these data and the remainder of the world's
data on these final states.

\section{Formalism}
\label{formalism}

\begin{figure}[htbp]
\vspace{5.5cm}
\includegraphics{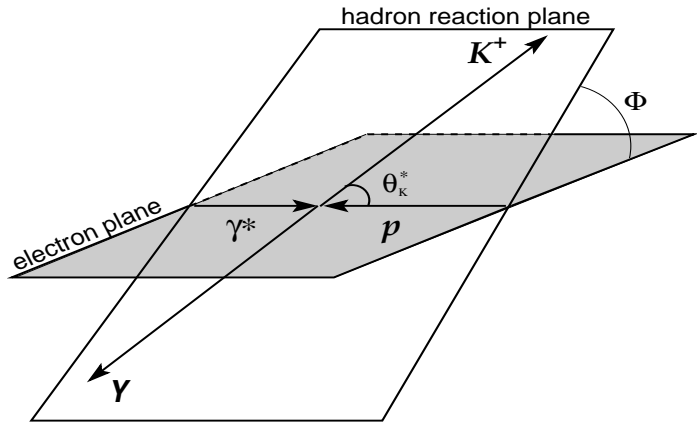}
\caption{\small{Kinematics for $K^+Y$ (where $Y$ is either a $\Lambda$ or
$\Sigma^0$) electroproduction defining the angles $\theta_K^*$ and $\Phi$
with respect to the center-of-mass reference frame.}} 
\label{fig-kin}
\end{figure}

In kaon electroproduction a beam of electrons with four-momentum 
$p_e = (E_e,\vec{p}_e\,)$ is incident upon a fixed proton target of 
mass $M_p$, and the outgoing scattered electron with momentum 
$p_{e'}=(E_{e'},\vec{p}_{e'}\,)$ and kaon with momentum 
$p_K=(E_K,\vec{p}_K)$ are measured.  The cross section for the exclusive 
$K^+$-hyperon state is then differential in the scattered electron momentum 
and kaon direction.  Under the assumption of single-photon exchange, where 
the photon has four-momentum $q=p_e-p_{e'}=(\nu,\vec{q}\,)$, this can be 
re-expressed as the product of an equivalent flux of virtual photons and 
the $\gamma^* p$ center-of-mass (c.m.) virtual photo-absorption cross 
section as:

\begin{equation}
\frac{d\sigma}{dE_{e'} d\Omega_{e'} d\Omega_K^*} = \Gamma
\frac{d\sigma_v}{d\Omega_K^*},
\end{equation}

\noindent
where the virtual photon flux factor $\Gamma$ depends upon only the
electron scattering process.  After integrating over the azimuthal angle 
of the scattered electron, the absorption cross section can be expressed 
in terms of the variables $Q^2$, $W$, $\theta_K^*$, and $\Phi$, where 
$q^2=-Q^2$ is the squared four-momentum of the virtual photon, 
$W=\sqrt{M_p^2+2M_p\nu-Q^2}$ is the total hadronic energy in the c.m. frame, 
$\theta_K^*$ is the c.m. kaon angle relative to the virtual photon direction, 
and $\Phi$ is the angle between the leptonic and hadronic production planes 
(see Fig.~\ref{fig-kin}).  After introducing the appropriate Jacobian, the 
form of the cross section can be written as:

\begin{equation}
\frac{d\sigma}{dQ^2 dW d\Omega_K^*} = \Gamma_v 
\frac{d\sigma_v}{d\Omega_K^*},
\end{equation}

\noindent
where

\begin{equation}
\label{eq:flux}
\Gamma_v = \frac{\alpha}{4\pi}\frac{W}{M_p^2 E^2}\frac{W^2-M_p^2}{Q^2}
\frac{1}{1-\epsilon}
\end{equation}

\noindent
is the flux of virtual photons, 

\begin{equation}
\epsilon=\left(1+2\frac{|\vec{q}\,|^2}{Q^2}\tan^2{\frac{\theta_{e'}}{2}}
\right)^{-1}
\end{equation}

\noindent
is the transverse polarization of the virtual photon, and $\theta_{e'}$
is the electron scattering angle in the laboratory frame.

After summing over the polarizations of the initial and final state
electrons and hadrons, the virtual photon cross section can be written as:

\begin{equation}
\frac {d\sigma_v}{d\Omega_K^*} = \sigma_T + \epsilon \sigma_L +
\epsilon \sigma_{TT} \cos 2\Phi + \sqrt{\epsilon(\epsilon+1)}\sigma_{LT} 
\cos \Phi .
\label{xsect}
\end{equation}

\noindent
In this expression, the cross section is decomposed into four structure
functions, $\sigma_T$, $\sigma_L$, $\sigma_{TT}$, and $\sigma_{LT}$, 
which are, in general, functions of $Q^2$, $W$, and $\theta_K^*$ only.  Note 
that this convention for the differential cross section is not used by all 
authors~\cite{convention}. 

Each of the structure functions is related to the coupling of the hadronic 
current to different combinations of the transverse and longitudinal 
polarization of the virtual photon.  
$\sigma_T = \frac{1}{2}(\sigma_\parallel + \sigma_\perp)$ is the 
differential cross section contribution for unpolarized transverse virtual 
photons.  In the limit $Q^2 \to 0$, this term must approach the cross 
section for unpolarized real photons which only have transverse polarization.
$\sigma_L$ is the differential cross section contribution for longitudinally 
polarized virtual photons.  $\sigma_{TT}$ and $\sigma_{LT}$ represent 
interference contributions to the cross section. 
$\sigma_{TT} = \frac{1}{2}(\sigma_\parallel - \sigma_\perp)$ is due to the 
interference of transversely polarized virtual photons and $\sigma_{LT}$ is 
due to the interference of transversely and longitudinally polarized virtual 
photons.  Here $\sigma_\parallel$ and $\sigma_\perp$ are the cross sections 
for virtual photons having their electric vector parallel and perpendicular 
to the hadronic production plane, respectively.  Note that the term
$\sigma_{TT}$ in electroproduction is related to the linearly polarized
photon beam asymmetry in photoproduction experiments, which is defined as
$\Sigma = -\sigma_{TT}/\sigma_T$.

For the remainder of this paper we will refer to 
$\sigma_U=\sigma_T + \epsilon \sigma_L$ as the ``unseparated'' part of 
the cross section. The further decomposition of these structure functions 
into response functions, which can then be expressed in terms of either 
complex amplitudes or as multipole expansions, is given, for example, in
Ref.~\cite{knochlein}.  In this paper we will compare theory to the
structure function terms introduced above.

\section{Previous Experimental Work}
\label{recent}

Hyperon electroproduction in the nucleon resonance region has remained 
largely unexplored.  Several low-statistics measurements were carried
out in the 1970's at Cambridge, Cornell, and DESY, and focussed mainly
on cross section measurements to explore differences in the production
dynamics between $\Lambda$ and $\Sigma^0$ hyperons.  The first experiment
was performed at the Cambridge Electron Accelerator~\cite{brown} using 
small-aperture spectrometers in kinematics spanning $Q^2$ below 1.2~GeV$^2$,
$W$ from 1.8 to 2.6~GeV, and forward kaon angles ($\theta_K^* < 28^{\circ}$).  
It was noted that the $K^+\Lambda$ channel dominated the $K^+\Sigma^0$ 
channel, with signs of a large longitudinal component in the $K^+\Lambda$ 
channel.  Subsequent results from Cornell~\cite{bebek1} for $Q^2 <$ 
2.0~GeV$^2$ and $W$=2.15 and 2.67~GeV confirmed this observation.  In this 
paper we show that $K^+\Lambda$ dominance only occurs at forward kaon 
angles, and the longitudinal strength is only important at forward angles 
and higher $W$.

An experiment from DESY~\cite{azemoon} used a large aperture spark-chamber 
spectrometer to measure both reaction channels at higher $W$ 
($1.9 < W < 2.8$~GeV) and lower $Q^2$ ($0.1 < Q^2 < 0.6$~GeV$^2$). That 
experiment managed the first $\sigma_{LT}$ and $\sigma_{TT}$ separations, 
albeit with large error bars, few data points, and considerable kinematic 
extrapolations to extract results at fixed values of $W$, $Q^2$, and 
$t=(q-p_K)^2$.  The results were consistent with zero for these interference 
cross sections due to the large uncertainties.  In this paper we show the
first measurements of the structure functions with enough precision to
determine non-zero interference terms.

Other measurements made at Cornell were reported~\cite{bebek2} for kaons 
produced at very small angles relative to the virtual photon 
($\theta_K^* < 15^{\circ}$).  The results for $W$ from 2.15 to 3.1~GeV 
included improved measurements of the $Q^2$ dependence of the differential 
cross sections over the range from 0.6 to 4.0~GeV$^2$, showing that the 
$K^+\Sigma^0$ cross section falls off much faster than the $K^+\Lambda$ 
cross section.  At that time this was explained by a vector meson dominance 
argument, or alternatively, as possible evidence for an isoscalar di-quark 
interaction that favors $\Lambda$ production over $\Sigma^0$ production off 
the proton~\cite{close}.  Another survey experiment from DESY~\cite{brauel} 
at $W$=2.2~GeV and $0.06 < Q^2 < 1.35$~GeV$^2$ that measured differential 
cross sections for the $K^+\Lambda$ and $K^+\Sigma^0$ final states, confirmed 
the measured $Q^2$ dependence of the cross sections, but with improved 
statistics. Our study of the $Q^2$ dependence shows that this same behavior 
for $\Sigma^0$ production relative to $\Lambda$ production also occurs for 
larger kaon angles.

The first $\sigma_L$/$\sigma_T$ separation via the Rosenbluth method was 
also made at Cornell~\cite{bebek2}, suggesting that $\sigma_L$ for the 
$K^+\Lambda$ channel is large but not dominant at forward kaon angles, as 
previously surmised, while it is vanishing for the $K^+\Sigma^0$ channel.  
At JLab two more recent results employing the Rosenbluth technique in parallel
kinematics ($\theta_K^*$=0$^{\circ}$) have been completed to separate 
$\sigma_L$ and $\sigma_T$.  The first result reported $\sigma_L$ and 
$\sigma_T$ for both the $K^+\Lambda$ and $K^+\Sigma^0$ final states using 
the small-aperture spectrometers in Hall C~\cite{mohring}.  Results 
were extrapolated to $W \sim$1.84~GeV for a range of $Q^2$ from 0.52 to 
2.00~GeV$^2$.  It showed, contrary to previous findings, that the ratio 
$\sigma_L$/$\sigma_T$ for the $\Sigma^0$ is not very different in the forward 
direction than for the $\Lambda$ over this $Q^2$ range.  The ratio for
both hyperons is about 0.4, albeit with large uncertainties.  The other 
existing measurement from JLab was performed in Hall A for $W$ in the range 
from 1.8 to 2.14~GeV with $Q^2$ values of 1.9 and 2.35~GeV$^2$~\cite{coman}.  
This measurement was only for the $K^+\Lambda$ final state and showed that 
the ratio of $\sigma_L/\sigma_T$ was consistent with the Hall C result.  

In a previous CLAS publication using the same data presented in this work, 
the polarization transfer from the virtual photon to the produced $\Lambda$ 
hyperon was reported~\cite{carman}.  These observables were expected 
theoretically to have strong sensitivity to the underlying resonance 
contributions.  Surprisingly, they seemed to have only a modest dependence on 
$W$.  The CLAS polarization data were also analyzed to extract the ratio 
$\sigma_L/\sigma_T$ at $\theta_K^*=0^{\circ}$~\cite{raue04}.  The measured
ratio was smaller than, but consistent with, that from the Hall C measurements,
providing an important cross-check on the extraction of $\sigma_L$ and 
$\sigma_T$ from a measurement with very different systematics.  

In this paper we present data for $\sigma_L$ and $\sigma_T$ in similar 
kinematics that can be compared to these data.  In addition we present the 
first available data for these structure functions for large $\theta_K^*$, 
away from parallel kinematics.  Here, the longitudinal and transverse 
structure functions are extracted using the standard Rosenbluth technique, 
as well as by a simultaneous $\epsilon$-$\Phi$ fit to our different beam 
energy data sets.  This analysis represents the first $\sigma_L$/$\sigma_T$
separation using the CLAS spectrometer.

In contrast to the sparse extant electroproduction data, there
exist several high-quality photoproduction data sets.  Recently, exclusive 
photoproduction of $K^+\Lambda$ and $K^+\Sigma^0$ final states have been 
investigated with the large-acceptance SAPHIR~\cite{saphir1,saphir2} and CLAS 
\cite{mcnabb,bradford1} detectors.  High statistics total cross sections, 
differential cross sections, and induced polarizations for the final state
hyperons have been measured that span the full nucleon resonance region.  In
addition, high statistics measurements from CLAS of the beam-recoil hyperon
polarization transfer have been completed for both the $K^+\Lambda$ and
$K^+\Sigma^0$ final states~\cite{bradford2}, and beam spin asymmetry
measurements have been made at LEPS for both $\Lambda$ and $\Sigma$ hyperons
\cite{leps1,leps2}.  The $W$ dependence of these data have been studied with
the aim to understand the underlying $s$-channel $N^*$ and $\Delta^*$ 
contributions.  Further information regarding interpretations of these data 
within different models is included in Section~\ref{sec:theory}.

Given this landscape of available data on the associated production of
hyperons, it is clear that the majority of the existing electroproduction 
data, while spanning similar ranges of $W$ and $Q^2$ as our data, only 
provide information for very forward kaon scattering angles.  This new data 
from CLAS represents a significant improvement in that it covers the full 
kaon scattering angular range, which will allow for an in-depth investigation 
of the contributing $s$-channel and $u$-channel diagrams in addition to the 
$t$-channel processes to these reactions.  The new CLAS data also provides 
full azimuthal coverage, which enables the first significant data sample to 
study the interference structure functions.  These structure functions provide 
new and unique information on interference between the underlying resonant and 
non-resonant amplitudes.  In addition, CLAS has made significant contributions 
to the data base with the photoproduction cross sections and polarization 
observables that have been published.  This new set of electroproduction data 
allows the study of the production dynamics as a function of the mass of the 
virtual photon, which provides an exciting complement to the real photon data.

\section{Theoretical Models}
\label{sec:theory}

At the medium energies used in this experiment, perturbative QCD is not yet 
capable of providing any analytical predictions for the differential cross 
sections or structure functions for kaon electroproduction.  In order to
understand the underlying physics, effective models must be employed 
that ultimately represent approximations to QCD.  This paper compares the
data to two different theoretical model approaches: hadrodynamic models 
and models based on Reggeon exchange.

\subsection{Hadrodynamic Models}

\begin{figure}[htpb]
\vspace{5.2cm}
\includegraphics{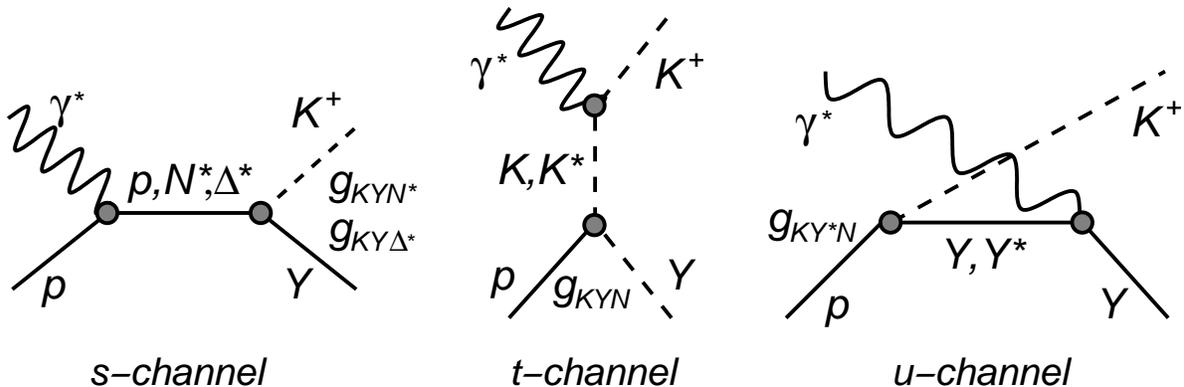}
\caption{\small{Feynman diagrams representing $s$-channel nucleon 
($p,N^*,\Delta^*$) exchange (left), $t$-channel kaon ($K,K^*$) exchange 
(middle), and $u$-channel hyperon ($Y,Y^*$) exchange (right) that 
contribute to the reaction models.  The vertex labels $g_{MBB}$ represent 
the strong coupling constants.}}
\label{feynman}
\end{figure}

Hadrodynamic models provide a description of the reaction based upon 
hadronic degrees of freedom.  In this approach, the strong interaction is 
modeled by an effective Lagrangian, which is constructed from tree-level 
Born and extended Born terms for intermediate states exchanged in the $s$, 
$t$, and $u$ reaction channels as shown in Fig.~\ref{feynman}.  Each 
resonance has its own strong coupling constants and strong decay widths.  A 
complete description of the physics processes requires taking into account 
all possible channels that could couple to the initial and final state 
measured, but the advantages of the tree-level approach are to limit 
complexity and to identify the dominant trends.  In the one-channel, 
tree-level approach, several dozen parameters must be fixed by fitting to 
the data, since they are poorly known and not constrained from other sources.  
Identification of the important intermediate states or resonances is guided 
by existing data and quark model predictions.  The coupling constants for 
each of the included resonances are extracted from global fits of the model 
calculations to the existing data base.  It is common practice to use 
phenomenological form factors to account for the extension of the point-like 
interactions at the hadronic vertices~\cite{janssen1}. Different models 
typically have different prescriptions for restoring gauge invariance.  The 
drawback of these models is the large number of exchanged hadrons that can 
contribute in the intermediate state of the reaction.  Depending on which 
set of resonances is included, very different conclusions about the strengths 
of the contributing diagrams may be reached.  As stated in 
Section~\ref{intro}, the models that are employed in this work have not 
been ``tuned'' to our data.  It should also be stated that due to the nature 
of these models, they have a much higher interpretative power than predictive 
power.  Therefore it will be the case that more definitive statements 
regarding the reaction dynamics and underlying resonant and background terms 
will only be possible after our data have been included in the model fits.

Two different hadrodynamic models are employed in this work.  The first is 
the model of Bennhold and Mart~\cite{bennhold} (referred to here as BM) and 
the second is model B of Janssen {\it et al.}~\cite{janssen1} (referred to 
here as JB).  In these models, the coupling strengths have been determined 
mainly by fits to existing $\gamma p \to K^+ Y$ data (with some older 
electroproduction data included in some cases) by adding the non-resonant 
Born terms with a number of resonance terms in the $s$, $t$, and $u$ reaction
channels, leaving the coupling constants as free parameters.  The coupling 
constants are required to respect the limits imposed by SU(3) allowing for a 
symmetry breaking at the level of about 20\%.  Both models have been compared 
against the existing photoproduction data from SAPHIR~\cite{saphir1,saphir2} 
and CLAS~\cite{mcnabb,bradford1} and provide a fair description of those 
results.  The model parameters are not based on fits to any CLAS data.  The 
specific resonances included within these calculations are listed in 
Table~\ref{tab-resonances}.  

\begin{table}[htbp]
\begin{center}
\begin{tabular} {||c|c|c||c|c||} \hline \hline
  & \multicolumn{2} {|c||} {$K^+\Lambda$} & \multicolumn{2} {|c||} 
{$K^+\Sigma^0$} \\ \hline
Resonance                    & BM & JB & BM & JB \\ \hline
$N^*$(1650) ($S_{11}$)       & *  & *  & *  & * \\ \hline
$N^*$(1710) ($P_{11}$)       & *  & *  & *  & * \\ \hline
$N^*$(1720) ($P_{13}$)       & *  & *  & *  & * \\ \hline
$N^*$(1895) ($D_{13}$)       & *  & *  &    &   \\ \hline\hline
$\Delta^*$(1900) ($S_{31}$)  &    &    & *  & * \\ \hline
$\Delta^*$(1910) ($P_{31}$)  &    &    & *  & * \\ \hline \hline
$K^*$(892)                   & *  & *  & *  & * \\ \hline
$K^*_1$(1270)                & *  & *  & *  &   \\ \hline\hline
$\Lambda^*$(1800) ($S_{01}$) &    & *  &    &   \\ \hline
$\Lambda^*$(1810) ($P_{01}$) &    & *  &    & *  \\ \hline
$\Sigma^*$(1880) ($P_{11}$)  &    &    &    & * \\ \hline \hline
\end{tabular}
\end{center}
\caption{\small{Resonances included in the hadrodynamic Bennhold and Mart 
model~\cite{bennhold} and the Janssen {\it et al.} model~\cite{janssen1} 
highlighted in this work for both the $K^+\Lambda$ and $K^+\Sigma^0$ final 
states.}}
\label{tab-resonances}
\end{table}

For $K^+\Lambda$ production, the BM model includes the Born terms as well as 
four baryon resonance contributions.  Near threshold, the steep rise of the 
cross section is accounted for with the $N^*$ states $S_{11}$(1650), 
$P_{11}$(1710), and $P_{13}$(1720).  To explain the broad bump in the energy 
dependence of the cross section seen by SAPHIR~\cite{saphir1} and CLAS
\cite{mcnabb,bradford1}, the BM model includes a spin-3/2 $D_{13}$(1895) 
resonance that was predicted in the relativized quark model of Capstick and 
Roberts~\cite{capstick} to have a strong coupling to the $K^+\Lambda$ channel, 
but which was not well established from existing pion-production data.  In 
addition this model includes $t$-channel exchange of the vector $K^*$(892) 
and pseudovector $K_1$(1270) mesons.  In this model the inclusion of hadronic 
form factors leads to a breaking of gauge invariance which is restored by the 
inclusion of counter terms following the prescription of Haberzettl
\cite{haber}.

For $K^+\Sigma^0$ production, the BM model includes the Born terms as well as 
the $N^*$ resonances $S_{11}$(1650), $P_{11}$(1710), and $P_{13}$(1720), and 
the $\Delta^*$ resonances $S_{31}$(1900) and $P_{31}$(1910).  The model also 
includes $K^*$(892) and $K_1$(1270) exchanges.  The modeling of hadronic form 
factors for the $\Sigma^0$ channel is handled as described above for the
$\Lambda$ channel.  The BM model does not include any $u$-channel diagrams 
for either $KY$ final state.

In this work, we also compare our data against model B of Janssen {\it et al.}
\cite{janssen1}, which counterbalances the strength from the Born terms 
by introducing hyperon resonances in the $u$-channel, where a destructive 
interference of the $u$-channel hyperon resonance terms with the other 
background terms occurs.  The authors of Ref.~\cite{janssen1,janssen2} claim 
that this is a plausible way to reduce the Born strength.  For the $K^+\Lambda$ 
calculations, the included $u$-channel resonances are the $\Lambda^*$ states 
$S_{01}$(1800) and $P_{01}$(1810).  For the $K^+\Sigma^0$ calculations, the 
included $u$-channel resonances are the $\Lambda^*$ $P_{01}$(1810) and the 
$\Sigma^*$ $P_{11}$(1880).  Ref.~\cite{janssen3} states that there is very 
little theoretical guidance on how to select the relevant resonances and how 
to determine realistic values for the associated coupling constants.  It is
stated that the same qualitative destructive interference effect was observed 
for other $u$-channel resonance choices, and that the introduced resonances 
should be interpreted more properly as ``effective'' particles that account 
for a larger set of hyperon resonances participating in the process.  The 
$s$-channel and $t$-channel resonances included in the JB model are nearly 
the same as in the BM model.  Hadronic form factors are included in the 
model with gauge invariance restoration based on the approach by Gross and 
Riska~\cite{gross}.

Different models have markedly different ingredients and fitted coupling 
constants.  Certainly not every available hadrodynamic model is discussed 
in this work.  However, it is worth mentioning that analysis of Saghai 
{\it et al.}~\cite{saghai}, using the same data set employed for the BM and 
JB models, have shown that by tuning the background processes involved in the 
$K^+\Lambda$ reaction in the form of additional $u$-channel resonances, the 
need to include the extra $D_{13}$ $N^*$ state was removed.  Another analysis 
including newer photo- and electroproduction data from JLab by Ireland 
{\it et al.}~\cite{ireland} has shown some evidence for the need for an 
additional $N^*$ state at about 1900~MeV (one or more of $S_{11}$, $P_{11}$, 
$P_{13}$, $D_{13}$), however they concluded that a more comprehensive data 
set would be required to make further progress.

A recent coupled-channels analysis by Sarantsev {\it et al.}~\cite{sarantsev} 
of the photoproduction data from SAPHIR and CLAS, as well as beam asymmetry 
data from LEPS for $K^+\Lambda$~\cite{leps1} and data from $\pi$ and $\eta$ 
photoproduction, reveals evidence for new baryon resonances in the high $W$ 
mass region.  In this analysis, the full set of data can only be 
satisfactorily fit by including a new $P_{11}$ state at 1840~MeV and two 
$D_{13}$ states at 1870 and 2130~MeV.  Of course these fits have certain 
ambiguities that can be resolved or better constrained by incorporating 
electroproduction data.

The CLAS and SAPHIR photoproduction experiments measured only the $\sigma_T$ 
term.  The more recent CLAS data~\cite{mcnabb,bradford1}, with higher 
statistical precision and finer binning compared to the SAPHIR data
\cite{saphir1,saphir2}, reveal that the strength and centroid of the $W$ 
structure near 1.9~GeV changes with angle, indeed pointing to the possible 
existence of more than one $s$-channel resonance as suggested by the analysis 
of Ref.~\cite{sarantsev}.  The interference structure functions, 
$\sigma_{TT}$ and $\sigma_{LT}$, which are accessible in the 
electroproduction data and presented in this work, will be useful in 
further constraining and testing models that include new $s$-channel 
resonance diagrams in this mass region.  In addition, the $\sigma_T$ and 
$\sigma_L$ structure functions will also provide crucial information to 
constrain the model parameters for the resonance and background diagrams.

\subsection{Regge Models}

In this work we also compare our results to a Reggeon-exchange model from
Guidal, Laget, and Vanderhaeghen~\cite{guidal} (referred to here as the GLV
model).  This calculation includes no baryon resonance terms at all, but is 
instead based only on gauge invariant $t$-channel $K$ and $K^*$ Regge 
trajectory exchange.  It therefore provides a complementary basis for 
studying the underlying dynamics of strangeness production.  It is important 
to note that the Regge approach has far fewer parameters compared to the 
hadrodynamic models.  These include the $K$ and $K^*$ form factors, which in 
the GLV model are assumed to be of a monopole form 
$F_{K,K^*} = [1 + Q^2/\Lambda_{K,K^*}^2]^{-1}$ with a mass scale 
$\Lambda_{K,K^*}$=1.5~GeV$^2$ chosen to reproduce the JLab Hall C $\sigma_L$, 
$\sigma_T$ data~\cite{mohring}.  In addition, the model employs values for 
the coupling constants $g_{KYN}$ and $g_{K^*YN}$ taken from photoproduction 
studies.

The model was fit to higher-energy photoproduction data where there is little 
doubt of the dominance of these kaon exchanges, and extrapolated down to JLab 
energies.  An important feature of this model is the way gauge invariance is 
achieved for the kaon $t$-channel exchange by Reggeizing the $s$-channel
nucleon pole contribution in the same manner as the kaon $t$-channel
diagram~\cite{guidal1}.  This approach has been noted as a possible reason 
why the Regge model, despite not including any $s$-channel resonances, was 
able to reproduce the JLab Hall C $\sigma_L/\sigma_T$ data at 
$Q^2 \ge 0.5$~GeV$^2$.  The stated reason is that due to gauge invariance, 
the $t$-channel kaon exchange and $s$-channel nucleon pole terms are 
inseparable and must be treated on the same footing.  In the GLV Regge model 
these terms are Reggeized in the same way and multiplied by the same 
electromagnetic form factor.  No counter terms need to be introduced to 
restore gauge invariance as is done in the hadrodynamic approach~\cite{guidal}.

\section{Experiment Description and Data Analysis}
\label{analysis}

\begin{figure}[htbp]
\vspace{6.5cm}
\includegraphics{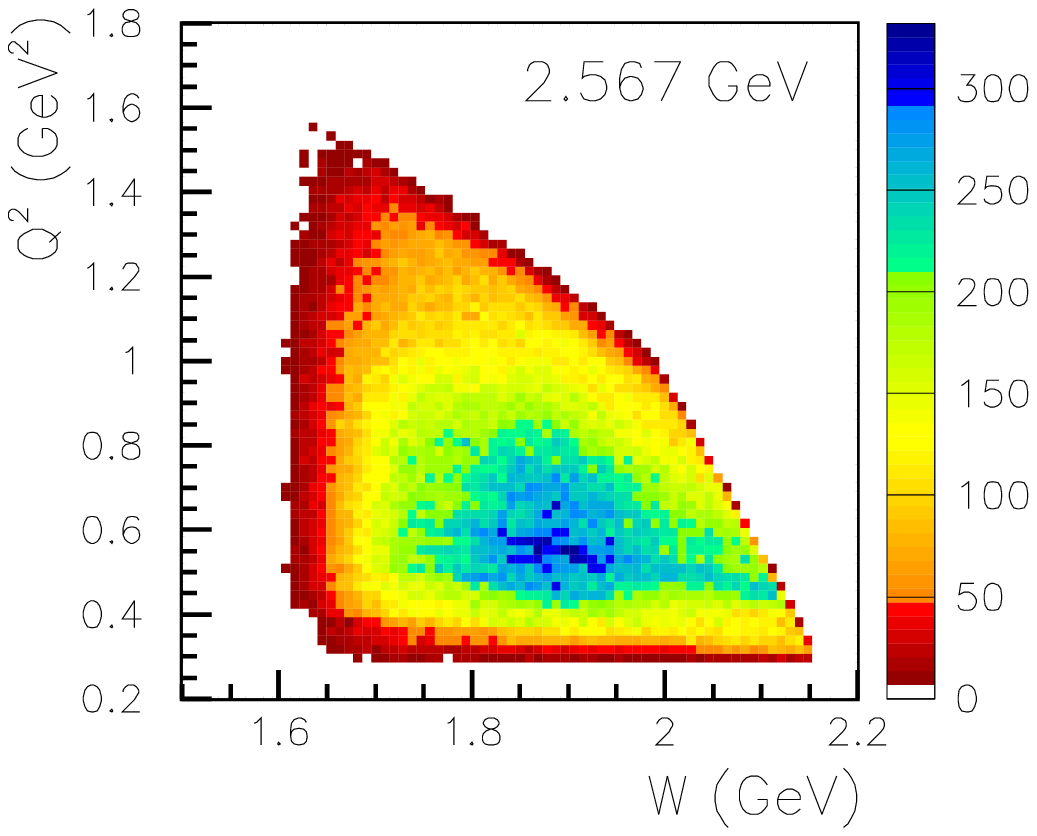}
\includegraphics{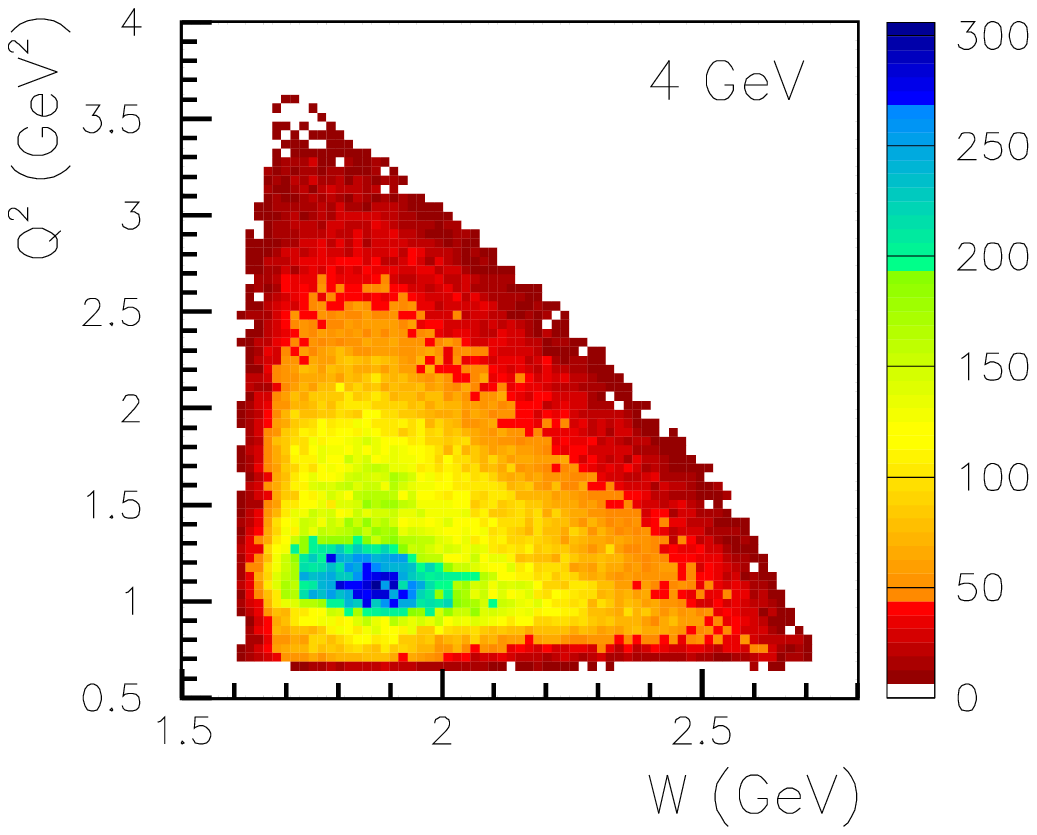}
\caption{\small{(Color online) CLAS kinematic coverage in terms of $Q^2$ 
vs. $W$ for $p(e,e'K^+)Y$ ($Y = \Lambda,\Sigma^0$) events for beam energies 
of 2.567~GeV (left) and 4~GeV (right).}}
\label{fig-Q2W}
\end{figure}

The experiment was performed using the electron beam at JLab and the CLAS 
detector in Hall B.  An electron beam of 5~nA current was incident upon a 
5-cm long liquid-hydrogen target, resulting in an average beam-target 
luminosity of $\mathcal{L} \approx 10^{34}~\mathrm{cm^{-2} s^{-1}}$.  
Data were taken with beam energies of 2.567, 4.056, and 4.247~GeV.  In
this analysis the 4.056 and 4.247~GeV data have been combined into a
single data set, referred to throughout this work as the 4~GeV data set.
The 2.567 data set has a live-time corrected luminosity of about 
1.32~fb$^{-1}$, while that for the 4.056 and 4.247~GeV data sets are about 
0.67~fb$^{-1}$ and 0.80~fb$^{-1}$, respectively.  The beam was effectively 
continuous with a 2.004~ns bunch structure. The large acceptance of CLAS 
enabled us to detect the final state electron and kaon over a broad range 
of momentum transfer $Q^2$ and invariant energy $W$ as shown in 
Fig.~\ref{fig-Q2W}.

CLAS is a large-acceptance detector~\cite{clas} used to detect
multi-particle final states from reactions initiated by either real photon 
or electron beams.  The central element of the detector is a six-coil 
superconducting toroidal magnet that provides a mostly azimuthal magnetic 
field, with a field-free region surrounding the target.  The integrated 
field strength varies from 2~T$\cdot$m for high momentum tracks at the most 
forward angles to about 0.5~T$\cdot$m for tracks beyond 90$^{\circ}$.  The 
field polarity was set to bend negatively charged particles toward the 
electron beam line.  Drift chambers (DC) situated before, within, and 
outside of the magnetic field volume provide charged-particle tracking with 
a momentum resolution of 1-2\% depending upon the polar angle within the 
six independent sectors of the magnet~\cite{dcnim}.  To protect the 
chambers from the charged electromagnetic background emerging from the 
target, a small normal-conducting ``mini-torus'' magnet was located just 
outside the target region.  The integral magnetic field of the mini-torus 
is about 5\% that of the main torus.  

The outer detector packages of CLAS that surround the magnet and drift 
chambers consist of large-volume gas {\v C}erenkov counters (CC) for 
electron identification~\cite{ccnim}, scintillators (SC) for triggering 
and charged particle identification via time-of-flight~\cite{scnim}, and a 
lead-scintillator electromagnetic shower counter (EC) used for electron/pion 
separation as well as neutral particle detection and identification
\cite{ecnim}.  An open trigger for scattered electrons formed from a 
coincidence of the CC and EC signals within a given sector gave event rates 
of about 2~kHz.  The total beam charge was integrated with a Faraday cup to 
an accuracy of better than 1\%.

The offline event reconstruction first identified a viable electron
candidate by matching a negatively charged track in the DC with hits
in the SC, CC, and EC counters. The hits in the CC and EC counters were
required to be within a fiducial region where the efficiency was large
and uniform.  The track was projected to the target vertex to estimate the 
event start time; the estimate was compared to the phase of the accelerator
radio-frequency (RF) signal to determine this time to better than 50~ps
($\sigma$).  In contrast to a straightforward subtraction of the electron
start time from the $K^+$ time, this use of the highly stable RF phase 
improved the hadronic time-of-flight measurements by almost a factor of 
$\sqrt{2}$.

\begin{figure}[htbp]
\vspace{9.2cm}
\includegraphics{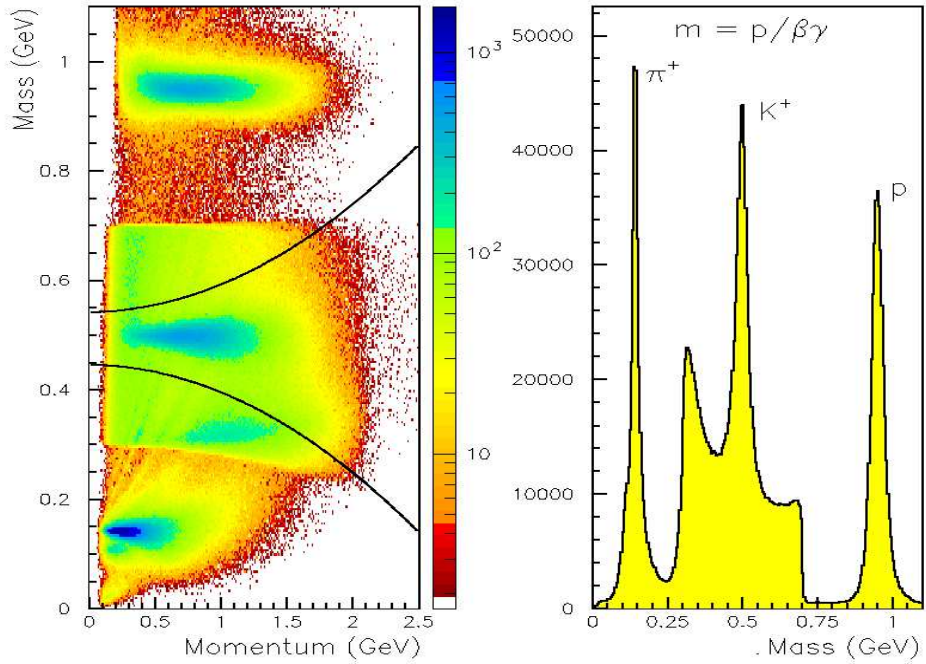}
\caption{\small{(Color online) Reconstructed mass for positively charged 
particles.  The left figure shows the mass plotted against the measured 
momentum. The lines show the mass cuts used to identify kaon candidates. 
A logarithmic yield density scale is employed.  The right figure shows 
the reconstructed mass.  These spectra were made from our kaon-filtered 
data files.}}
\label{kmass}
\end{figure}

A positively charged kaon candidate was identified as an out-bending track
found in the DC that spatially matched to a SC hit that projected back to
the target.  The measured time-of-flight of the track and the fitted path 
length were used to calculate the velocity of the particle.  This velocity 
and the measured momentum were used to calculate the mass of each charged 
hadron. For the data discussed here, the kaon momentum range was between 
300~MeV (software cut) and $\approx$3~GeV (kinematic limit), with a typical 
flight path of 4.5~m.  The measured mass resolution was primarily due to the 
reconstructed time-of-flight resolution, which was 190~ps ($\sigma$) on 
average, and also included contributions from the 1.5\% momentum resolution 
and 0.5~cm path-length uncertainty.  A loose cut around the reconstructed 
kaon mass was used to initially select the kaon candidates (a data filtering 
condition), however a large background of positively charged pions and protons 
still remained.  A momentum-dependent mass cut was used to select the $K^+$ 
events for the final analysis as shown in Fig.~\ref{kmass} for our
filtered data files.

Corrections to the electron and kaon momenta were devised to correct for
reconstruction inaccuracies.  These arise due to relative misalignments of 
the drift chambers in the CLAS magnetic field, as well as for uncertainties 
in the magnetic field map employed during charged track reconstructions.  
These corrections were typically less than 1\%.

\begin{figure}[htbp]
\vspace{9.2cm}
\includegraphics{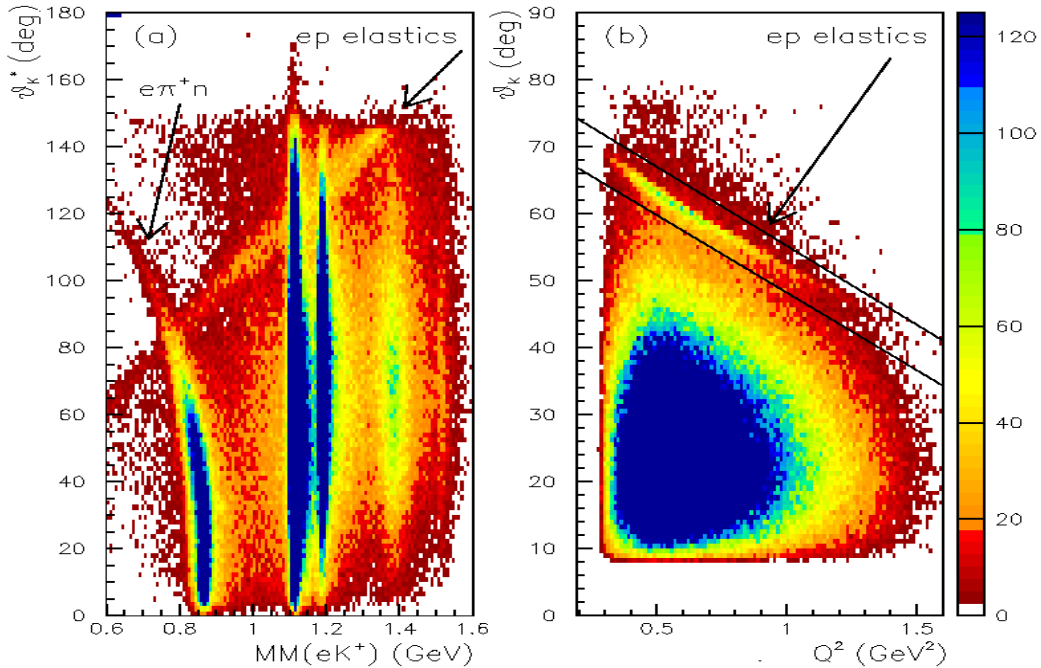}
\caption{\small{(Color online) (a) $\theta_K^*$ versus $p(e,e'K^+)Y$ 
missing mass showing $ep$ elastic events and $e\pi^+ n$ events. The 
vertical bands correspond to ground state $\Lambda(1116)$, $\Sigma^0(1193)$, 
and $\Lambda(1405)/\Sigma^0(1385)$ hyperons. (b) $\theta_K$ versus $Q^2$ 
for $p(e,e'K^+)Y$ showing the $ep$ elastic events and the cut used to 
remove them.}}
\label{fig-elastics}
\end{figure}

\begin{figure}[htbp]
\vspace{6.5cm}
\includegraphics{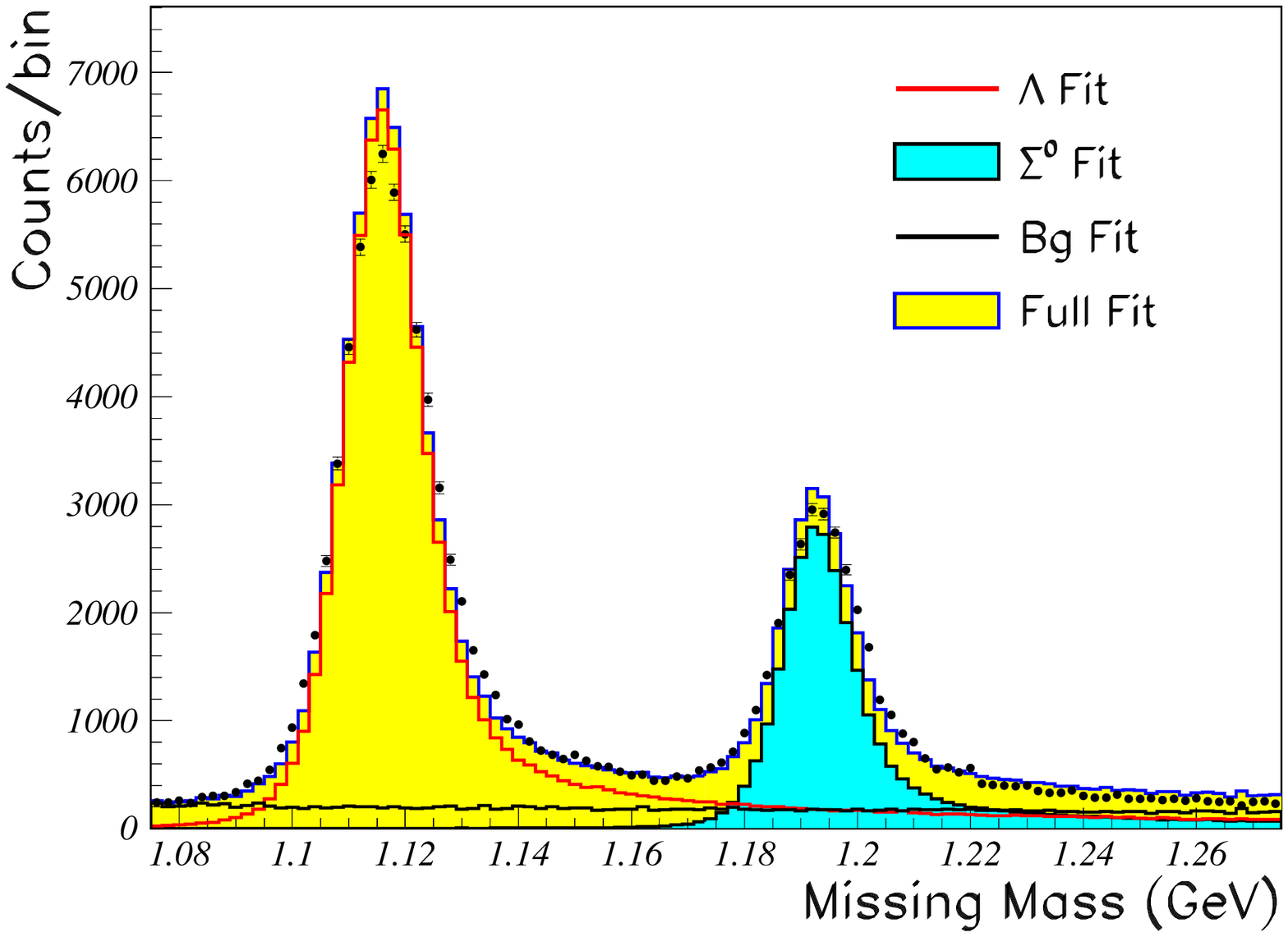}
\includegraphics{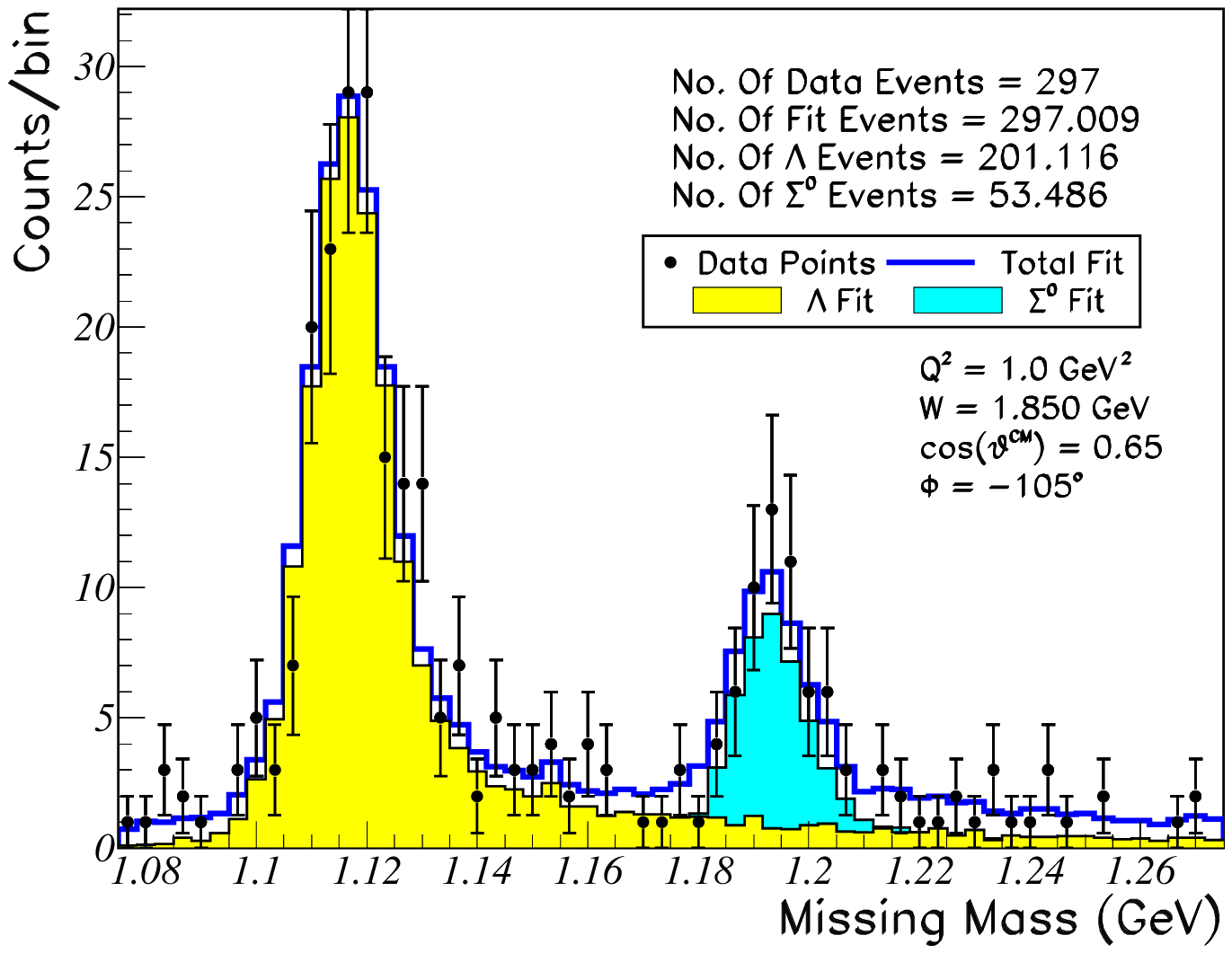}
\caption{\small{(Color online) Signal and background fits from the 2.567~GeV 
data for the $e'K^+$ missing mass spectrum summed over all kinematics (left) 
and for a typical $\cos \theta_K^*$/$\Phi$ bin (right) at $Q^2$=1.0~GeV$^2$ 
and $W$=1.85~GeV to demonstrate the typical fit quality in our data.}}
\label{sigbck}
\end{figure}

Using the four-momenta of the incident electron, scattered electron, and 
$K^+$ candidate, the missing mass, corresponding to the mass of the 
recoiling hyperon, was calculated. The missing-mass distribution contains 
a background that includes a continuum beneath the hyperons that arises 
due to multi-particle final states where the candidate $K^+$ results from a 
misidentified pion or proton, as well as events from $ep$ elastic scattering 
(protons misidentified as kaons) and events from $\pi^+ n$ final states 
(pions misidentified as kaons).  The elastic events are kinematically 
correlated and show up clearly in plots of $\theta_K^*$ (c.m. angle) versus 
missing mass and $\theta_K$ (lab angle) versus $Q^2$ 
(Figs.~\ref{fig-elastics}a and b, respectively). A cut on the elastic band 
in the $\theta_K$ versus $Q^2$ space removes this contribution with a small 
loss of hyperon yield that is later accounted for with our Monte Carlo 
generated acceptance function.  The $\pi^+ n$ events are removed with a 
simple missing-mass cut in which the detected kaon candidates are assumed to 
be pions. Typical missing-mass distributions showing clear $\Lambda(1116)$ 
and $\Sigma^0(1193)$ peaks are shown in Fig.~\ref{sigbck}.

The continuum from the multi-particle final states that lies beneath the
$\Lambda$ and $\Sigma^0$ hyperon peaks in our mass spectra is accounted
for by a fitting process in which identified pions and protons in our
unfiltered data files are assumed to be kaons.  Missing-mass distributions 
are generated for each assumption in each of our different bins in $Q^2$, 
$W$, $\cos \theta_K^*$, and $\Phi$.  The resulting distributions, along with 
template shapes for the $\Lambda$ and $\Sigma^0$ hyperons determined from 
Monte Carlo simulations, are fit to the missing-mass spectra using a 
maximum-log-likelihood method appropriate for the low statistical samples in 
our four-dimensional bins.  The template shapes for the hyperons were produced 
from a simulation that included radiative processes and was matched to the 
detector resolution.  Typical fits of the missing-mass distributions are shown 
in Fig.~\ref{sigbck}.  The final yields in each kinematic bin were determined 
by taking the number of counts determined from the fits that fell within a 
mass window around the $\Lambda$ (1.095 to 1.165~GeV) and $\Sigma^0$ (1.165 
to 2.3~GeV) peaks.  Hyperon events in the tails of the distributions that 
fell outside of our mass windows were accounted for by our acceptance 
correction function.  After removal of all backgrounds, a total of 
$1.4\times10^5$ $K^+\Lambda$ and $6.7\times10^4$ $K^+\Sigma^0$ final state 
events were obtained across the entire kinematic range for the 2.567~GeV 
data set, while $9.7\times10^4$ $K^+\Lambda$ and 
$4.7\times 10^4$ $K^+\Sigma^0$ events were obtained for the 4~GeV data set.

\begin{table}[htbp]
\begin{center}
\begin{tabular} {||c|c|l|c|l||c|l||} \hline \hline
                  & \multicolumn{4} {c||} {2.567 GeV}  & \multicolumn{2} {c||} {4~GeV} \\ \hline\hline
Variable          & N$_{bins}$ & Range & N$_{bins}$ & Range & N$_{bins}$ & Range \\ \hline
$Q^2$             & 2 & 0.5 -- 0.8             &   &                 & 4 & 0.9 -- 1.3  \\
(GeV$^2$)         &   & 0.8 -- 1.3             &   &                 &   & 1.3 -- 1.8  \\
                  &   &                           &   &                 &   & 1.8 -- 2.3  \\
                  &   &                           &   &                 &   & 2.3 -- 2.8  \\ \hline
$W$               & 8 & 1.60 -- 1.70           & 5 & 1.60 -- 1.70~& 8 & 1.6 -- 1.7  \\
(GeV)             &   & 1.70 -- 1.75           &   & 1.70 -- 1.80~&   & 1.7 -- 1.8  \\
                  &   & 1.75 -- 1.80           &   & 1.80 -- 1.90~&   & 1.8 -- 1.9  \\
                  &   & 1.80 -- 1.85           &   & 1.90 -- 2.00~&   & 1.9 -- 2.0  \\
                  &   & 1.85 -- 1.90           &   & 2.00 -- 2.10~&   & 2.0 -- 2.1  \\
                  &   & 1.90 -- 1.95           &   &                 &   & 2.1 -- 2.2  \\
                  &   & 1.95 -- 2.00           &   &                 &   & 2.2 -- 2.3  \\
                  &   & 2.00 -- 2.10           &   &                 &   & 2.3 -- 2.4  \\ \hline
$\cos \theta_K^*$ & 6 & -0.8 -- -0.4           &   &                 & 6 & -0.8 -- -0.4\\
                  &   & -0.4 -- -0.1           &   &                 &   & -0.4 -- -0.1\\
                  &   & -0.1 -- 0.2            &   &                 &   & -0.1 -- 0.2 \\
                  &   & ~0.2 -- 0.5            &   &                 &   & ~0.2 -- 0.5 \\
                  &   & ~0.5 -- 0.8            &   &                 &   & ~0.5 -- 0.8 \\
                  &   & ~0.8 -- 1.0            &   &                 &   & ~0.8 -- 1.0 \\ \hline
$\Phi$            & 8 & $\cos\theta_K^*<-0.1$ &   &                     & 8 & $\cos\theta_K^*<0.2$ \\
                  & 12& $\cos\theta_K^*>-0.1$ &   &                     & 12& $\cos\theta_K^*>0.2$  \\ \hline \hline
\end{tabular}
\end{center}
\caption{\small{The number of bins and the bin sizes employed for the 
2.567 and 4~GeV analyses in this work.  For the 2.567~GeV data analysis
two different binning choices were made for $W$ for each bin in $Q^2$, 
$\cos \theta_K^*$, and $\Phi$.}}
\label{tab:bins}
\end{table}

The data were binned in a four-dimensional space of the independent kinematic
variables, $Q^2$, $W$, $\cos \theta_K^*$, and $\Phi$. Table~\ref{tab:bins} 
lists the kinematic bin definitions used in the analysis. The data at 
2.567~GeV consisted of data sets taken with two settings of the main CLAS 
torus field that were combined together.  As mentioned above, our 4~GeV
data set consisted of data acquired at beam energies of 4.056 and 4.247~GeV 
at the same torus field setting.  When combining the data at 4.056 and 
4.247~GeV, we evolve the cross sections at 4.247~GeV to the bin center of 
the 4.056~GeV data using our model of the cross section.  As the two data 
sets are close in energy and $\epsilon$, there is little systematic 
uncertainty involved in this procedure.

In this analysis the number of $\Phi$ bins was 8 for the two backward-most 
bins of $\theta_K^*$ and 12 for the four forward-most $\theta_K^*$ bins.  
The larger number of $\Phi$ bins for forward and central $\theta_K^*$ 
increased the reliability of the $\Phi$-fits in the presence of the forward 
beam ``hole'' of the spectrometer, an area of depleted acceptance 
corresponding to tracks with small laboratory angles. Bins significantly 
overlapping this forward hole were excluded from our analysis.  

The average differential cross section for each hyperon final state in each 
bin $i$ was computed using the form:

\begin{equation}
\frac{d \sigma_v^i}{d\Omega_K^*} = \frac{1}{\Gamma_v} 
\left(\frac{1}{\Delta Q^2 \Delta W \Delta\cos\theta^*_K \Delta\Phi}\right)
\left(\frac{N_i}{R_i A_i}\right) \left(\frac{1}{N_0 (N_A \rho t/A_w)}\right),
\end{equation}

\noindent
where $N_i$ is the hyperon yield, $A_i$ is the acceptance, $N_0$ is the 
live-time corrected incident electron flux, $R_i$ is the radiative correction 
factor, $N_A$ is Avogadro's number, $\rho$ is the target density
($\langle \rho \rangle$ = 0.072~g/cm$^3$), $t$ is the target length, and $A_w$ 
is the atomic weight of hydrogen (1.00794~g/mol).  The product
$\Delta Q^2 \Delta W \Delta\cos\theta^*_K \Delta\Phi$ represents the
volume of the $i^{th}$ bin corrected for kinematic limits.  

The geometric acceptance and reconstruction efficiencies were calculated
using a standard model of the CLAS detector based upon a GEANT simulation
\cite{geant}.  To reduce the model dependence of the computed CLAS
acceptance, it is important to match the distributions of accepted data and 
Monte Carlo events as a function of the relevant kinematic variables $Q^2$, 
$W$, $\cos \theta_K^*$, and $\Phi$ for both the $K^+\Lambda$ and $K^+\Sigma^0$ 
final states.  This match must be ensured at all beam energies and torus field 
settings employed in the analysis.  

\begin{figure}[htpb]
\vspace{8.0cm}
\includegraphics{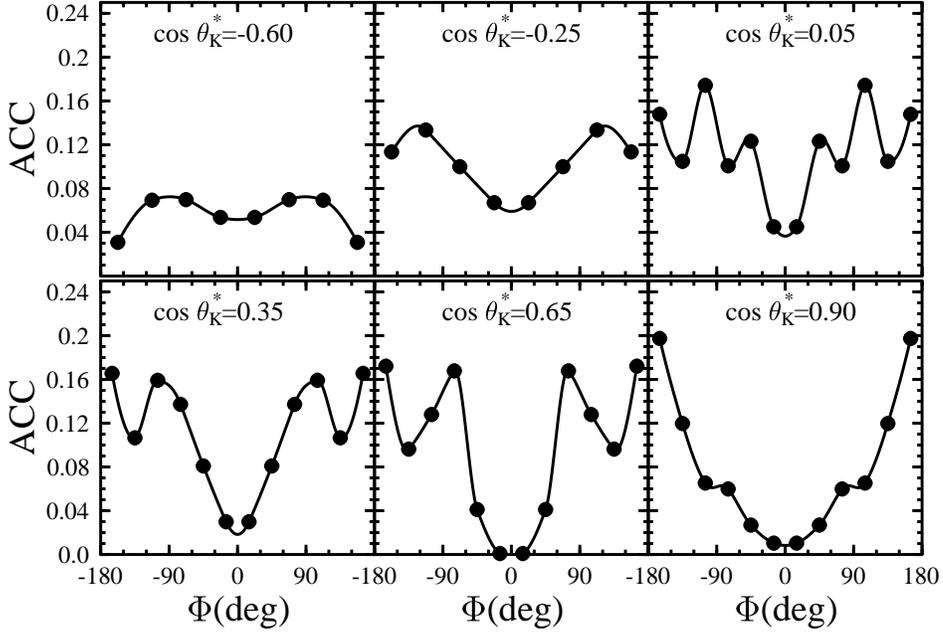}
\caption{\small{Distribution of the computed $K^+\Lambda$ acceptance for
CLAS as a function of $\cos \theta_K^*$ and $\Phi$ for the $W$=1.85~GeV 
and $Q^2$=0.65~GeV$^2$ bin. The depleted area near $\Phi$ of $0^\circ$ for 
forward angles is the forward ``hole'' in CLAS due to the beam pipe.  The
number of bins in $\Phi$ is different for forward and backward 
$\cos \theta_K^*$.  The statistical error bars from the Monte Carlo are 
smaller than the symbol size on this plot.  The curves on each plot serve 
only to guide the eye.}}
\label{accep}
\end{figure}

A variety of reaction models (see e.g. Ref.~\cite{bennhold}) were 
employed as input event generators for the simulated events.  Because none 
of the models agreed particularly well with our $K^+\Lambda$ or $K^+\Sigma^0$
data, we developed our own models that were able to match the data 
reasonably well over our full kinematic phase space. We developed two 
different ad hoc models for the $K^+\Lambda$ analysis that both represented 
our data equally well, although with slightly different dependencies on 
$Q^2$, $W$, and $\cos \theta_K^*$.  One ad hoc model was developed for the
$K^+\Sigma^0$ analysis. These models were used as input to determine our
detector acceptance function, radiative corrections, and bin-centering 
correction factors.  In addition we developed an event generator based on
fits to our $K^+\Lambda$ data.  Differences between our two ad hoc models and 
our data-fitted model for $K^+\Lambda$ were used to estimate the model 
dependence of our results.  Further details are included in 
Section~\ref{systematics}.  Figure~\ref{accep} shows the dependence of the 
$K^+\Lambda$ acceptance upon $\theta_K^*$ and $\Phi$ for a bin in $Q^2$ and 
$W$.  Typical acceptances of CLAS for the $e'K^+$ final state were in the 
range of 1 to 30\% depending on kinematics.  Note the strong variation in 
acceptance as a function of both $\cos \theta_K^*$ and $\Phi$ due to the
geometry of CLAS.
 
In the detector simulation, particles generated at the target were
propagated through the CLAS magnetic field and were permitted to interact 
with materials and to undergo decay.  These tracks then generated simulated 
detector hits.  Hits corresponding to the known dead areas in the detectors 
were removed, and the hits were smeared according to the known detector 
resolution effects.  These simulated events were passed through the same 
analysis chain as the real data.  Geometrical fiducial cuts were applied to 
both the data and simulated events to eliminate areas of inefficient detector 
response or where the response was not well modeled.  These areas were 
typically within a few degrees of the magnet coils and near the edges of the 
{\v C}erenkov detector.

The radiative correction for each kinematic bin was computed from the ratio
of the model cross sections with and without radiative effects.  We used 
two very different methods to compute this correction factor.  

The first method used an acceptance-rejection technique, where events were 
generated uniformly in $W$, $Q^2$, $\cos\theta^*_K$, and $\Phi$, with a weight 
determined via the cross section model of Ref.~\cite{bennhold}.  The energies 
of externally radiated photons from the incident electron, and from the emitted 
electron and kaon in the region of the target proton, were generated according 
to the formulas of Mo and Tsai~\cite{motsai}.  The weight of the event was
adjusted to account for hard and soft internal radiative effects, and the 
post-radiation kinematic variables were calculated to identify the bin into 
which the event fell.  While lacking in computational efficiency, this method 
benefited from being able to compute cross sections as well as the expected event 
distributions, both with and without radiative effects, thus providing a consistent 
event sample to use for the acceptance studies.  The second method used the same 
formula for calculating the radiated cross section from the assumed non-radiated 
model.  However, it integrated the resulting six-dimensional cross section over the 
two unseen dimensions that corresponded to a radiated photon from either the initial 
or scattered electron.  The ratio of the integrated radiated cross section (now 
corresponding to a four-dimensional space) to the unradiated cross section yielded 
the radiative correction factor.  We chose to use this second method because of its 
superior computational speed.  It was extensively checked versus the 
``{\tt EXCLURAD}'' code of Afanasev~\cite{afanasev}.  Differences between the two 
methods allowed us to estimate the size of any residual uncertainty in the radiative 
correction procedure (see Section~\ref{systematics}).

In order to do a full separation into four structure functions, we can fit 
our full set of data including the differential cross sections from both 
beam energies with a function of the form $f(\Phi,\epsilon)$; the fitted 
parameters being the values of $\sigma_T$, $\sigma_L$, $\sigma_{TT}$, and 
$\sigma_{LT}$ at some fixed point in $Q^2$, $W$, and $\cos \theta_K^*$.  
Alternatively, we can extract $\sigma_U$ in each $Q^2$, $W$, and 
$\cos \theta_K^*$ bin from a $\Phi$ fit for each beam energy separately, and 
then do a linear $\epsilon$ fit to separately extract $\sigma_T$ and 
$\sigma_L$; this being the well-known Rosenbluth separation technique.
In either case, in order to do these fits, we must first define the cross 
sections at a specific {\em fixed point} within the bin, and not merely as 
an average over a given bin volume.  This is especially true when the bins 
are large and event-weighted average values of kinematic variables can be 
different for different $\Phi$ bins.  Using an integration over our model 
cross section, we calculate the cross section at the fixed point given its 
average over the bin volume.  This correction is referred to as a 
``bin-centering correction'', or more accurately, as a ``finite bin size 
correction''.  By using our model event generators, we simply calculated 
the ratio of the cross section evaluated at the assigned bin center to the 
average cross section integrated over the bin to obtain the finite bin
size correction factor.  Systematic uncertainties associated with these
corrections were extensively studied (see Section~\ref{systematics}).

\begin{figure}[htpb]
\vspace{10.0cm}
\includegraphics{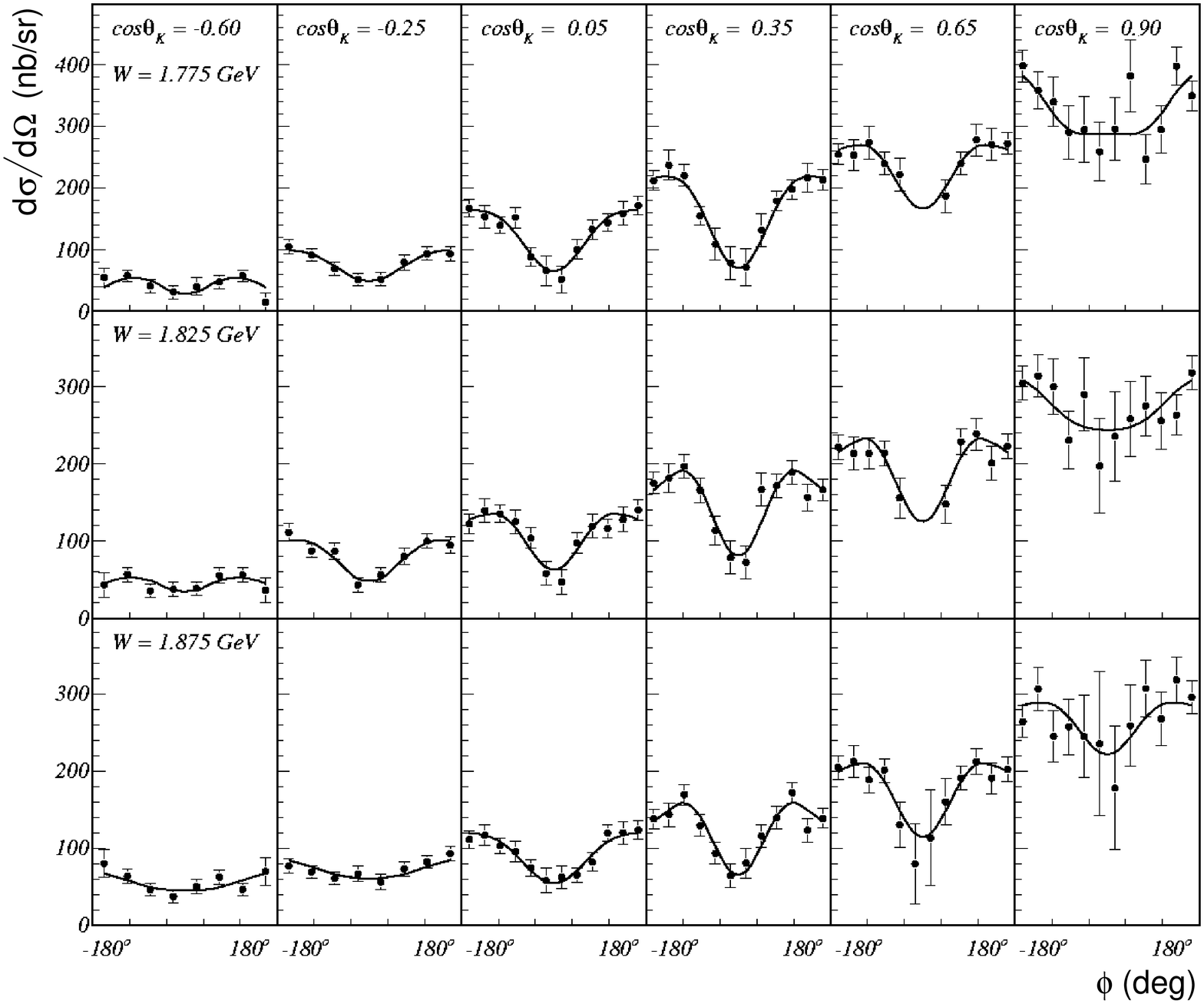}
\caption{\small{$\Phi$-dependent differential cross sections (nb/sr) and 
fits for $K^+\Lambda$ events from our 2.567~GeV data at $Q^2$=0.65~GeV$^2$ 
for each of our six $\cos \theta_K^*$ bins (labeled at the top of each
column) for three different $W$ bins (labeled on the left of each row).
The curves represent fits to the $\Phi$-dependent differential cross
sections.}}
\label{phifits}
\end{figure}

In Fig.~\ref{phifits} we show a sample of the $\Phi$-dependent differential
cross sections for the $K^+\Lambda$ final state at 
representative kinematic points.  The different shapes of the differential 
cross sections vs. $\Phi$ in each of our bins in $Q^2$, $W$, and 
$\cos \theta_K^*$ reflect differences of the interference terms, 
$\sigma_{TT}$ and $\sigma_{LT}$.  The differences in scale reflect the 
differences in $\sigma_U$.  

\section{Structure Function Extraction}
\label{sfsep}

The full set of differential cross sections $d\sigma_v/d\Omega_K^*$ 
included in this work for each hyperon final state consists of 156 bins 
in $Q^2$, $W$, and $\cos \theta_K^*$ for the 2.567~GeV data (accounting 
for both $W$ binning scenarios given in Table~\ref{tab:bins}) and 192 bins 
for the 4~GeV data.  This amounts to 1664 data points in $Q^2$, $W$, 
$\cos \theta_K^*$, and $\Phi$ for the 2.567~GeV data and 1920 data points 
at 4~GeV for each hyperon final state.  In this section we provide details 
regarding the structure function extraction.  In Section~\ref{3struct} we 
focus on the separation of the structure functions $\sigma_U$, $\sigma_{TT}$, 
and $\sigma_{LT}$.  In Section~\ref{ltsect} we present the extraction 
of $\sigma_T$ and $\sigma_L$ separately from $\sigma_U$ using a Rosenbluth 
fit and also a simultaneous $\epsilon - \Phi$ fit of our data at 2.567 and 
4~GeV.

\subsection{Extraction of $\sigma_U$, $\sigma_{TT}$, and $\sigma_{LT}$}
\label{3struct}

The differential cross sections plotted in Fig.~\ref{phifits} are
actually the mean values within the finite size of the $\Phi$ bins
and therefore do not necessarily reflect the value at the bin center.
Thus directly fitting these data with eq.(\ref{xsect}) to extract the
structure functions $\sigma_U = \sigma_T + \epsilon \sigma_L$,
$\sigma_{TT}$, and $\sigma_{LT}$ would be inappropriate.  Integrating 
eq.(\ref{xsect}) over the finite bin size, $\Delta\Phi=\Phi_u-\Phi_l$, 
where $\Phi_u$ and $\Phi_l$ are the upper and lower limits of the bin, 
respectively, gives:

\begin{eqnarray}
\bar{\sigma}_0&\equiv &\frac{1}{\Delta\Phi}\int_{\Phi_l}^{\Phi_u}
\left(\sigma_U + \epsilon\sigma_{TT}\cos 2\Phi + 
\sqrt{\epsilon(\epsilon+1)}\sigma_{LT}\cos\Phi\right)d\Phi \nonumber \\
&=&\frac{1}{\Delta\Phi}\left(\sigma_U\Delta\Phi +
\frac{\epsilon}{2}\sigma_{TT}\left(\sin 2\Phi_u-\sin 2\Phi_l\right) +
\sqrt{\epsilon(\epsilon+1)}\sigma_{LT}\left(\sin\Phi_u-\sin\Phi_l\right) 
\right).
\label{eq-csecfit}
\end{eqnarray}

\noindent
$\bar{\sigma}_0$ now represents the value of the measured bin-averaged 
cross section in a given $\Phi$ bin and fitting the data with 
eq.(\ref{eq-csecfit}) yields the separated structure functions for a
given bin in $Q^2$, $W$, and $\cos\theta_K^*$.  The ``$\epsilon$''
pre-factors were evaluated at the bin center and divided out.

Prior to the $\Phi$ fits, the statistical uncertainty of each cross 
section bin was combined linearly with that portion of the systematic 
uncertainty arising from the yield extraction procedures (see 
Section~\ref{systematics} for details).  A few points were removed from 
the fits based upon their low acceptance in CLAS, in order to prevent bins 
with a very small acceptance from distorting the extracted structure 
functions. A point was rejected if its acceptance at 2.567~GeV (4~GeV) was 
less than 2.0\% (1.0\%) or less than 10.0\% (5.0\%) of the average 
acceptance over all bins at the same $Q^2$, $W$, and $\cos\theta_K^*$.

In reporting the final results from our $\Phi$ fits, several
$\cos \theta_K^*$ bins have been discarded.  In general, these bins were
near the edge of our kinematic acceptance and had limited $\Phi$ coverage.
In addition, the statistical uncertainties were large on the points in these 
bins that survived the acceptance criteria described above.  Typically, the 
missing points were near $\Phi=\pm \pi$ or $\Phi=0$ -- exactly where points 
are needed to constrain the interference structure functions.  The resulting
$\Phi$ fits for these bins had $\chi^2/\nu$ values, where $\nu$ represents
the number of degrees of freedom, that were uniformly too small considering 
the expected $\chi^2/\nu$ distributions. In other words, a three-parameter 
fit of these bins had too many parameters, given the low number of data 
points and the large uncertainties, to give unambiguous solutions for the
structure functions.  We also examined the $\chi^2/\nu$ distributions for 
the remaining fits and found that they were well represented by their 
expected probability distributions, which instills confidence in the 
quality of the data, the assigned uncertainties, and the fits.

\subsection{Separation of $\sigma_T$ and $\sigma_L$}
\label{ltsect}

The extraction procedure detailed in Section~\ref{3struct} yielded the
bin-centered structure functions $\sigma_U$, $\sigma_{TT}$, and $\sigma_{LT}$.  
To further separate $\sigma_U$ into its component parts, $\sigma_T$ and 
$\sigma_L$, we have two options.  The first is the standard Rosenbluth 
separation technique, in which $\sigma_U$ is determined for two different 
beam energies (or different $\epsilon$ values) but for the same point in 
$Q^2$, $W$, and $\cos \theta_K^*$, and fit as a linear function of $\epsilon$.  
An alternative approach is to simultaneously fit the data from the two 
energies as a function of $\epsilon$ and $\Phi$, this time explicitly 
replacing $\sigma_U$ in eq.(\ref{eq-csecfit}) with $\sigma_T+\epsilon\sigma_L$.
This method has the advantage of constraining the individual parameters, 
$\sigma_T$, $\sigma_L$, $\sigma_{TT}$, and $\sigma_{LT}$ to have the same 
value for the two different beam energies, as they must since they are 
explicit functions of $Q^2$, $W$, and $\theta_K^*$ only.  This approach 
represents an important systematic check as the forward beam hole of CLAS 
affects the acceptance function differently at 2.567~GeV relative to 4.056 
and 4.247~GeV.

\begin{figure}[htpb]
\vspace{11.0cm}
\includegraphics{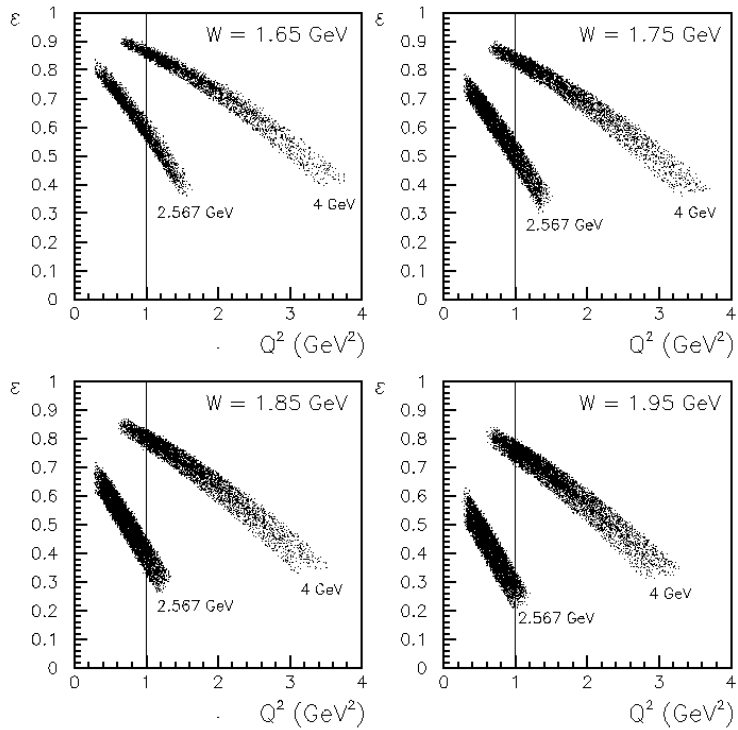}
\caption{\small{CLAS acceptance of the scattered electron in terms of
$\epsilon$ vs. $Q^2$ at 2.567~GeV (lower bands) and 4~GeV (upper bands)
for four 100-MeV $W$ bins centered from 1.65 to 1.95~GeV.  The vertical 
line at $Q^2$=1.0~GeV$^2$ marks where we have performed the separation 
of $\sigma_T$ and $\sigma_L$.}}
\label{overlap}
\end{figure}

The separation of the structure functions $\sigma_T$ and $\sigma_L$ can
only be performed in ($Q^2$, $W$, $\cos \theta_K^*$) bins where the 
2.567~GeV and 4~GeV data overlap.  Figure~\ref{overlap} shows plots of 
$\epsilon$ vs. $Q^2$ for four different 100-MeV wide bins in $W$ from 1.65~GeV 
to 1.95~GeV, and highlights the kinematic coverage of CLAS.  The cut-off 
at low $Q^2$ is due to the minimum $\theta_e$ detectable by CLAS and the low
$\epsilon$ cut-off is due to the maximum $\theta_e$ detectable by the 
{\v C}erenkov detectors.  For this analysis, the data overlap only for a
rather narrow $Q^2$ region at about 1~GeV$^2$.  We have performed a separation 
of $\sigma_T$ and $\sigma_L$ for $Q^2$=1.0~GeV$^2$ for the $K^+\Lambda$ final 
state for $W$=1.65, 1.75, 1.85, and 1.95~GeV, and at values of $W$=1.75, 1.85, 
and 1.95~GeV for the $K^+\Sigma^0$ final state.  In this two beam energy 
separation, we have typical differences in $\epsilon$ of about 0.4.  Of central 
importance in this analysis is the fact that this separation is performed for 
the {\em first time} away from the condition of parallel kinematics (i.e. 
$\theta_K^*$=0$^{\circ}$ or along the virtual photon direction).

Before the separation of $\sigma_U$ could proceed, we first had to
account for the binning differences between the 2.567~GeV and 4~GeV
data sets.  Due to consideration of statistics in the two separate
data sets, the 2.567~GeV data were sorted in 50-MeV wide $W$ bins for 
the extraction of $\sigma_U$, $\sigma_{TT}$, and $\sigma_{LT}$, while the 
data at 4~GeV were sorted in 100-MeV wide $W$ bins (see 
Table~\ref{tab:bins}).  In order to perform either the Rosenbluth fit or 
the simultaneous $\epsilon-\Phi$ fit, the 2.567~GeV data had to be resorted 
into $W$ bins that were 100-MeV wide.  In computing the cross sections for 
the 100-MeV wide $W$ bins at 2.567~GeV, the hyperon yield fits were redone 
and all other factors associated with computing the cross section were 
recalculated using Monte Carlo based on the 100-MeV wide bin.  

\begin{figure}[htpb]
\vspace{10.5cm}
\includegraphics{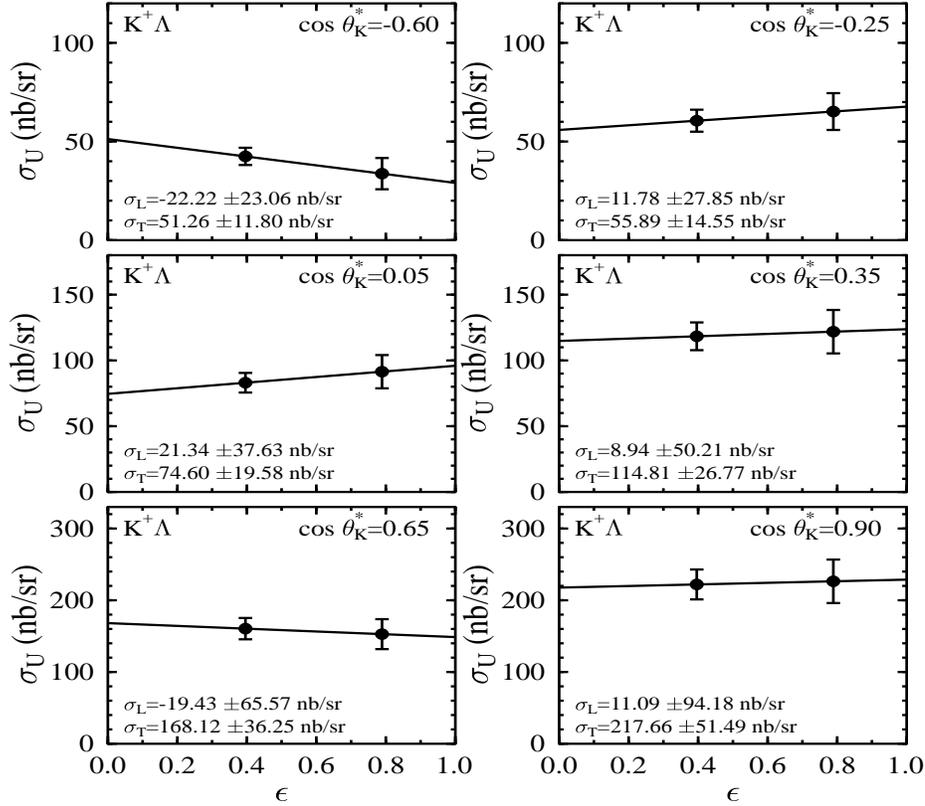}
\caption{\small{Representative Rosenbluth separation plots of $\sigma_U$ 
(nb/sr) vs. $\epsilon$ for our $K^+\Lambda$ data at $Q^2$=1.0~GeV$^2$ and 
$W$=1.85~GeV for our six $\cos \theta_K^*$ bins.  The lines represent the 
fits that determine the slope parameter ($\sigma_L$) and the intercept 
parameter ($\sigma_T$), which are printed on each plot.  The error bars 
on the data points and the errors listed for $\sigma_L$ and $\sigma_T$ on
each plot represent the combined statistical and systematic uncertainties.}}
\label{rosen}
\end{figure}

The Rosenbluth extraction procedure is a standard technique to separate 
$\sigma_T$ and $\sigma_L$.  The error bars on these structure functions 
result from the statistical and systematic uncertainties on the two 
$\sigma_U$ cross section points used in the extraction.  With only two data 
points, the slope parameter ($\sigma_L$) and the intercept parameter 
($\sigma_T$) along with their associated uncertainties, can be computed 
analytically.  Figure~\ref{rosen} shows a representative plot of the 
$\sigma_U$ cross sections for the $K^+\Lambda$ final state at $W$=1.85~GeV 
for each of our six $\cos \theta_K^*$ bins.  This plot also serves to
indicate the typical $\epsilon$ values and spread for the two data sets.
The data points at 2.567~GeV and 4~GeV have each been evolved using our 
ad hoc models to the point $Q^2$=1.0~GeV$^2$.  The analysis employs the
highest $Q^2$ bin from our 2.567~GeV data set and the lowest $Q^2$ bin
from our 4~GeV data set.

\begin{figure}[htpb]
\vspace{11.3cm}
\includegraphics{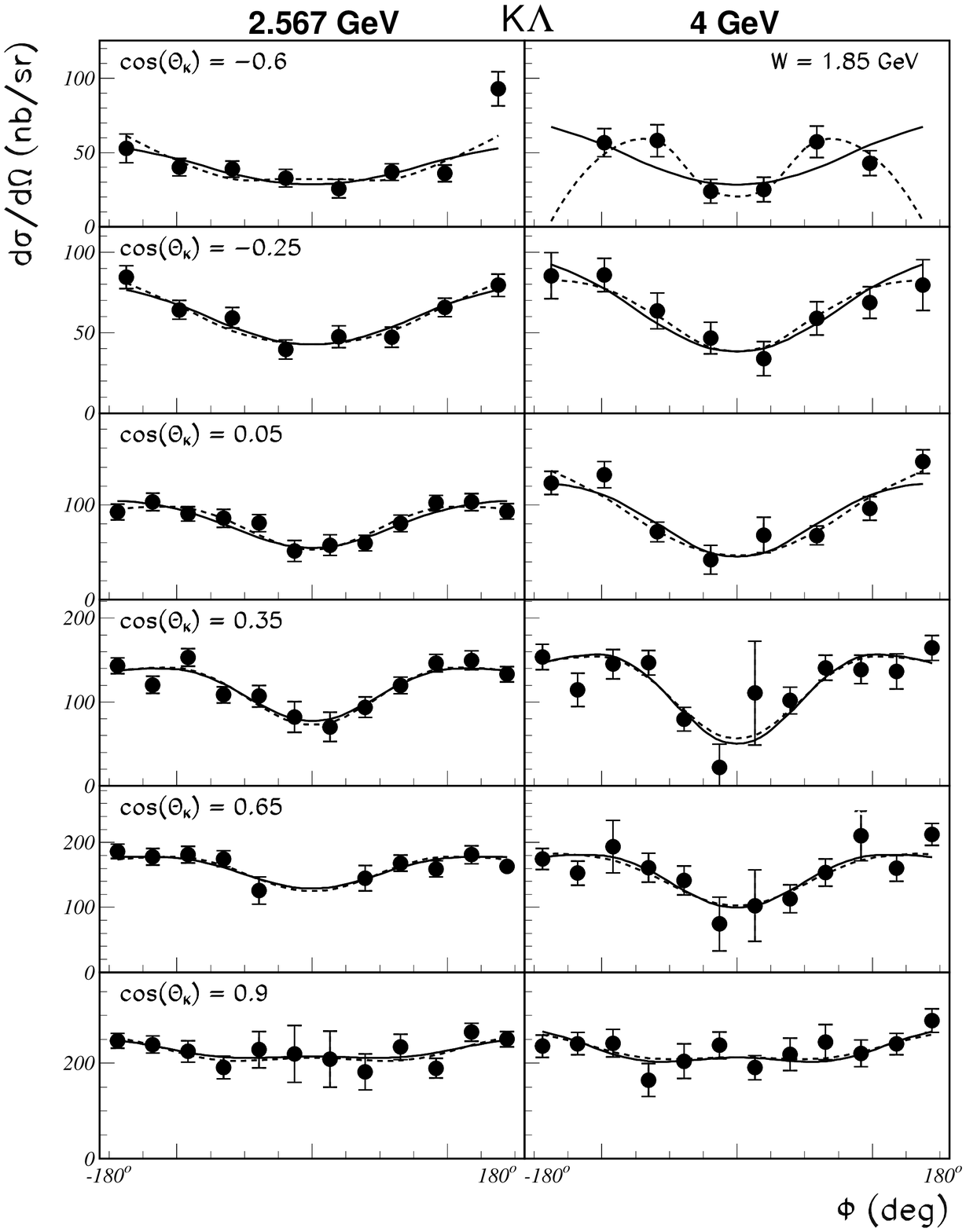}
\includegraphics{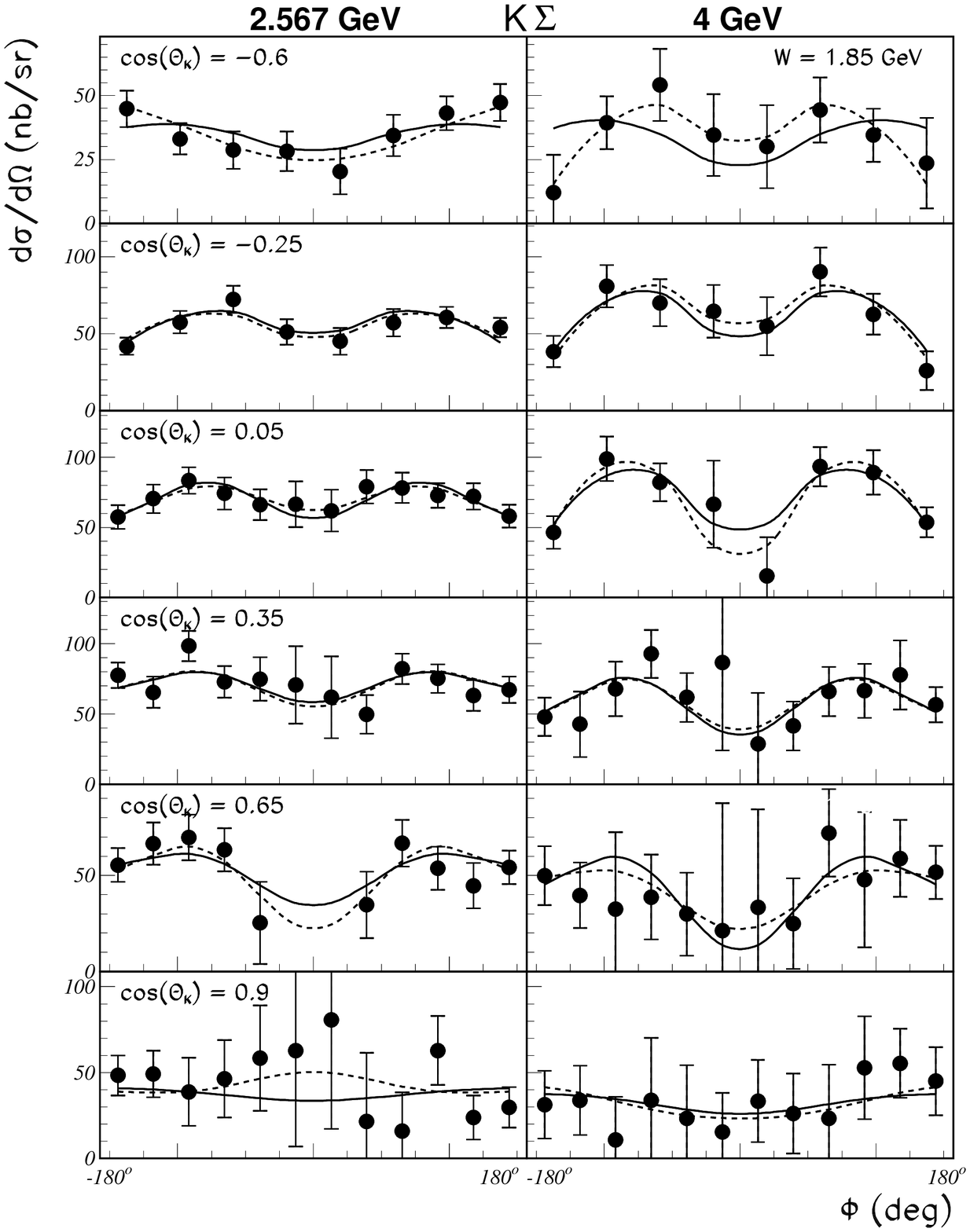}
\caption{\small{Comparison of $\Phi$ fits to the differential cross 
sections performed for two different algorithms.  In the first approach
(dashed curves) the cross sections at 2.567 and 4~GeV are fit separately.
In the second approach (solid curves) the two different beam energy data
sets are fit simultaneously.  These plots are for kinematics with 
$Q^2$=1.0~GeV$^2$ and $W$=1.85~GeV for the $K^+\Lambda$ (left columns) and 
$K^+\Sigma^0$ (right columns) final states at different values of 
$\cos \theta_K^*$ as indicated.}}
\label{phiepsilon}
\end{figure}

An example of the comparison between the separate $\Phi$ fits for the 
2.567~GeV and 4~GeV data and the simultaneous fit for both energies 
for the $K^+\Lambda$ and $K^+\Sigma^0$ reactions is shown in 
Fig.~\ref{phiepsilon} for $Q^2$=1.0~GeV$^2$ and $W$=1.85~GeV.  The differences 
between the differential cross sections for the 2.567 and 4~GeV data for a 
given bin (see Fig.~\ref{phiepsilon}) in $\cos \theta_K^*$ are due not only to 
the beam energy dependent $\epsilon$ pre-factors (defined in eq.(\ref{xsect})), 
but also to the different systematic variations associated with the acceptance 
functions of CLAS at these energies. Of importance is that the simultaneous 
fits differ from the single beam energy fits only where the single beam energy 
fits have large error bars (e.g. 4~GeV back-angle $K^+\Sigma^0$ bin) or are 
missing $\Phi$ points due to our minimum acceptance cut-off criteria (e.g. 
4~GeV back-angle $K^+\Lambda$ bin).  In these cases the simultaneous fit 
procedure leads to extracted structure function with reduced uncertainties 
compared to the single beam energy fits.

\section{Systematic Uncertainties}
\label{systematics}

\subsection{Overview}

To obtain a virtual photoabsorption cross section, we reconstruct events
with an outgoing electron and $K^+$, and then fit the missing-mass spectra
for each of our bins in $Q^2$, $W$, $\cos \theta_K^*$, and $\Phi$ to obtain 
the yields for the reactions $K^+\Lambda$ and $K^+\Sigma^0$.  The yields are 
corrected for the acceptance function of CLAS, radiative corrections, and 
finite bin size effects.  Finally, we divide by the virtual photon flux 
factor at the bin center, the bin volume corrected for kinematic limits, 
and the beam-target luminosity to yield the cross section.  Each of these 
procedures is subject to systematic uncertainty.  We typically estimate the 
size of systematic uncertainties by repeating a procedure in a slightly 
different way (e.g. by varying a cut parameter within reasonable limits or 
by employing a slightly different algorithm) and noting how the results 
change.  The difference in the results is then used as a measure of the 
systematic uncertainty.  In this section we describe our main sources of 
systematics.  

With respect to their effect on our results, there are three types of 
systematic effects: uncertainties that affect the yield extraction in a 
seemingly random fashion where the systematic uncertainty is proportional 
to the size of the statistical uncertainty, ``scaling'' uncertainties that 
affect both the cross sections and structure functions by a simple scale 
factor, and $\Phi$-dependent uncertainties such as using an event 
generator with a $\Phi$-dependence that does not quite match the data.

These uncertainties are handled in different ways.  Because the size of the 
``yield extraction'' uncertainty depends on the size of the statistical 
uncertainty, we take the random-type systematic uncertainties into account 
by enlarging the statistical uncertainty before the $\Phi$ fit to extract 
the structure functions.  These fractional systematic scaling uncertainties 
are multiplied by the value of the cross section or the structure function 
in question to get the absolute uncertainty.  The remaining uncertainties, 
which can in general have $\Phi$ or other kinematic dependencies, are 
estimated by extracting the $\Phi$-dependent structure functions for two, 
similar procedures.  This method gives an absolute estimate for a structure 
function uncertainty.

The primary sources of systematic uncertainty for this experiment came from
the Monte Carlo model dependence (acceptance, radiative corrections, finite 
bin size corrections), detector efficiency, and the yield extraction. With 
the very large acceptance and a four-dimensional kinematic space, systematic 
uncertainties were studied on a bin-by-bin basis.  Table~\ref{syserror} 
summarizes our estimates of the average systematic uncertainties on the
differential cross sections associated with various effects.  The different 
types of systematic uncertainties mentioned above are referred to as 
``stat.'', ``scaling'', and ``$\Phi$-dep.'' in the column labeled ``Type'' 
in Table~\ref{syserror}.

\begin{table}[htbp]
\begin{center}
\begin{tabular} {||l|c|c|c||} \hline \hline
~~~~~~~~{\bf Category}         & {\bf Type} & {\bf Sources}                  & {\bf Avg. Size} \\ \hline
{\bf i). Event Reconstruction} & scaling    & Trigger+tracking efficiency    & 1\%  \\ 
                               & $\Phi$-dep.& Electron fiducial cut          & 3.6\% \\
                               & $\Phi$-dep.& Kaon fiducial cut              & 4.1\% \\
                               & scaling    & Electron PID efficiency        & 1.5\% \\ 
                               & scaling    & Kaon PID efficiency            & 1.0\% \\
                               & scaling    & CC efficiency                  & 2-5\% \\ 
                               & scaling    & CLAS forward angle response    & 1-10\% \\ \hline
{\bf ii). Yield Extraction}    & stat.      & Signal templates               & 25\% $\cdot$ stat \\ 
                               &            & Background removal             & \\ \hline
{\bf iii). Model Dependence}   & $\Phi$-dep.& Acceptance calculations,~~~~~~~~~~~~ & 8.0\% \\ 
                               &            & Radiative corrections, \&~~~~~ & \\ 
                               &            & ~~~~~~Finite bin size corrections   & \\ \hline
{\bf iv). RadCorr: Theory}     & scaling    & Integration vs. {\tt EXCLURAD}& 3.4\% \\ \hline
{\bf v). Photon Flux-factor}   & scaling    & Momentum and angle uncertainties     & 3.0\% \\ \hline
{\bf vi). Luminosity}          & scaling    & Live time correction          & 0.5\% \\ 
                               & scaling    & Faraday cup accuracy          & 1.0\% \\ 
                               & scaling    & Hydrogen target thickness     & 3.0\%   \\ \hline \hline
\end{tabular}
\caption{\small{Sources, types, and average sizes of systematic uncertainties on 
the differential cross sections.}}
\label{syserror}
\end{center}
\end{table}

The main categories of systematic uncertainty in this analysis include:
(i). event reconstruction efficiency ($\delta \sigma_{ER}$), (ii). yield 
extraction ($\delta \sigma_{YE}$), (iii). model dependence 
($\delta \sigma_{MOD}$), (iv). radiative correction theory uncertainty 
($\delta \sigma_{RC}$), (v). virtual photon flux ($\delta \sigma_{flux}$), 
and (vi). luminosity ($\delta \sigma_{\cal{L}}$).  Each of these categories 
is explained in more detail in the next section.  The final systematic 
uncertainty assignment to our extracted structure functions is explained 
fully in Section~\ref{finalerr}.  While the yield extraction systematic 
uncertainty, as explained below, is treated as an effective increase in 
our statistical uncertainty, the remaining systematic sources are added 
in quadrature to arrive at our final uncertainty assignment as:

\begin{equation}
\label{toterr}
\delta \sigma_{sys} = \left( \delta \sigma_{ER}^2 + \delta \sigma_{MOD}^2 
+ \delta \sigma_{RC}^2 + \delta \sigma_{flux}^2 + \delta \sigma_{\cal{L}}^2 
\right)^{1/2}.
\end{equation}

\subsection{Systematic Uncertainty Categories}

(i). Event reconstruction efficiency:  This efficiency is a convolution of 
the charged particle track reconstruction efficiency in CLAS, the efficiency 
of our particle identification algorithms for the electron and kaon, and the 
triggering efficiency.  The CLAS trigger and tracking efficiency (which are
essentially 100\%) have been studied and represent small contributions to our 
systematics.  The definitions of the electron and kaon fiducial cut boundaries 
(which cut $\sim$10\% of our event sample) and the particle identification 
(PID) cuts (which cut $\sim$15\% of our event sample) have been varied within 
reasonable limits to determine their effect on the resulting cross sections.  
Each of these systematic sources is relatively small, and overall they 
contribute about 6\% to our total systematic uncertainty.  Each source is 
independent of the kinematics of the final state particles.

There are two additional sources of systematic uncertainty in this category 
that have a value that depends on the final state kinematics.  One of these 
sources accounts for unphysical small-scale fluctuations in the measured 
efficiency function of the {\v C}erenkov detector (which has typical 
efficiencies of 95\% in our fiducial region), which were much more apparent 
at forward angles in CLAS.  This ``CC efficiency'' systematic has been 
assigned as 5\% for the lowest $Q^2$ bin for each beam energy data set
($Q^2$=0.65~GeV$^2$ at 2.567~GeV and $Q^2$=1.0~GeV$^2$ at 4~GeV) where the
electrons populate smaller angles in CLAS.  For all other bins the systematic 
has been assigned to be 2\%.  The other kinematics-dependent systematic 
arises due to the fact that our $\sigma_L/\sigma_T$ extraction was performed 
in a region with only modest kinematic overlap between the 2.567~GeV and 
4~GeV data sets, namely $Q^2$=1.0~GeV$^2$ (see Fig.~\ref{overlap}).  The 
electrons in the 2.567~GeV data sample populate a well understood and well 
modeled portion of the CLAS detector.  However the electron sample in the 
4~GeV data populate the forward-most portion of CLAS where the acceptance 
is difficult to model due to the forward beam hole of CLAS and where the 
{\v C}erenkov efficiency varies rapidly due to the mirror geometry of the 
detector~\cite{ccnim}.  With the 4~GeV data at $Q^2$=1.0~GeV$^2$, we have 
assigned a $W$-dependent systematic uncertainty that is 10\% at $W$=1.65~GeV, 
5\% at $W$=1.75~GeV, 2\% at $W$=1.85~GeV, and 1\% at $W$=1.95~GeV.

(ii). Yield extraction: As discussed in Section~\ref{analysis}, we use 
Monte Carlo templates that have been matched to the data for the $\Lambda$ 
and $\Sigma^0$ peaks and background forms based on the spectra of 
misidentified pions and protons in order to fit the hyperon missing-mass 
spectra.  We studied various changes to our procedures such as changing the 
histogram bin size in the fitting procedure and using different forms for the 
background shape (e.g. using both misidentified pions and protons, only 
misidentified pions, and only misidentified protons) and conclude that all 
systematic effects get larger in direct proportion to the size of the 
statistical uncertainty.  We estimated that any remaining systematic 
uncertainty due to the yield extraction is roughly equal to 25\% of the size 
of the statistical uncertainty in any given bin.  We added these correlated 
uncertainties linearly with the statistical uncertainties on our 
differential cross sections before performing the $\Phi$ fits.

(iii). Model dependence: We have studied the systematic uncertainty 
associated with the model dependence of the convolution of the CLAS 
acceptance correction, the radiative corrections, and the finite bin size 
correction together because they are correlated, especially by their 
sensitivity to the underlying physics model that we use for the Monte Carlo 
event generator.  Specifically we studied the overall model dependence by 
varying the physics model used in our Monte Carlo program and stepping 
through the full analysis chain from yields to cross sections to structure 
function extraction.

\begin{figure}[htpb]
\vspace{9.0cm}
\includegraphics{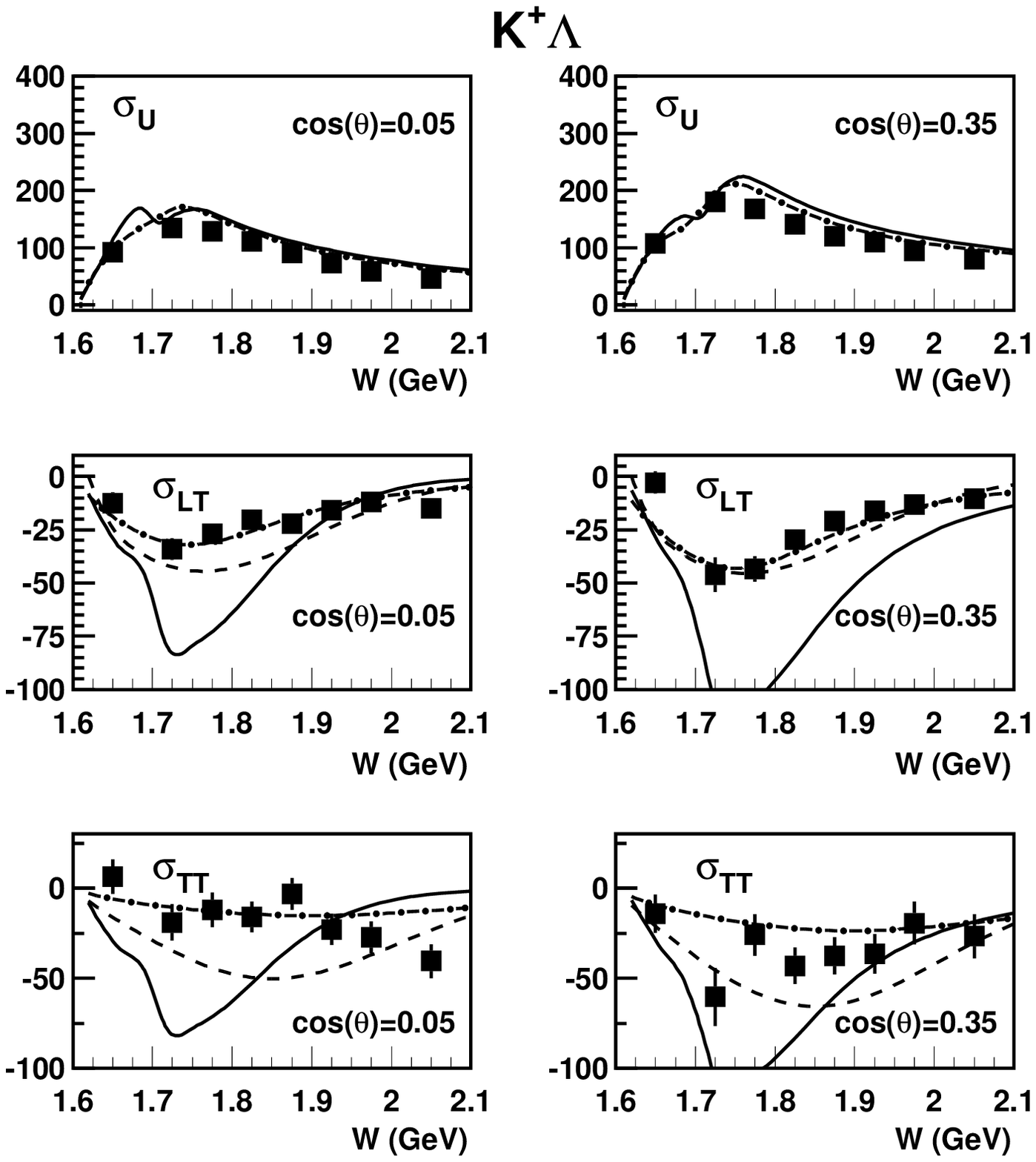}
\includegraphics{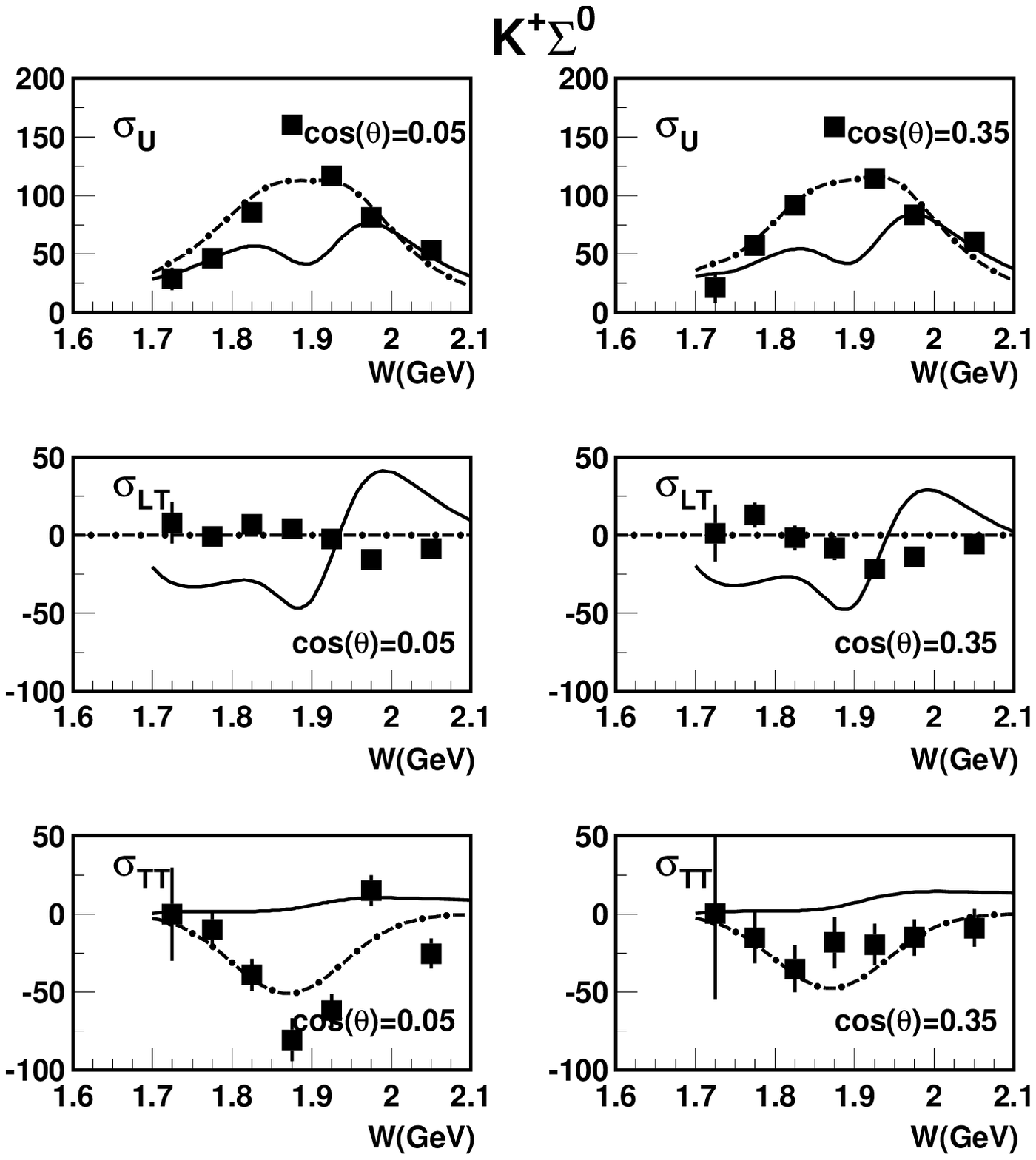}
\caption{\small{Cross section models for the $K^+\Lambda$ (left) and 
$K^+\Sigma^0$ (right) structure functions (in units of nb/sr) vs. $W$ 
compared to our CLAS data (square points) at $\cos \theta_K^*=0.05$ and 0.35.
The ad hoc models employed (dashed and dot-dashed curves -- discussed in
Section~\ref{analysis}) were based on the Bennhold-Mart model~\cite{bm_calc} 
(solid curve) as a starting point. The plots are for the 2.567~GeV data set 
at $Q^2$=0.65~GeV$^2$.}}
\label{modelcomp}
\end{figure}

We tried a number of existing hadrodynamic models, but found the agreement 
with our data to be unsatisfactory.  Ultimately we employed the model of
Bennhold and Mart~\cite{bennhold} and adjusted the parameters in an ad hoc 
fashion to get a better match to our measured $K^+\Lambda$ and $K^+\Sigma^0$ 
cross sections as a function of $Q^2$, $W$, $\cos \theta_K^*$, and $\Phi$ 
(see discussion in Section~\ref{analysis}).  In Fig.~\ref{modelcomp} we show 
comparisons of our ad hoc event generator models to our initial model from
Bennhold and Mart~\cite{bennhold} and to our data.  Our studies showed that 
the event-generator model dependence introduced an average systematic 
uncertainty on our differential cross sections of 8\%.

(iv). Radiative correction theoretical uncertainty: The radiative correction 
factor was calculated using a multi-dimensional integral approach (see 
Section~\ref{analysis}).  To calculate the theoretical uncertainty, our 
results were compared to the exact one-loop calculations from the 
{\tt EXCLURAD} code~\cite{afanasev}.  The average deviation was approximately 
3.4\% over all kinematical bins.

(v). Virtual photon flux factor: We estimated uncertainties on the average 
virtual photon flux factor across our kinematics by propagating through the 
flux definition (see eq.(\ref{eq:flux})) the uncertainties associated with 
$W$ and $Q^2$ that arise from the absolute uncertainty in the reconstructed 
electron momentum and angles.  The uncertainty in the flux factor was 
determined to be less than 3\%.   

(vi). Luminosity: The uncertainty in our luminosity is based on the
uncertainty in our electron flux, target thickness, and measured live
time.  The total systematic uncertainty from these sources is assigned
as 3.2\%.

\subsection{Final Systematic Uncertainty Assignments}
\label{finalerr}

\begin{figure}[htbp]   
\vspace{14.6cm}
\includegraphics{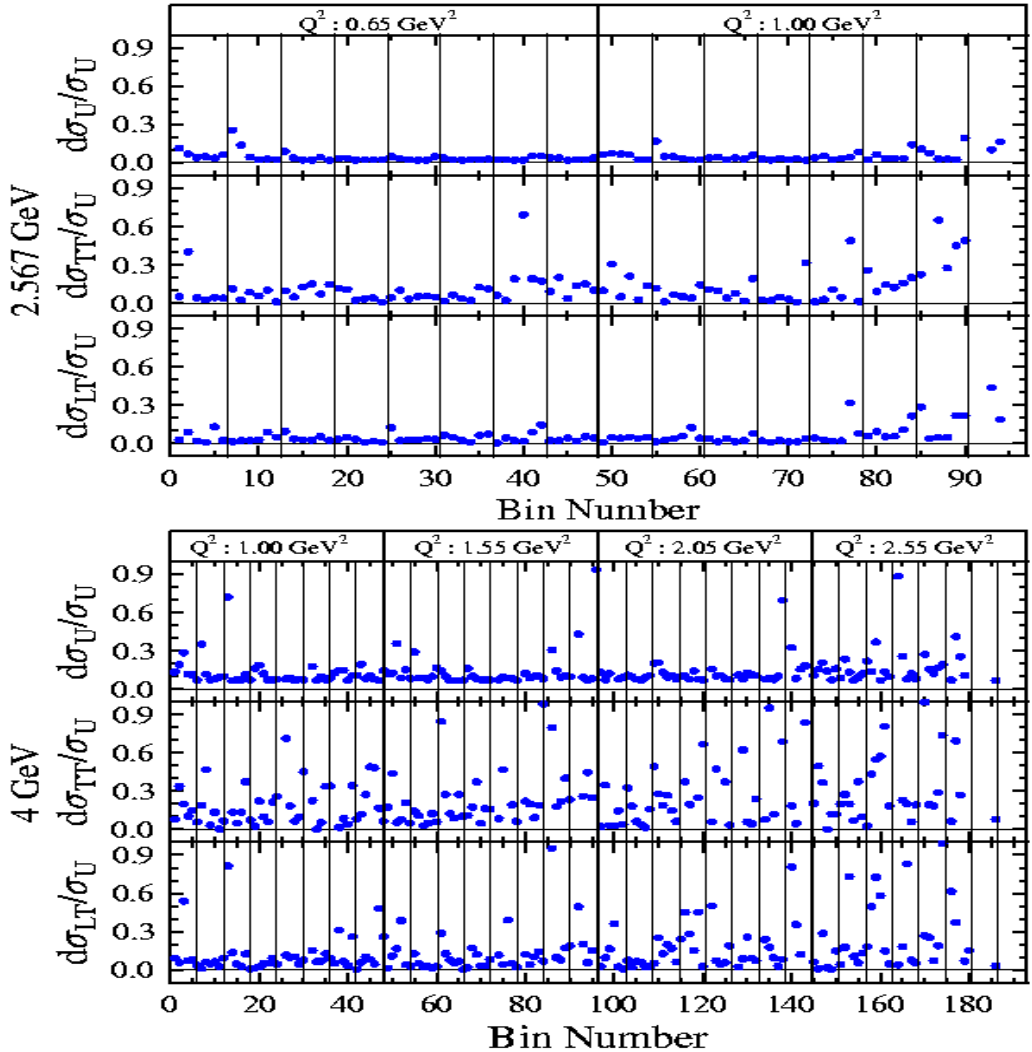}
\caption{\small{(Color online) Total kinematic-independent systematic 
uncertainties from eq.(\ref{toterr}) on the structure functions $\sigma_U$, 
$\sigma_{TT}$, and $\sigma_{LT}$ normalized to $\sigma_U$ for the $K^+\Lambda$ 
data as a function of bin number for the 2.567~GeV (top) and 4~GeV (bottom) 
data sets.  The wide vertical boundaries indicate the $Q^2$ bins, the narrow 
vertical boundaries indicate the $W$ bins within each $Q^2$ range, and the six 
points in each $W$ bin represent the angle bins from $\cos \theta_K^*$=-0.65 
to 0.90.}}
\label{systot}
\end{figure}

The relative systematic uncertainties on the interference structure functions
$\delta\sigma_{TT}/\sigma_{TT}$ and $\delta\sigma_{LT}/\sigma_{LT}$ must be
interpreted with some caution as both of these interference structure
functions are frequently small in our kinematics.  In this regard, defining a
relative uncertainty is mathematically meaningless.  We have chosen instead 
to quote all systematic uncertainties relative to $\sigma_U$.  
Fig.~\ref{systot} shows that the kinematic-independent systematic 
uncertainties on each of the structure functions $\sigma_U$, $\sigma_{TT}$, 
and $\sigma_{LT}$ relative to $\sigma_U$ are reasonably independent of $Q^2$, 
$W$, $\cos \theta_K^*$, and $\Phi$.  For this reason we have decided to quote 
the relative systematic uncertainty as the mean of these distributions for 
each beam energy.  This eliminates the fluctuations in the determination of 
the systematic uncertainties associated with low statistics portions of our 
phase space.  From these distributions we compute the mean and then add in 
quadrature the systematics associated with the {\v C}erenkov detector 
efficiency (a $Q^2$-dependent systematic) and the forward angle response of 
CLAS (a $W$-dependent systematic) to get the final total systematic 
uncertainty assigned to our data points.  The same systematics determined 
from the analysis of the $K^+\Lambda$ final state are assigned to the data 
for the $K^+\Sigma^0$ final state as the $K^+\Lambda$ data has smaller 
statistical uncertainties.  The final total systematic uncertainty 
assignments relative to $\sigma_U$ for our three structure function 
separations are given in Table~\ref{errtab}.  

\begin{table}[htbp]
\begin{center}
\begin{tabular} {||c|c|r|c|c||} \hline \hline
            &            & \multicolumn{3} {|c|} {Systematic Uncertainty} \\ \hline
Beam Energy & Term & \multicolumn{2} {|c|} {$Q^2$=0.65~GeV$^2$} & $Q^2$=1.00~GeV$^2$ \\ \hline
2.567~GeV   & $\delta \sigma_U/\sigma_U$   & \multicolumn{2} {|c|} { 9.6\%} & 8.4\% \\ \cline{2-5}
            & $\delta \sigma_{TT}/\sigma_U$& \multicolumn{2} {|c|} {11.7\%} & 10.8\% \\ \cline{2-5}
            & $\delta \sigma_{LT}/\sigma_U$&  \multicolumn{2} {|c|} {7.8\%}  & 6.3\% \\ \hline \hline
Beam Energy & Term & \multicolumn{2} {|c|} {$Q^2$=1.00~GeV$^2$} & $Q^2$=1.55, 2.05, 2.55~GeV$^2$ \\ \hline 
4~GeV     & $\delta \sigma_U/\sigma_U$    & $W$=1.65~GeV & 13.9\% & 8.4\% \\
          &                               &     1.75~GeV & 10.8\% & \\
          &                               &     $\ge 1.85$~GeV  & 10\% & \\ \cline{2-5}
          & $\delta \sigma_{TT}/\sigma_U$ & $W$=1.65~GeV & 15.4\% & 10.8\% \\
          &                               &     1.75~GeV & 12.7\% & \\
          &                               &     $\ge 1.85$~GeV & 12\% & \\ \cline{2-5}
          & $\delta \sigma_{LT}/\sigma_U$ & $W$=1.65~GeV & 12.7\% & 6.3\% \\
          &                               &     1.75~GeV & 9.3\% & \\
          &                               &     $\ge 1.85$~GeV & 8\% & \\ \hline \hline
\end{tabular}
\end{center}
\caption{\small{Total systematic uncertainties assigned to our structure function 
measurements for both the $K^+\Lambda$ and $K^+\Sigma^0$ final states as a function
of kinematics.  Uncertainties for $\sigma_U$, $\sigma_{TT}$, and $\sigma_{LT}$ are
all quoted relative to $\sigma_U$.}}
\label{errtab}
\end{table}

The systematic uncertainty analysis on the separated structure functions 
$\sigma_T$ and $\sigma_L$ was carried out only for the Rosenbluth separation 
method.  In order to be conservative, the same systematic uncertainty is 
assigned to $\sigma_T$ and $\sigma_L$ extracted from the simultaneous 
$\epsilon-\Phi$ fit.  This was done as we could not fully disentangle the
point-to-point and scale-type systematic uncertainties between the two beam
energy data sets in the $\epsilon-\Phi$ fits.  In this analysis, we simply use 
the two different techniques as a way to perform a consistency check on our 
extracted structure functions. Our analysis shows very good agreement between 
the two techniques giving us confidence in our assigned systematics.

We performed several consistency checks on our data.  The most important were 
that cross sections at 2.567~GeV taken with two different magnetic field 
settings and our cross sections taken at 4.056 and 4.247~GeV agreed within 
the quoted systematics.  This tested the accuracy of our knowledge of the 
acceptance because it varied strongly with field setting and beam energy.  
The other check was to fit the two beam energy data sets simultaneously in 
each of our bins to verify that the relative normalization factor between the 
two data sets was consistent with unity.

\section{Results}
\label{results}

\subsection{Three Structure Function Separation}

\subsubsection{Angular Dependence}

\begin{sidewaysfigure}[htbp]
\vspace{14.0cm}
\includegraphics{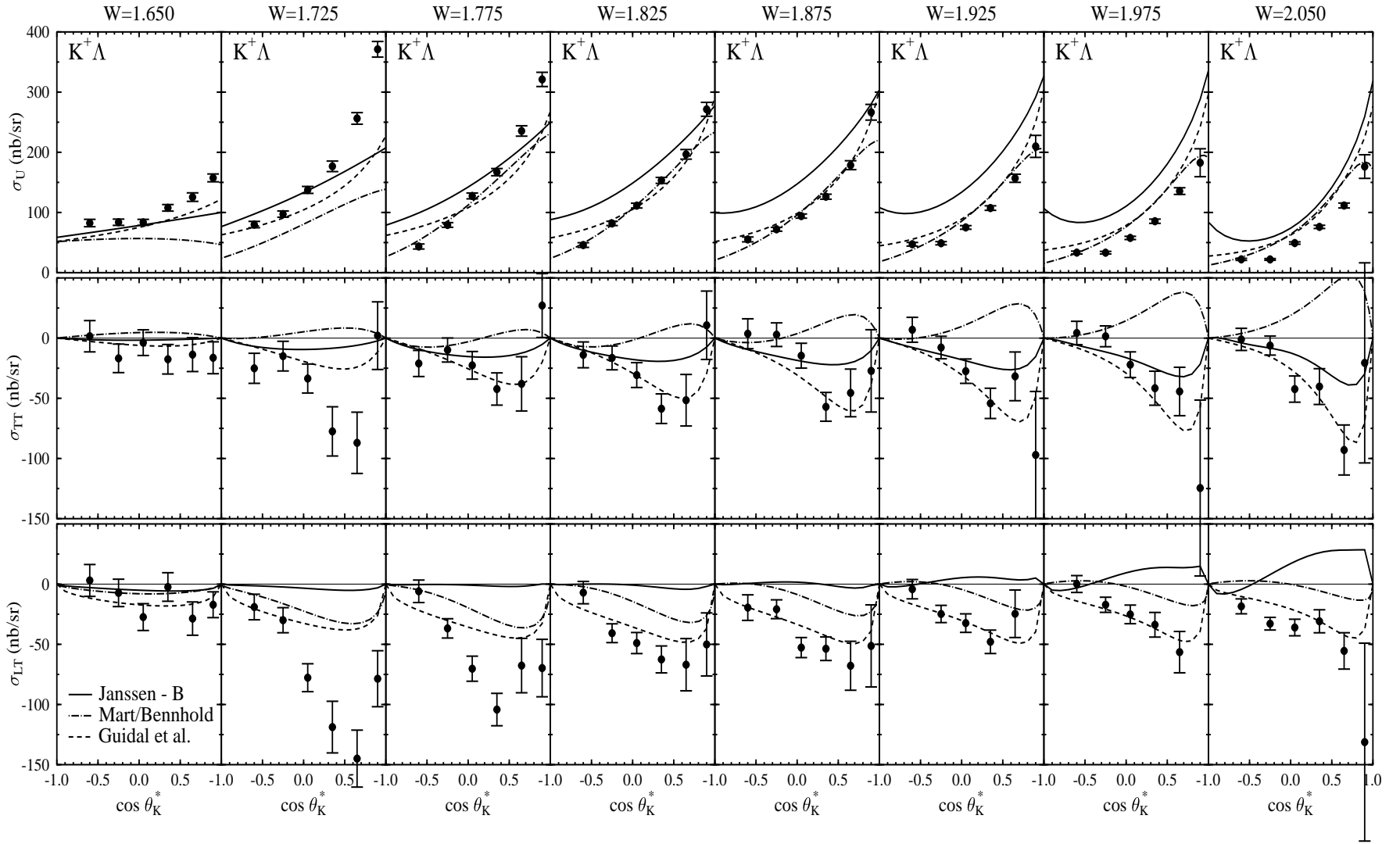}
\caption{\small{Structure functions $\sigma_U$, $\sigma_{TT}$, and 
$\sigma_{LT}$ (in nb/sr) for $K^+\Lambda$ production vs. $\cos \theta_K^*$ 
at 2.567~GeV for $Q^2$=0.65~GeV$^2$ and $W$ from 1.650 to 2.050~GeV. The 
error bars represent the statistical uncertainties only. The relative 
systematic uncertainties to $\sigma_U$ are given in Table~\ref{errtab}.  
The curves shown are from the model calculations of Bennhold and Mart (BM) 
~\cite{bm_calc} (dot-dashed), Janssen {\it et al.} (JB)~\cite{jb_calc} 
(solid), and Guidal {\it et al.} (GLV)~\cite{glv_calc} (dashed).  The models 
are described in the text.}}
\label{master_l1} 
\end{sidewaysfigure}

\begin{sidewaysfigure}[htbp]
\vspace{14.0cm}
\includegraphics{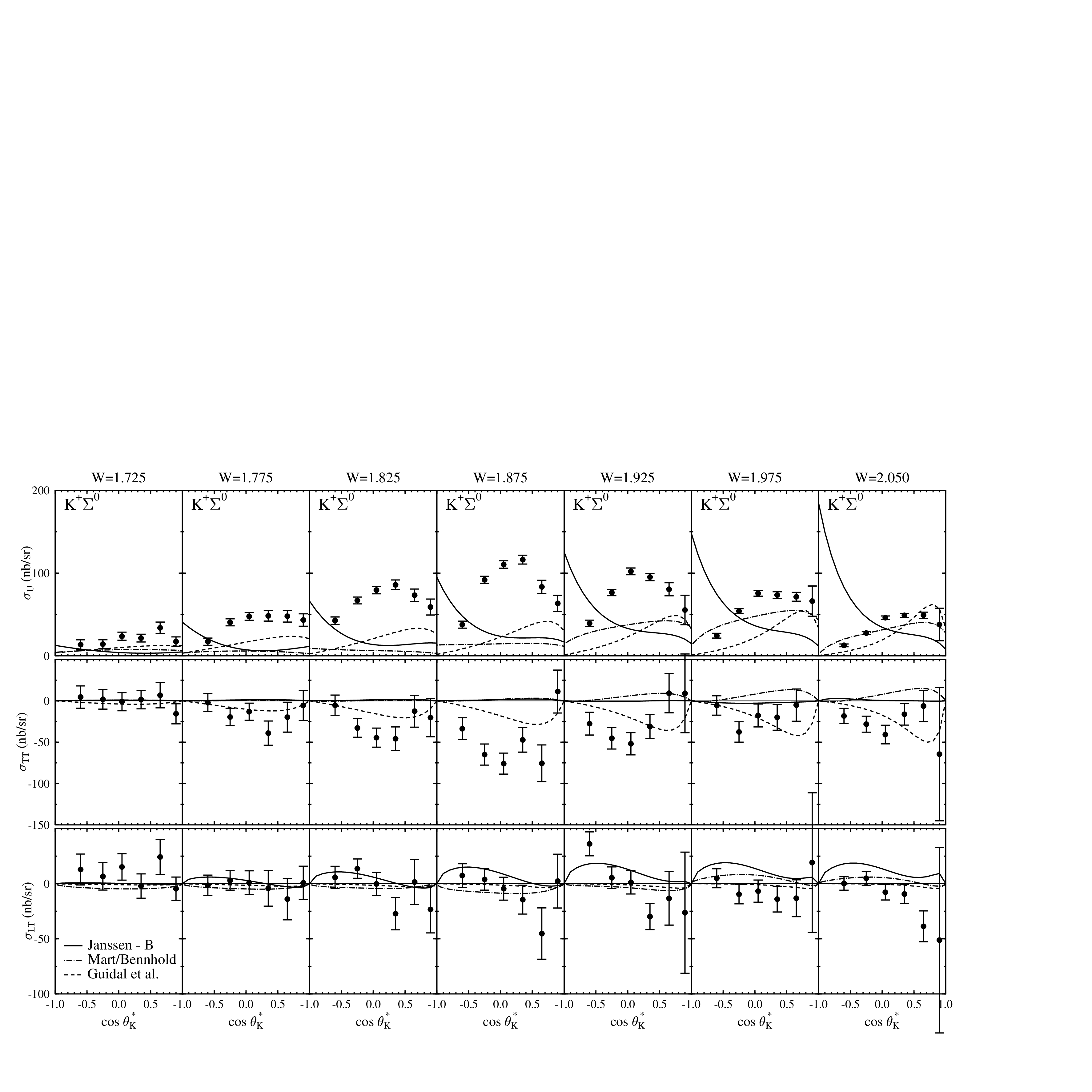}
\caption{\small{Structure functions $\sigma_U$, $\sigma_{TT}$, and 
$\sigma_{LT}$ (in nb/sr) for $K^+\Sigma^0$ production vs. $\cos \theta_K^*$ 
at 2.567~GeV for $Q^2$=0.65~GeV$^2$ and $W$ from 1.725 to 2.050~GeV. The 
error bars represent the statistical uncertainties only. The relative 
systematic uncertainties to $\sigma_U$ are given in Table~\ref{errtab}.  
The curves shown are from the model calculations of Bennhold and Mart (BM)
~\cite{bm_calc} (dot-dashed), Janssen {\it et al.} (JB)~\cite{jb_calc} 
(solid), and Guidal {\it et al.} (GLV)~\cite{glv_calc} (dashed).  The models 
are described in the text.}}
\label{master_s1} 
\end{sidewaysfigure}

In Figs.~\ref{master_l1} and \ref{master_s1} we show the extracted 
structure functions $\sigma_U$, $\sigma_{TT}$, and $\sigma_{LT}$ versus 
$\cos \theta_K^*$ for $K^+\Lambda$ and $K^+\Sigma^0$ for different $W$ 
points at $Q^2$=0.65~GeV$^2$ from our 2.567~GeV data set.  Although we 
focus on the $\cos \theta_K^*$ dependence of our low $Q^2$ data set at 
2.567~GeV, the general conclusions that can be drawn from studying the 
angular dependence are similar for our other data sets.  However, the full 
set of our data is available in Ref.~\cite{database}.  In these plots, the 
data are sorted into $W$ bins 50-MeV wide, except for the first and last 
$W$ bins which are 100-MeV wide (see Table~\ref{tab:bins}).  All data points 
have been evolved to the given $Q^2$, $W$, and $\cos \theta_K^*$ bin centers.  
The curves shown are from the hadrodynamic models of Bennhold and Mart (BM)
\cite{bm_calc} (dot-dashed curves) and of Janssen {\it et al.} (JB)
\cite{jb_calc} (solid curves), and the Reggeon-exchange model of Guidal 
{\it et al.} (GLV)~\cite{glv_calc} (dashed curves).

A number of observations can be made independent of the model calculations. 
First, we observe that the $K^+\Lambda$ and $K^+\Sigma^0$ electroproduction 
dynamics are very different. The data in Figs.~\ref{master_l1} and 
\ref{master_s1} reveal that $\sigma_U$ is more forward-peaked in 
$K^+\Lambda$ production than for $K^+\Sigma^0$ production across our full 
range in $W$.  With regard to the interference structure functions, 
$\sigma_{TT}$ for $K^+\Lambda$ is roughly one-fourth of $\sigma_U$ and 
always negative and very similar in structure and magnitude to $\sigma_{LT}$, 
while $\sigma_{TT}$ for $K^+\Sigma^0$ is generally smaller in magnitude than 
for $K^+\Lambda$ with a peaking at more mid-range angles.  The $K^+\Lambda$ 
reaction has a significant $\sigma_{LT}$ component in the forward 
direction compared to $\sigma_U$, while $\sigma_{LT}$ for the $K^+\Sigma^0$ 
reaction is everywhere consistent with zero.  

The forward-peaking of $\sigma_U$ and $\sigma_{LT}$ for $K^+\Lambda$ compared 
to $K^+\Sigma^0$ can be qualitatively explained by the effect of the 
longitudinal coupling of the virtual photons.  We note that the two channels 
are of nearly equal strength at $Q^2$=0~GeV$^2$~\cite{mcnabb,bradford1}, 
while here at $Q^2$=0.65~GeV$^2$ the $K^+\Lambda$ channel is stronger than 
the $K^+\Sigma^0$ channel at forward angles by a factor of 2 to 3.  For 
transverse (real) photons, the $t$-channel mechanism at low $t$ is dominated 
by vector $K^{*+}$ exchange, which relates directly to the relative 
magnitudes of the $g_{K^*YN}$ coupling constants to the $g_{KYN}$ constants.  
As $Q^2$ rises from zero, the photon can acquire a longitudinal polarization 
and the importance of pseudoscalar $K^+$ exchange increases.  Given that 
$g_{K\Lambda N}^2 \gg g_{K\Sigma^0 N}^2$~\cite{adel_saghai,deswart}, this 
effect increases the cross section for $K^+\Lambda$ relative to 
$K^+\Sigma^0$.  This argument was already noted in the earliest reports of 
hyperon electroproduction~\cite{brown}, and is strengthened by our 
observation of a sizeable $\sigma_{LT}$ for $K^+\Lambda$ and a $\sigma_{LT}$ 
consistent with zero for $K^+\Sigma^0$.  It should also be the case that
since $g_{K^*\Sigma N} \gg g_{K\Sigma N}$, $K^*$ exchange should dominate the 
$\Sigma^0$ channel.  Because $K^*$ exchange must vanish at forward angles due 
to angular momentum conservation, the $\Sigma^0$ cross section should also 
decrease at forward angles~\cite{guidal}.

None of the three different models shown is particularly successful at 
describing all of the data. In general the models better agree with the
$K^+\Lambda$ data than with the $K^+\Sigma^0$ data.  The three 
models tend to reproduce the qualitative fall-off in $\cos \theta_K^*$ of 
$\sigma_U$ for the $K^+\Lambda$ data but do not include sufficient 
forward-angle strength for $W < 1.8$~GeV.  At higher $W$ the BM model 
generally reproduces $\sigma_U$, while the JB model consistently is too 
large at forward and backward kaon angles.  The GLV model goes above our 
data as $\cos \theta_K^* \to 1$, but describes the structure of the 
$K^+\Lambda$ data surprisingly well considering that it has no built-in 
$s$-channel resonances.  $\sigma_U$ for the $K^+\Sigma^0$ data is poorly 
described by all models, especially the JB model, which includes too much 
$u$-channel strength, while the BM and GLV models generally include too 
little strength or miss the broad peaking about $\cos \theta_K^* \sim 0$.

Within the GLV Regge model, the functions $\sigma_{TT}$ and $\sigma_{LT}$
arise from the interference of the $K$ and $K^*$ Regge trajectories.
This modeling is sufficient to qualitatively reproduce the 
behavior of both the $K^+\Lambda$ and $K^+\Sigma^0$ data over our full
kinematic phase space.  The quality of the comparisons
of the hadrodynamic models to the $\sigma_{TT}$ and $\sigma_{LT}$ data are
much less favorable.  For the JB model, $\sigma_{TT}$ for $K^+\Lambda$ has
the correct sign, but its strength is too small and the angular dependence
does not match the data.  For $K^+\Sigma^0$, the JB model predicts
$\sigma_{TT} \sim 0$ everywhere, in strong disagreement with the data.
For the BM model, $\sigma_{TT}$ for $K^+\Lambda$ has a strength and angle
dependence that qualitatively matches the data, but has the wrong sign.  
For $K^+\Sigma^0$, the BM model has the wrong sign for $\sigma_{TT}$ 
and doesn't match the angular distribution of the data.  From the JB model,
$\sigma_{LT}$ for $K^+\Lambda$ is consistent with zero at low $W$, but
increases in strength for higher $W$, where the model has the wrong sign
compared to the data.  For $K^+\Sigma^0$, the JB model has both the wrong 
sign and angular dependence.  For the BM model, $\sigma_{LT}$ follows the 
trends of the $K^+\Lambda$ data but has overall too little strength, while 
it is reasonably consistent with the $K^+\Sigma^0$ data.

\subsubsection{Energy Dependence}

\begin{sidewaysfigure}[htbp]
\vspace{14.0cm}
\includegraphics{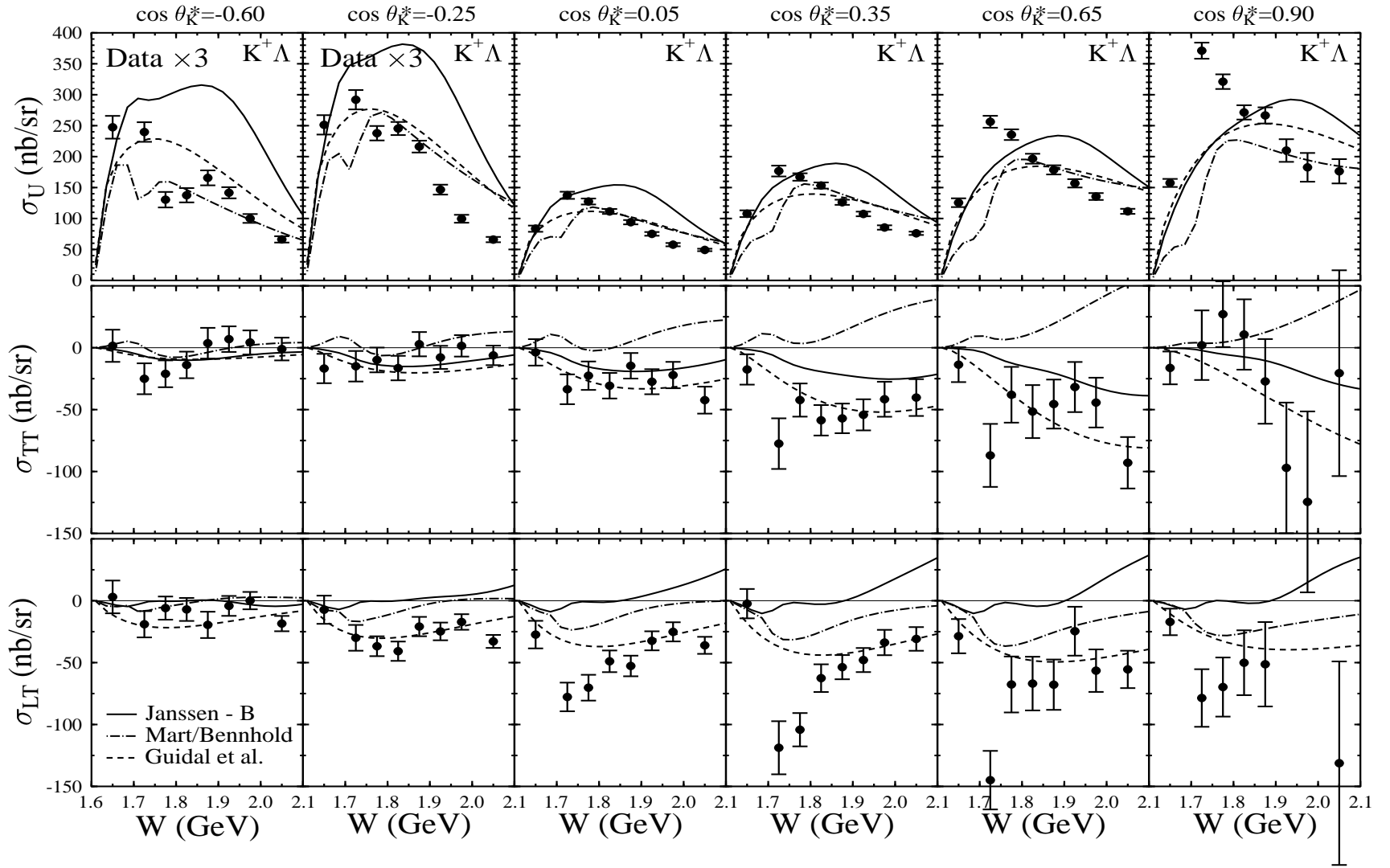}
\caption{\small{Structure functions $\sigma_U$, $\sigma_{TT}$, and 
$\sigma_{LT}$ (in nb/sr) for $K^+\Lambda$ production vs. $W$ at 2.567~GeV 
for $Q^2$=0.65~GeV$^2$ for our 6 $\cos \theta_K^*$ bins.  The error bars 
represent the statistical uncertainties only.  The relative systematic 
uncertainties to $\sigma_U$ are given in Table~\ref{errtab}.  The curves 
shown are from the model calculations of Bennhold and Mart (BM)~\cite{bm_calc} 
(dot-dashed), Janssen {\it et al.} (JB)~\cite{jb_calc} (solid), and Guidal 
{\it et al.} (GLV)~\cite{glv_calc} (dashed).  The models are described in 
the text.  The data (and calculations) for $\sigma_U$ for the two back-angle 
points have been scaled by a factor of 3 for clarity.}}
\label{master_l2} 
\end{sidewaysfigure}

\begin{sidewaysfigure}[htbp]
\vspace{14.0cm}
\includegraphics{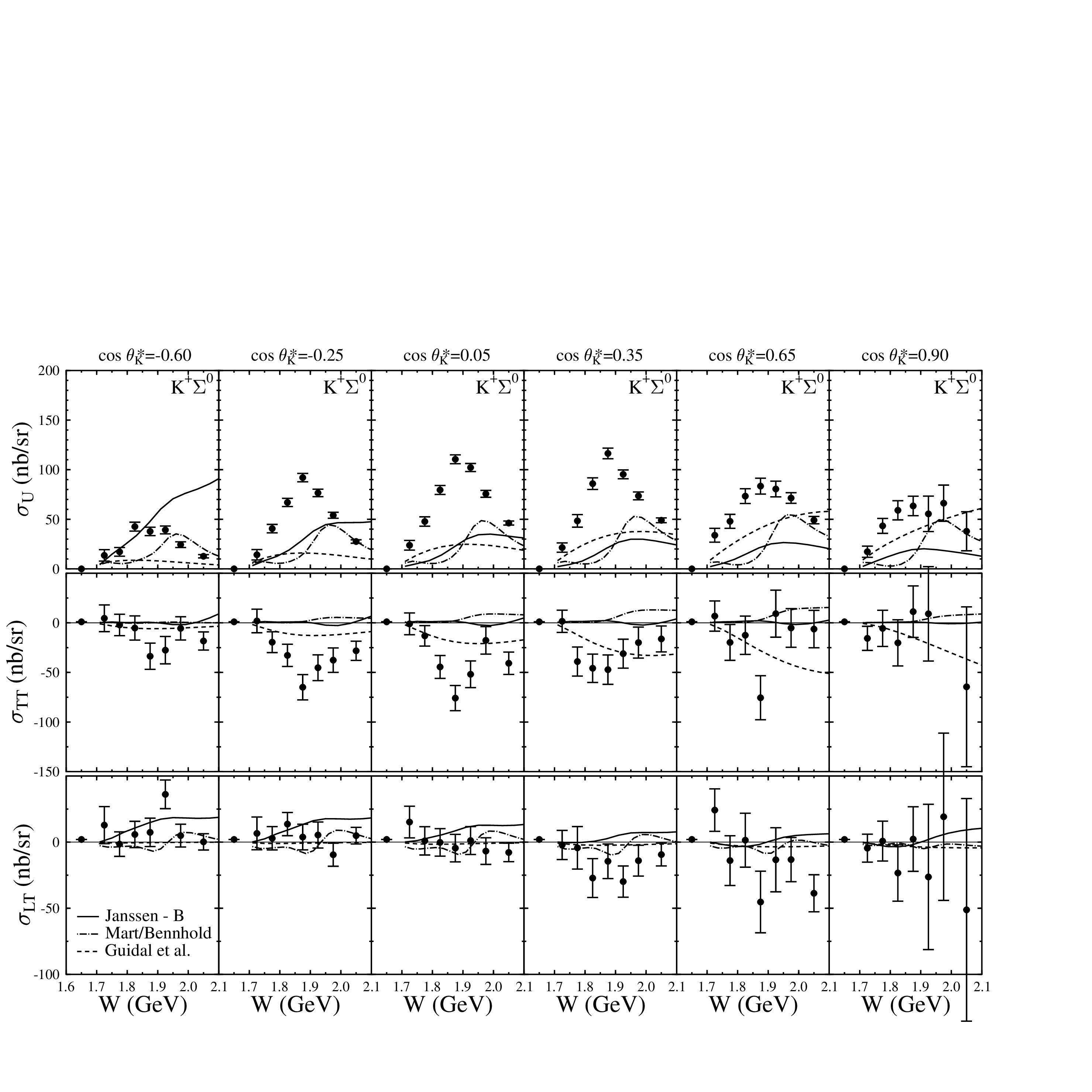}
\caption{\small{Structure functions $\sigma_U$, $\sigma_{TT}$, and 
$\sigma_{LT}$ (in nb/sr) for $K^+\Sigma^0$ production vs. $W$ at 2.567~GeV for 
$Q^2$=0.65~GeV$^2$ for our 6 $\cos \theta_K^*$ bins.  The error bars represent 
the statistical uncertainties only. The relative systematic uncertainties to 
$\sigma_U$ are given in Table~\ref{errtab}.  The curves shown are from the 
model calculations of Bennhold and Mart (BM)~\cite{bm_calc} (dot-dashed), 
Janssen {\it et al.} (JB)~\cite{jb_calc} (solid), and Guidal {\it et al.} 
(GLV)~\cite{glv_calc} (dashed).  The models are described in the text.}}
\label{master_s2} 
\end{sidewaysfigure}

Even if $\Lambda$ production for forward-going $K^+$ mesons is dominated by 
$t$-channel exchange, there is still room for $s$-channel resonance 
contributions at more central angles and at all angles for the $\Sigma^0$. 
To more directly look for $s$-channel resonance evidence, the extracted 
structure functions are presented as a function of the center-of-mass energy 
$W$ for our six bins in $\cos \theta_K^*$.  Figures~\ref{master_l2} and
\ref{master_s2} show the results for our 2.567~GeV data at $Q^2$=0.65~GeV$^2$ 
for the contiguous angle bins centered at $\cos \theta_K^*$ = -0.6, -0.25, 
0.05, 0.35, 0.65, and 0.90.

Several characteristics of the data stand out.  For $K^+\Lambda$ production, 
$\sigma_U$ shows a broad peak at about 1.7~GeV at forward angles, and two 
peaks separated by a dip at about 1.75~GeV for our two backward 
angle points.  Across our phase space, $\sigma_{TT}$ and $\sigma_{LT}$ for 
$K^+\Lambda$ production are predominantly negative and about one-third the 
size of $\sigma_U$.  Where the statistical uncertainties on our data are
reasonable (away from the most forward-angle point), $\sigma_{TT}$ and
$\sigma_{LT}$ seem to be similar in shape to $\sigma_U$, but opposite in
sign.  The $K^+\Sigma^0$ structure functions have a different set of 
features.  Both $\sigma_U$ and $\sigma_{TT}$ exhibit a broad bump at about 
1.85~GeV, while $\sigma_{LT}$ is consistent with zero everywhere.  The 
$\Sigma^0$ shapes are similar for both forward and back-angle production, 
with a strong peaking at central angles.

We argue here that our spectra likely reflect the existence of a few 
underlying $s$-channel resonances along with $t$-channel processes, but 
acknowledge that the physical interpretation is not straightforward and 
will require detailed modeling.   The $W$-dependence of the $K^+\Lambda$ 
data for $\sigma_U$ near threshold shows more structure than a model based 
upon only $t$-channel exchanges provides (GLV model - dashed curves) and is 
probably evidence of resonance activity. In this range of $W$, the 
$S_{11}(1650)$ is believed to be dominant in the $s$-channel~\cite{bennhold}.  
There are also a number of known $N^*$ resonances near 1.7~GeV that can 
contribute to the $K^+\Lambda$ and $K^+\Sigma^0$ final states, in particular, 
the $P_{11}(1710)$ and $P_{13}(1720)$. The effect of these resonances can be 
seen in the hadrodynamic model calculations (BM model - dot-dashed curves, 
JB model - solid curves), though clearly their strengths at the measured $Q^2$ 
are not correct.

The double-peaking of $\sigma_U$ for $K^+\Lambda$ production at backward
$\theta_K^*$ angles as seen in Fig.~\ref{master_l2}, corroborates a similar 
structure seen in recent photoproduction results
\cite{saphir1,saphir2,mcnabb,bradford1}. Within existing hadrodynamic models,
the structure just above the threshold region is typically accounted for by 
the known $S_{11}$(1650), $P_{11}$(1710), and $P_{13}$(1720) nucleon 
resonances.  However there is no consensus as to the origin of the bump 
feature at $\sim$1.9~GeV that was first seen in the $K^+\Lambda$ 
photoproduction data from SAPHIR~\cite{saphir1}.  It is tempting to speculate 
that this is evidence for a previously ``missing'', negative-parity $J=3/2$ 
resonance at 1.96~GeV predicted in the quark model of Capstick and Roberts
\cite{capstick}.  This explanation was put forward in the work of Bennhold 
and Mart~\cite{bennhold}, in which they postulated the existence of a 
$D_{13}$ state at 1.9~GeV.  However, other groups have shown that the same 
data can also be explained by accounting for $u$-channel hyperon exchanges
\cite{saghai1} or with an additional $P$-wave resonance~\cite{janssen2}.
From our data, the $W$ spectra of the interference terms, $\sigma_{TT}$ 
and $\sigma_{LT}$, show no clear structures in the region about 1.9~GeV, 
whereas an $s$-channel resonance would likely be reflected in the structure 
of the interference terms, particularly $\sigma_{TT}$.  Note that the BM and 
JB models include a $D_{13}$(1895) resonant state whose coupling strength 
was determined from fits to the SAPHIR $K^+\Lambda$ total cross section 
data~\cite{saphir1,saphir2}.  Clearly the differences between both models 
and our data indicate that either the resonance parameters are not accurate, 
that more resonant terms are required, or that the bump at 1.9~GeV in our 
$W$ spectra has a non-resonant origin.  We conclude that the $W$ dependence 
of $K^+\Lambda$ production provides suggestive evidence for baryon 
resonance activity within the reaction mechanism, but that the data in 
comparison to present models does not allow any simple statement to be made.

In the $K^+\Sigma^0$ channel, $\sigma_U$ is peaked at about 1.85~GeV, which 
also matches the photoproduction result~\cite{saphir2,mcnabb}. In addition, 
$\sigma_{TT}$ shows a broad feature in this same region. These features are
consistent with a predominantly $s$-channel production mechanism.  In this 
region, beyond the specific $N^*$ resonances believed to contribute to
$K^+\Lambda$ production (and hence are strong candidates to contribute to
$K^+\Sigma^0$ production), there are a number of known $\Delta^*$ resonances 
near 1.9~GeV~\cite{pdg} that can contribute to the $K^+\Sigma^0$ final state, 
particularly the $\Delta^*$(1900) and $\Delta^*$(1910).  These $\Delta^*$ 
states are forbidden to couple to the $K^+\Lambda$ state due to isospin 
conservation.  Current hadrodynamic models seem to indicate that both 
$N^*$ and $\Delta^*$ states (see Table~\ref{tab-resonances}) are necessary 
to describe the existing photo- and electroproduction data.

The comparison of the hadrodynamic model calculations to the data clearly
indicates that significant new constraints on the model parameters will be
brought about when these new data are included in the fits.  The models do
not reproduce $\sigma_U$, $\sigma_{TT}$, or $\sigma_{LT}$ at any level,
especially for the $K^+\Sigma^0$ data.  The Regge model tends to 
underpredict the strength of $\sigma_U$ across the full angular range, 
which is suggestive of $s$-channel contributions to this reaction.  Again 
the trends of $\sigma_{TT}$ and $\sigma_{LT}$ are reasonably well 
reproduced with the inclusion of only the $K$ and $K^*$ Regge trajectories.

\subsubsection{$Q^2$ Dependence}

\begin{figure}[htpb]
\vspace{12.5cm}
\includegraphics{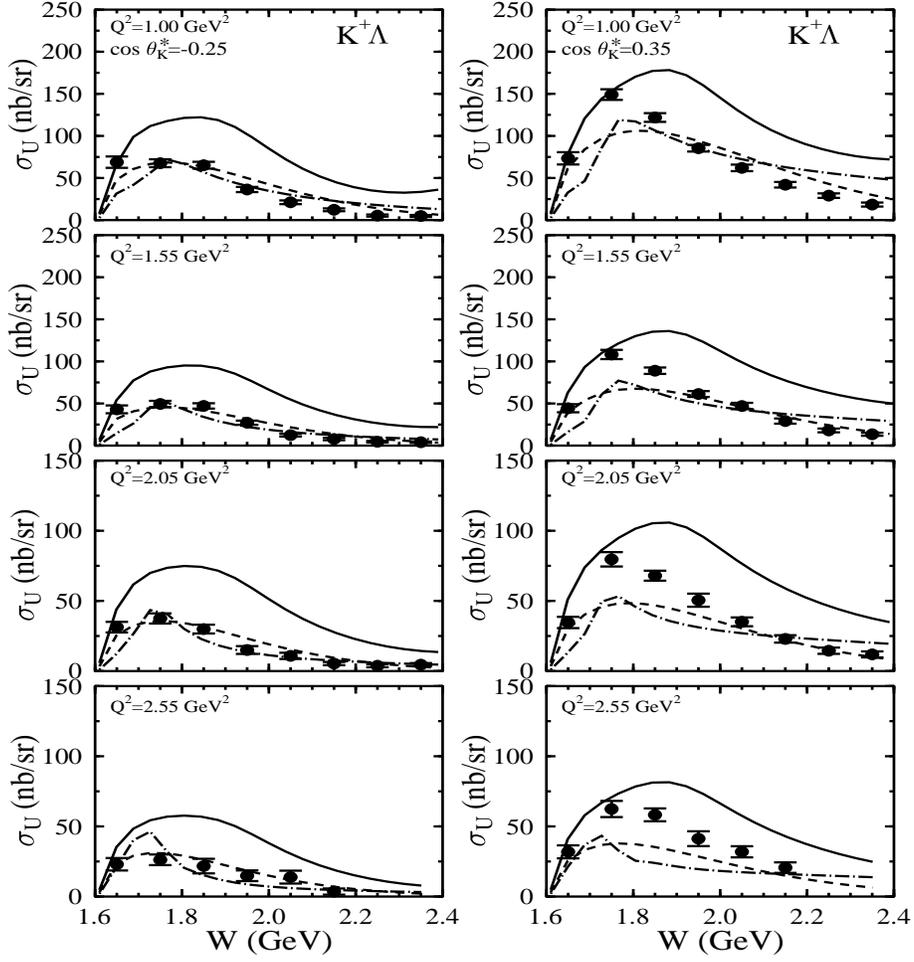}
\caption{\small{$W$ distributions of $\sigma_U$ for the $K^+\Lambda$ final 
state from our 4~GeV data set (evolved to 4.056~GeV) for each of our four 
points in $Q^2$ for $\cos \theta_K^*$=-0.25 (left) and 0.35 (right).  The 
relative systematic uncertainties to $\sigma_U$ are given in 
Table~\ref{errtab}.  The curves shown are from the model calculations
of Bennhold and Mart (BM)~\cite{bm_calc} (dot-dashed), Janssen {\it et al.}
(JB)~\cite{jb_calc} (solid), and Guidal {\it et al.} (GLV) (dashed).}}
\label{4sigu_lamp}
\end{figure}

\begin{figure}[htpb]
\vspace{12.5cm}
\includegraphics{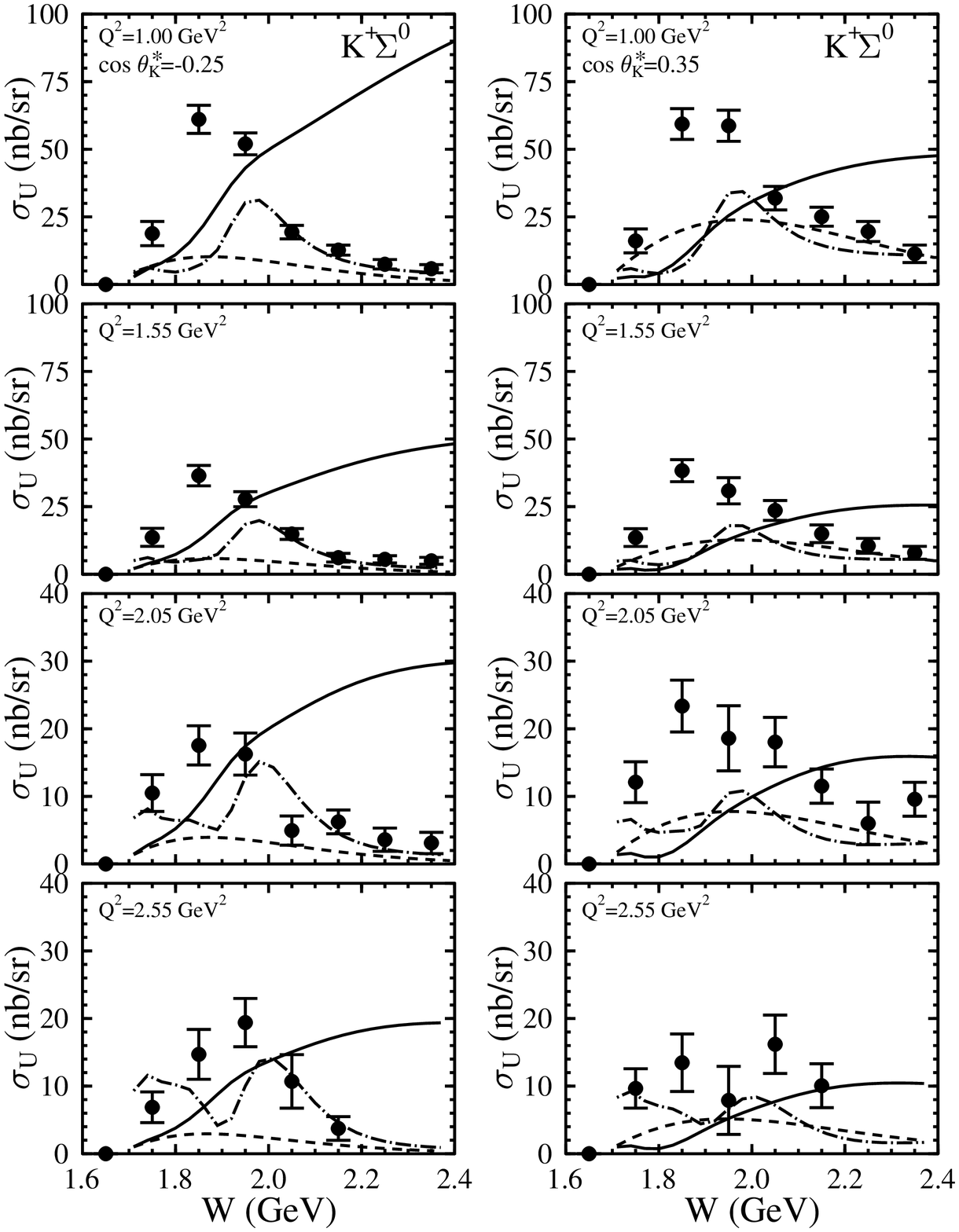}
\caption{\small{$W$ distributions of $\sigma_U$ for the $K^+\Sigma^0$ final 
state from our 4~GeV data set (evolved to 4.056~GeV) for each of our four 
points in $Q^2$ for $\cos \theta_K^*$=-0.25 (left) and 0.35 (right).  The 
relative systematic uncertainties to $\sigma_U$ are given in 
Table~\ref{errtab}.  The curves shown are from the model calculations
of Bennhold and Mart (BM)~\cite{bm_calc} (dot-dashed), Janssen {\it et al.}
(JB)~\cite{jb_calc} (solid), and Guidal {\it et al.} (GLV) (dashed).}}
\label{4sigu_sigp}
\end{figure}

The data shown in Figs.~\ref{master_l2} and \ref{master_s2} were obtained 
from our 2.567~GeV data set at $Q^2$=0.65~GeV$^2$.  Our data set at 4~GeV 
provides a much larger $Q^2$ reach and it is instructive to study the $W$ 
spectra for increasing values of $Q^2$.  These data are shown in 
Figs.~\ref{4sigu_lamp} and \ref{4sigu_sigp} for one of our backward-angle 
points ($\cos \theta_K^*$=-0.25) and a more forward-angle point 
($\cos \theta_K^*$=0.35).  The interference structure functions (not shown) 
do not have a strong $Q^2$ dependence, while $\sigma_U$ shows a smooth 
fall-off.  Note that at $Q^2$=1.0~GeV$^2$, our back-angle data do not show 
the double-peaked structure that was evident at $Q^2$=0.65~GeV$^2$ (see 
Fig.~\ref{master_l2}) and also in our $Q^2$=1.0~GeV$^2$ data at 2.567~GeV 
(not shown).  This could be due to our increased $W$ bin width at 4~GeV 
(100~MeV compared to 50~MeV at 2.567~GeV), or could imply a strong 
$\epsilon$ dependence to the resonance strength.

None of the models reproduces the $K^+\Lambda$ data in detail.  Both
hadrodynamic models are very poor matches to these data, while the GLV
Regge model tends to underpredict the strength in our more forward angle
point, although it is in fair agreement with the data in our more backward
angle point.  For the $K^+\Sigma^0$ data none of the models shown reproduce
even the qualitative aspects of the data.

\begin{figure}[htpb]
\vspace{8.5cm}
\includegraphics{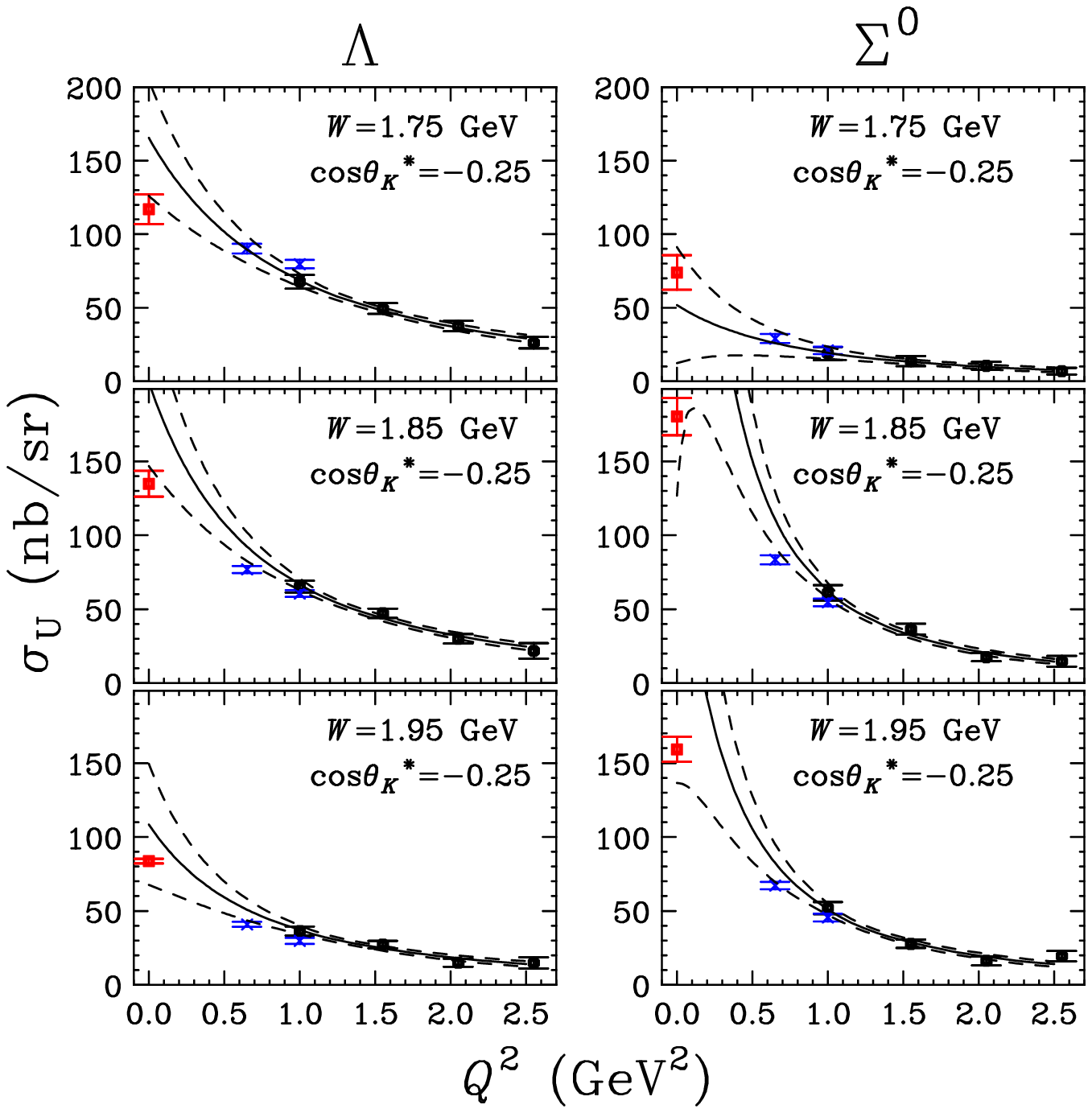}
\includegraphics{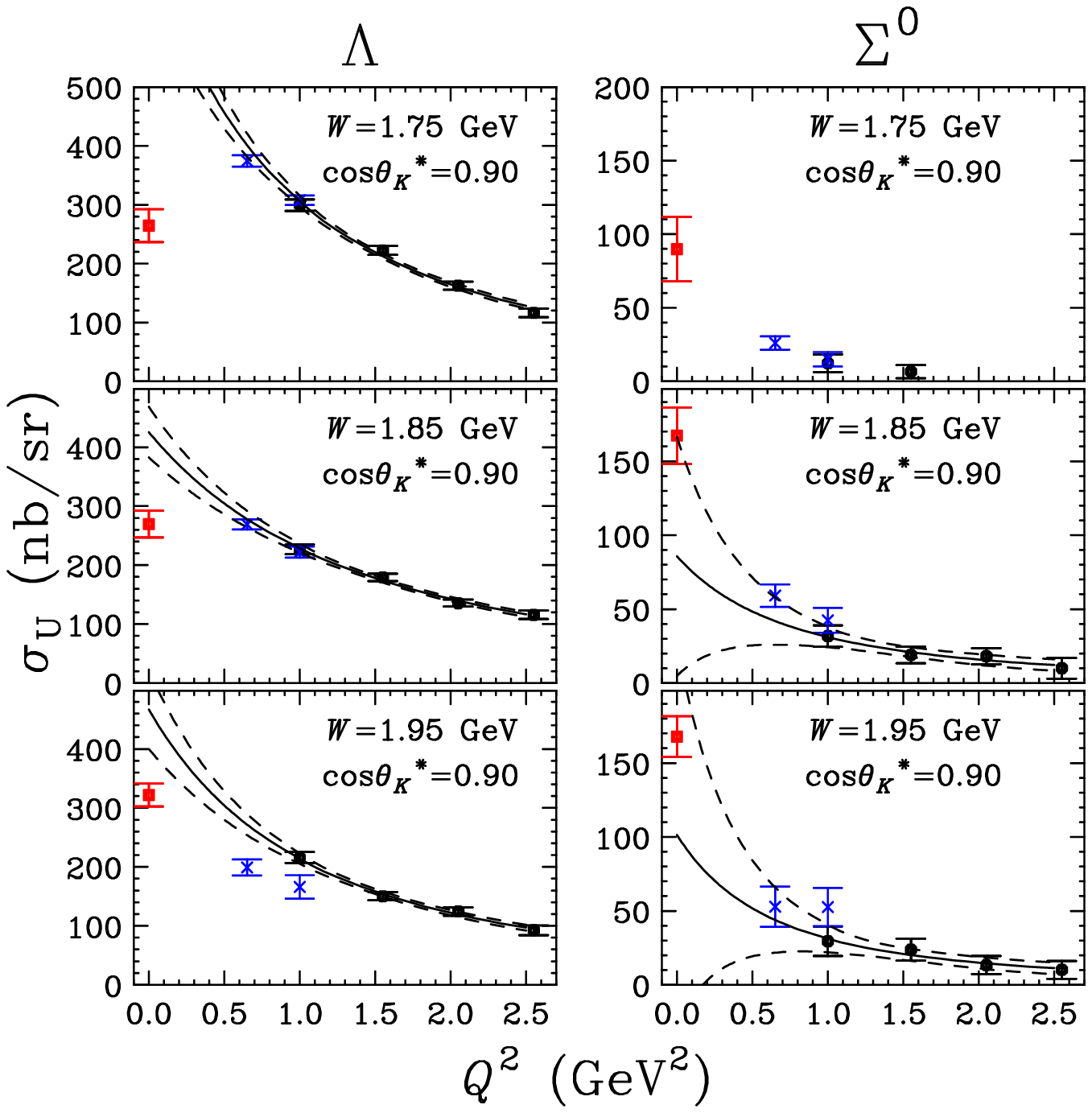}
\caption{\small{(Color online) $Q^2$ distributions of $\sigma_U$ for the 
$K^+\Lambda$ and $K^+\Sigma^0$ final states from our 4~GeV data set (dark
filled circles -- evolved to 4.056~GeV) at $\cos \theta_K^*$=-0.25 (left) 
and 0.90 (right) for $W$=1.750~GeV (top), 1.850~GeV (middle), and 1.950~GeV 
(bottom). The solid curves are from a dipole mass fit to the 4~GeV data
of the form $C(Q^2 + M^2)^{-2}$.  The $Q^2$=0 points (solid squares) come 
from Bradford {\it et al.}~\cite{bradford1} and the two data points from 
our 2.567~GeV data (light crosses) are not included in the fits.  The 
dashed lines represent the error bands from the fits.}}
\label{q2fits}
\end{figure}

The $Q^2$ dependence of $\sigma_U$ for $K^+\Lambda$ and $K^+\Sigma^0$ 
can be studied within our 4~GeV data set as shown in Fig.~\ref{q2fits}.  
The data shown are from our points at $\cos \theta_K^*$=-0.25 and 0.90 
for three different $W$ values across the nucleon resonance region.  Also 
included on these plots are our two data points from the 2.567~GeV data 
set at $Q^2$=0.65~GeV$^2$ and 1.00~GeV$^2$, along with the CLAS $\sigma_T$ 
data from photoproduction at $Q^2$=0 from Bradford {\it et al.}
\cite{bradford1}.  No clear features are apparent here, with the data 
showing a smooth fall-off with respect to the photon point with
increasing $Q^2$ for both final states except for the forward-angle 
$K^+\Lambda$ data at $W$=1.75~GeV.  In order to compare more directly with 
the existing measurements from the 1970's (taken for 
$\theta_K^* < 15^{\circ}$, $0.5 < Q^2 < 4.0$~GeV$^2$, and evolved to 
$W$=2.15~GeV) compiled by Bebek {\it et al.} in Ref.~\cite{bebek2}, we have 
fit our 4~GeV $\sigma_U$ data with the dipole form $C/(Q^2 + M^2)^2$ (where 
$C$ is an arbitrary constant and the CLAS photoproduction data are not 
included in the fits) and compared the mass terms to those extracted from 
the fits in Ref.~\cite{bebek2}.  Fits to the older data suggested that the 
$K^+\Sigma^0$ data with $M^2=(0.785\pm0.095)$~GeV$^2$ fell off more rapidly 
with increasing $Q^2$ than the $K^+\Lambda$ data with 
$M^2=(2.67\pm0.28)$~GeV$^2$.  The results from our fits are contained in 
Table~\ref{massterms} and shown in Fig.~\ref{q2fits}.  Our extracted mass 
terms, even for backward angles where $s$-channel and $u$-channel 
contributions are expected to be more important relative to $t$-channel 
kaon exchange, are consistent with the fits of Ref.~\cite{bebek2} extracted 
from forward kaon angle data.  These results highlight the fact that the 
production mechanisms for $K^+\Lambda$ and $K^+\Sigma^0$ are quite different.

\begin{table}[htbp]
\begin{center}
\begin{tabular} {||c|c|c|c||} \hline \hline
                  &           & $K^+\Lambda$    & $K^+\Sigma^0$ \\ \cline{3-4}
$\cos \theta_K^*$ & $W$ (GeV) & $M^2$ (GeV$^2$) & $M^2$ (GeV$^2$) \\ \hline
-0.25 & 1.75 & $1.81\pm0.48$ & $1.58\pm1.18$ \\ \hline
-0.25 & 1.85 & $1.34\pm0.37$ & $0.41\pm0.22$ \\ \hline
-0.25 & 1.95 & $1.41\pm0.54$ & $0.64\pm0.28$ \\ \hline
 0.90 & 1.75 & $1.75\pm0.21$ & -- \\ \hline
 0.90 & 1.85 & $2.75\pm0.38$ & $1.51\pm1.46$ \\ \hline
 0.90 & 1.95 & $2.09\pm0.35$ & $1.25\pm1.45$ \\ \hline \hline
\end{tabular}
\end{center}
\caption{\small{Mass terms from the fit to our $\sigma_U$ structure functions 
vs. $Q^2$ (not including the photoproduction points).  A dipole form of 
$C(Q^2 + M^2)^{-2}$ is employed.}}
\label{massterms}
\end{table}

It is interesting to see that the $Q^2$ fits for the $K^+\Lambda$ data
significantly overshoot the photon point for our forward angle data.  In 
the absence of other knowledge, one might speculate that this is entirely 
due to a significant contribution to the cross section from $\sigma_L$.  
However, when the points from the 2.567~GeV data set are included on the 
plot, we see that they fall near the curve fit to our 4~GeV data for all
points for $K^+\Lambda$ and $K^+\Sigma^0$ except our highest $W$ point for 
$K^+\Lambda$.  This suggests a small contribution from $\sigma_L$.  Indeed, 
the $\sigma_L$/$\sigma_T$ separations shown in the next section verify this.  
Janssen {\it et al.}~\cite{janssen1} have calculated a fall off in $\sigma_T$ 
near $Q^2$=0, as indicated by our data. This was accomplished by including a 
$Q^2$ dependence to the kaon and proton form factors.

\subsection{$\sigma_T$ and $\sigma_L$ Separation}
\label{seplt}

\begin{figure}[htpb]
\vspace{10.0cm}
\includegraphics{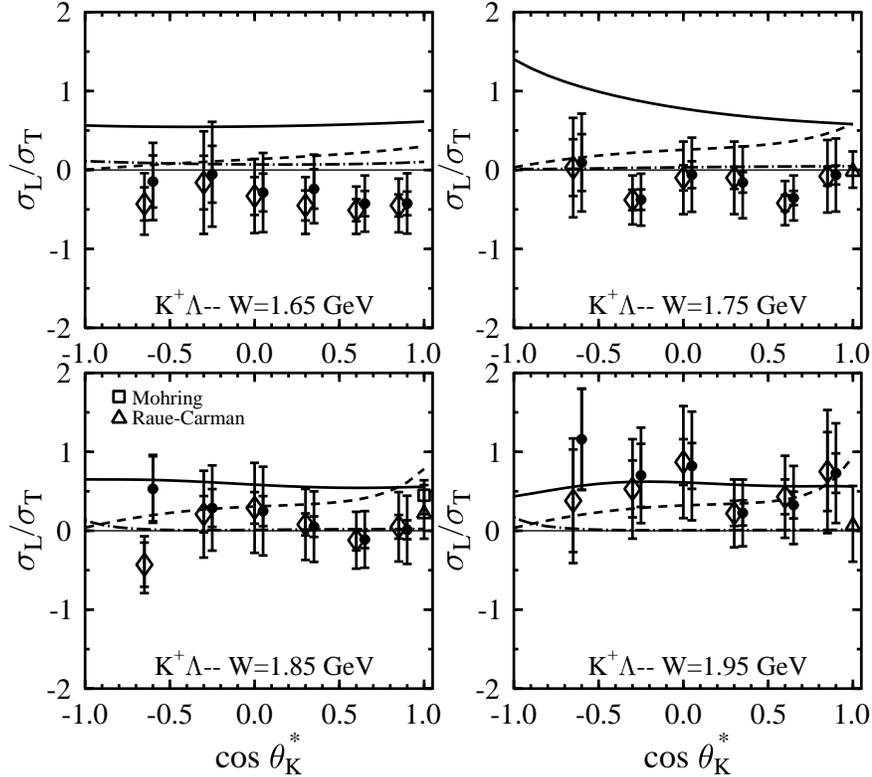}
\caption{\small{Results for the ratio $R=\sigma_L/\sigma_T$ for the 
$K^+\Lambda$ reaction for the Rosenbluth technique (diamonds) and the 
simultaneous $\epsilon-\Phi$ fit (filled circles).  The Rosenbluth results 
have been offset in angle for clarity.  The data are plotted versus 
$\cos \theta_K^*$ for our four $W$ points at $Q^2$=1.0~GeV$^2$.  The inner 
error bars are statistical only and the outer error bars are combined 
statistical and systematic.  The curves shown are from the model calculations
of Bennhold and Mart (BM)~\cite{bm_calc} (dot-dashed), Janssen {\it et al.}
(JB)~\cite{jb_calc} (solid), and Guidal {\it et al.} (GLV) (dashed). The 
parallel kinematics data points come from Mohring {\it et al.}~\cite{mohring} 
(open square) and Raue-Carman~\cite{raue04} (open triangles).}}
\label{rat_lam}
\end{figure}

\begin{figure}[htpb]
\vspace{10.0cm}
\includegraphics{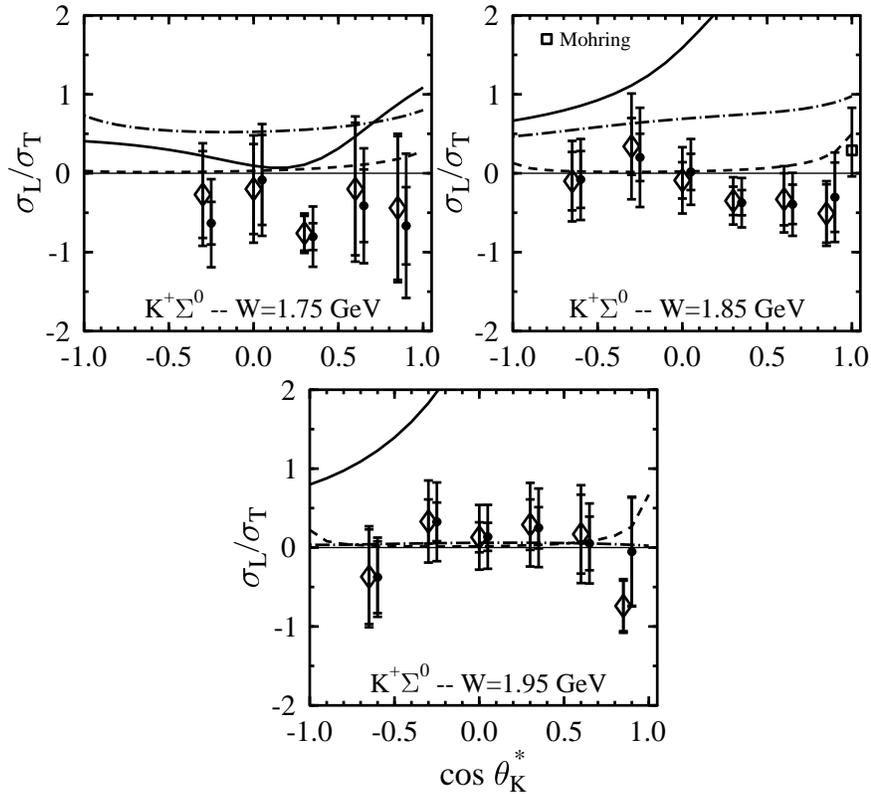}
\caption{\small{Results for the ratio $R=\sigma_L/\sigma_T$ for the
$K^+\Sigma^0$ reaction for the Rosenbluth technique (diamonds) and the
simultaneous $\epsilon-\Phi$ fit (filled circles).  The Rosenbluth results 
have been offset in angle for clarity.  The data are plotted versus 
$\cos \theta_K^*$ for our three $W$ points at $Q^2$=1.0~GeV$^2$.   The inner 
error bars are statistical only and the outer error bars are combined 
statistical and systematic.  The curves shown are from the model calculations
of Bennhold and Mart (BM)~\cite{bm_calc} (dot-dashed), Janssen {\it et al.}
(JB)~\cite{jb_calc} (solid), and Guidal {\it et al.} (GLV) (dashed).  The 
parallel kinematics data point comes from Mohring {\it et al.}~\cite{mohring} 
(open square).}}
\label{rat_sig}
\end{figure}

Our analysis results for $\sigma_T$ and $\sigma_L$ are presented in
Fig.~\ref{rat_lam} for the $K^+\Lambda$ final state and in Fig.~\ref{rat_sig} 
for the $K^+\Sigma^0$ final state.  The data are shown here in terms of the 
ratio $R=\sigma_L/\sigma_T$ as a function of $\cos \theta_K^*$ for our
different $W$ values.  For the $K^+\Lambda$ final state our analysis includes 
$W$ points from 1.65 to 1.95~GeV, and for the $K^+\Sigma^0$ final state our 
analysis includes $W$ points from 1.75 to 1.95~GeV. Note that the statistical 
quality of our data did not allow us to separate $\sigma_L$ and $\sigma_T$ at 
$W$=2.05~GeV.  The figures show the ratio extraction using both the Rosenbluth 
and the simultaneous $\epsilon-\Phi$ fit techniques, and the error bars show 
both statistical and total statistical and systematic uncertainties.  The 
discussion of systematic uncertainties on these quantities is included 
in Section~\ref{systematics}.  

The agreement between the Rosenbluth and simultaneous $\epsilon-\Phi$ fits 
is generally very good across our full $W$ and $\cos \theta_K^*$ phase space 
at $Q^2$ = 1.0~GeV$^2$.  The ratio of $\sigma_L/\sigma_T$ for both the 
$K^+\Lambda$ and $K^+\Sigma^0$ final states shows $\sigma_L$ to be consistent
with zero over our full kinematic range except in our highest $W$ point for 
the $K^+\Lambda$ reaction, where the value of $R$ varies between 0.5 and 1 
depending on kaon angle.  While several of the extracted values for $R$ are 
negative and might be considered "unphysical", the majority of these points 
are consistent with zero within the combined statistical and systematic 
uncertainties.

Our data at $W$=1.85~GeV and $\cos \theta_K^*$=0.90 are consistent with the 
parallel kinematics measurement of Mohring {\it et al.}~\cite{mohring} from 
Hall C which found $R=0.45^{+0.19}_{-0.16}$ for $K^+\Lambda$ and 
$R=0.29^{+0.54}_{-0.33}$ for $K^+\Sigma^0$ at $W$=1.84~GeV, as well as the 
recent Hall B results of Raue and Carman~\cite{raue04} for $K^+\Lambda$ 
which found $R$=0.005$\pm$0.228, 0.239$\pm$0.343, and 0.088$\pm$0.480 for 
$\theta_K^*$=0$^{\circ}$ and $W$=1.72, 1.84, and 1.98~GeV, respectively.  
Note that the quoted uncertainties on $R$ from Mohring {\it et al.}
\cite{mohring} and Raue and Carman~\cite{raue04} given here represent the 
total statistical and systematic uncertainties.

The predictions of the hadrodynamic models for the ratio $R$ are very
sensitive to the dynamics included in the calculation.  All models shown
for the $K^+\Lambda$ final state predict $R$ to be less than unity.
Given the size of the error bars on the data, all of the models can be
said to be roughly consistent with the data.  However it is clear that the
JB model~\cite{jb_calc} predicts too much longitudinal strength at low $W$.
Also for the highest $W$ point, where $R$ begins to increase, the BM
model~\cite{bm_calc} predicts too little longitudinal strength.  The GLV
model~\cite{glv_calc} is in very good agreement with the data over the full
kinematic range shown.

For the $K^+\Sigma^0$ final state, the JB model~\cite{jb_calc} disagrees
with the measured ratio $R$, which will be shown to result from too 
little transverse strength.  The BM model~\cite{bm_calc} also suffers 
from too little transverse strength for our two lowest $W$ points, but 
is consistent with the data at $W$=1.95~GeV.  Again, the GLV model
\cite{glv_calc} agrees well with the data over the full kinematic range 
shown.

Clearly these data, even with their sizeable statistical and systematic 
uncertainties, can provide for important constraints on the underlying 
dynamics and production models for both the $K^+\Lambda$ and 
$K^+\Sigma^0$ final states.

\begin{figure}[htpb]
\vspace{10.0cm}
\includegraphics{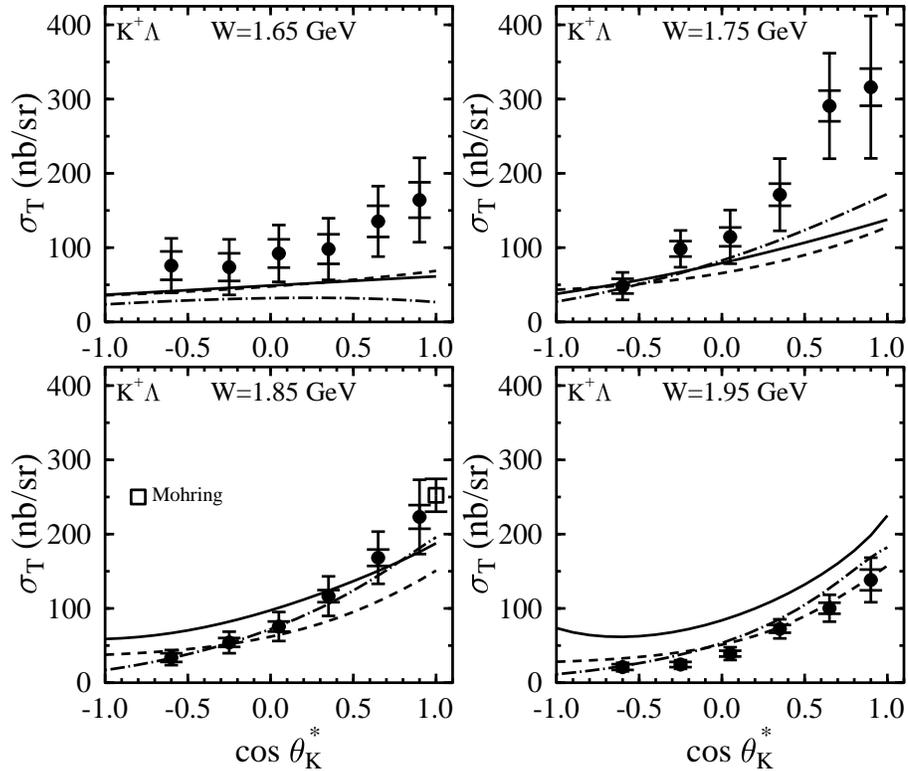}
\caption{\small{Structure function $\sigma_T$ vs. $\cos \theta_K^*$ for
the $K^+\Lambda$ final state for our different $W$ points at $Q^2$=1.0~GeV$^2$
from the $\epsilon-\Phi$ fit.  The inner error bars are statistical only and 
the outer error bars are combined statistical and systematic.  The curves 
shown are from the model calculations of Bennhold and Mart (BM)~\cite{bm_calc} 
(dot-dashed), Janssen {\it et al.} (JB)~\cite{jb_calc} (solid), and Guidal 
{\it et al.} (GLV) (dashed).  The parallel kinematics data point comes from 
Mohring {\it et al.}~\cite{mohring} (open square).}}
\label{t_lam}
\end{figure}

\begin{figure}[htpb]
\vspace{10.0cm}
\includegraphics{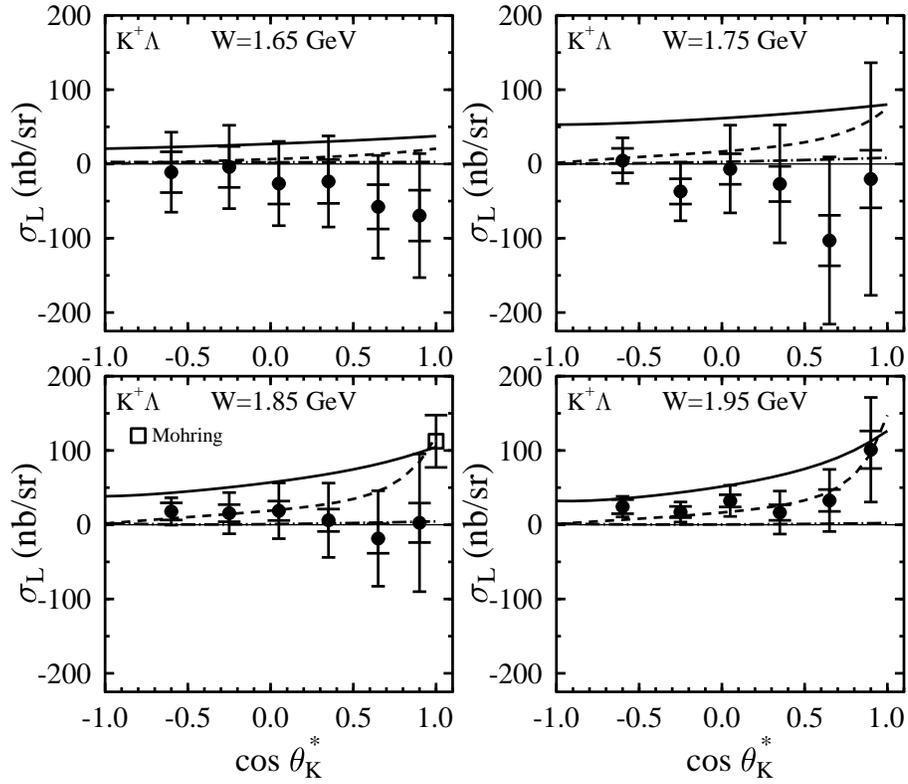}
\caption{\small{Structure function $\sigma_L$ vs. $\cos \theta_K^*$ for
the $K^+\Lambda$ final state for our different $W$ points at $Q^2$=1.0~GeV$^2$
from the $\epsilon-\Phi$ fit.  The inner error bars are statistical only and 
the outer error bars are combined statistical and systematic.  The curves 
shown are from the model calculations of Bennhold and Mart (BM)~\cite{bm_calc} 
(dot-dashed), Janssen {\it et al.} (JB)~\cite{jb_calc} (solid), and Guidal 
{\it et al.} (GLV) (dashed).  The parallel kinematics data point comes from 
Mohring {\it et al.}~\cite{mohring} (open square).}}
\label{l_lam}
\end{figure}

The structure functions $\sigma_T$ and $\sigma_L$ are plotted separately 
in Figs.~\ref{t_lam} and \ref{l_lam} for $K^+\Lambda$ as a function of 
$\cos \theta_K^*$.  Here $\sigma_T$ has a similar trend in angle for
all $W$ points, peaking at forward kaon angles and falling off smoothly
as the angle increases.  For $\sigma_L$, the strength is consistent with 
zero for our points in $W$ from 1.65 to 1.85~GeV.  At $W$=1.95~GeV, 
$\sigma_L$ is comparable with $\sigma_T$ in its angular dependence and 
its strength.  The data for $\sigma_T$ and $\sigma_L$ are consistent with 
the existing parallel kinematics measurement at $W$=1.85~GeV of Mohring 
{\it et al.}~\cite{mohring} from Hall C at JLab.

The comparison of the models to $\sigma_T$ for the $K^+\Lambda$ final
state shows that they underpredict the data and the strength of the
forward-angle rise for our two lowest $W$ points.  For our two highest
$W$ points, the calculations (with the exception of the JB model
\cite{jb_calc} at $W$=1.95~GeV) are in good agreement with the data.
For $\sigma_L$, the calculations from the BM model~\cite{bm_calc} and
the GLV model~\cite{glv_calc} are in reasonable agreement with the
data given the error bars.  However, the JB model~\cite{jb_calc}
predicts too much longitudinal strength for the full kinematic range
shown.

\begin{figure}[htpb]
\vspace{10.0cm}
\includegraphics{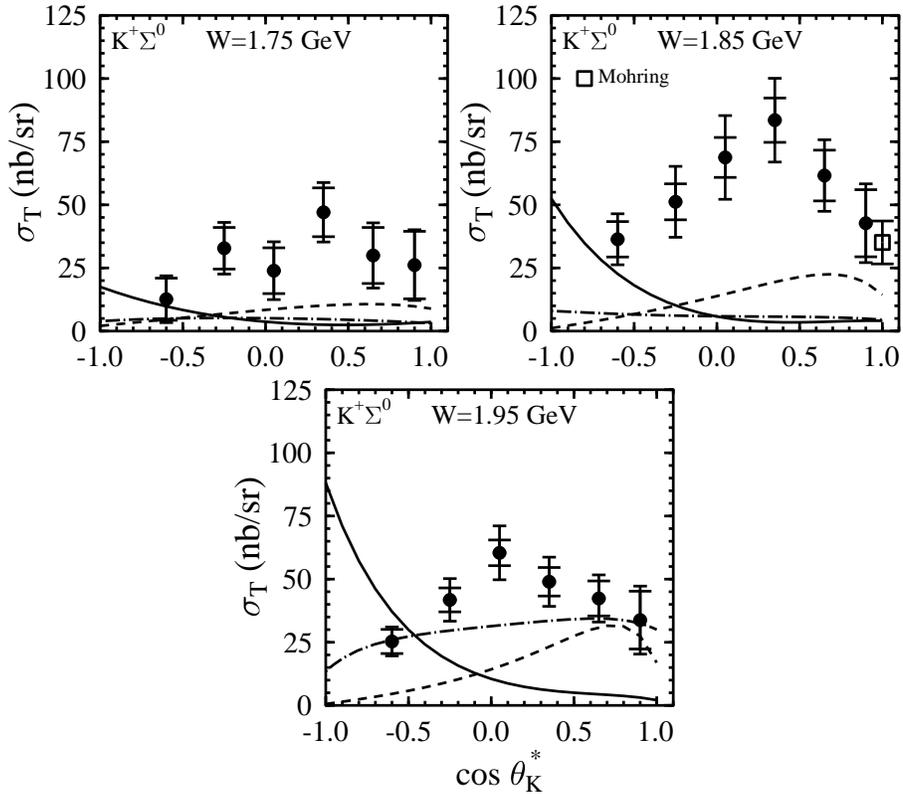}
\caption{\small{Structure function $\sigma_T$ vs. $\cos \theta_K^*$ for
the $K^+\Sigma^0$ final state for our different $W$ points at 
$Q^2$=1.0~GeV$^2$ from the $\epsilon-\Phi$ fit.  The inner error bars are 
statistical only and the outer error bars are combined statistical and 
systematic.  The curves shown are from the model calculations of Bennhold 
and Mart (BM)~\cite{bm_calc} (dot-dashed), Janssen {\it et al.} (JB)
\cite{jb_calc} (solid), and Guidal {\it et al.} (GLV) (dashed).  The 
parallel kinematics data point comes from Mohring {\it et al.}~\cite{mohring} 
(open square).}}
\label{t_sig}
\end{figure}

\begin{figure}[htpb]
\vspace{10.0cm}
\includegraphics{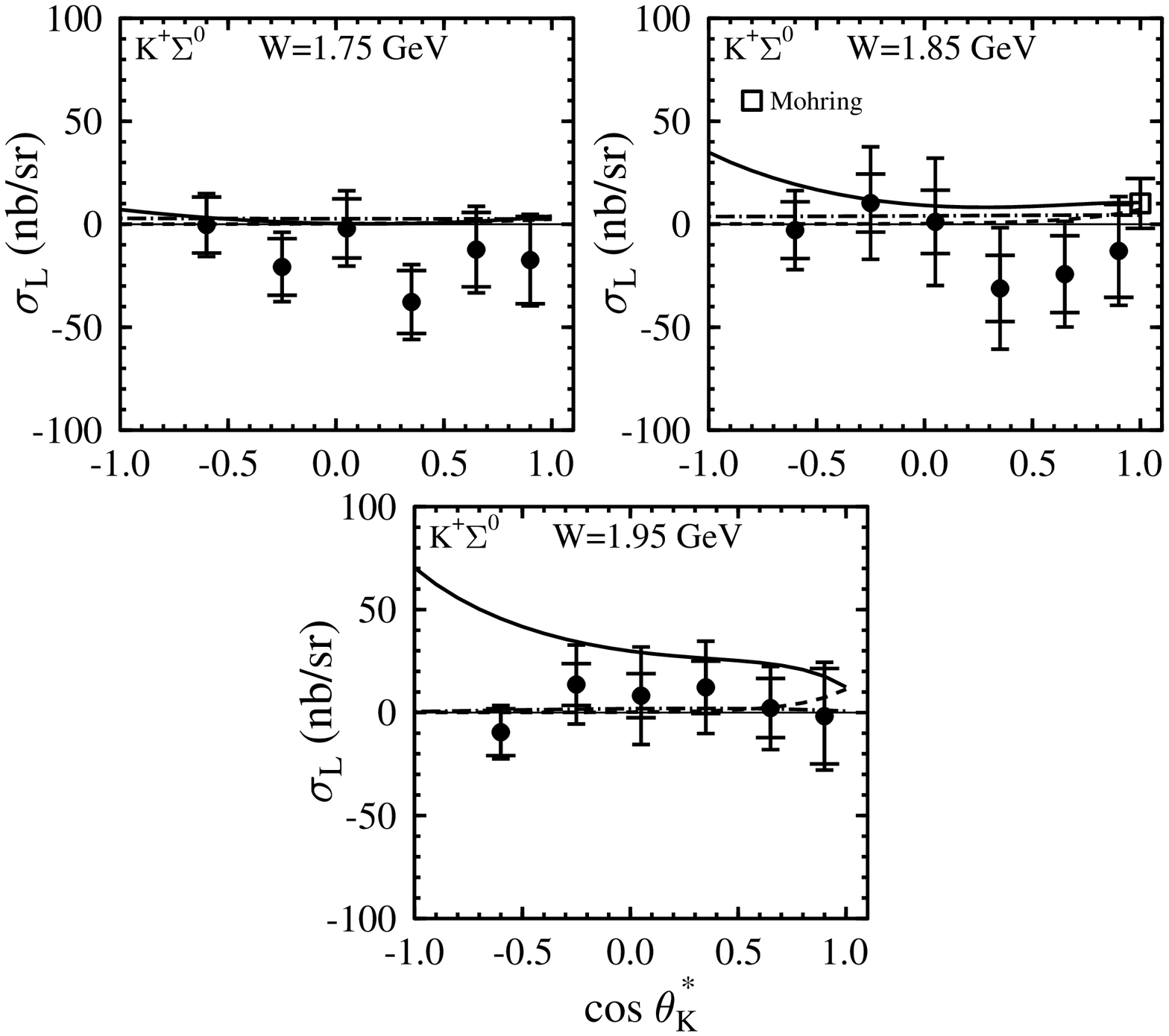}
\caption{\small{Structure function $\sigma_L$ vs. $\cos \theta_K^*$ for
the $K^+\Sigma^0$ final state for our different $W$ points at
$Q^2$=1.0~GeV$^2$ from the $\epsilon-\Phi$ fit.  The inner error bars are 
statistical only and the outer error bars are combined statistical and 
systematic.  The curves shown are from the model calculations of Bennhold 
and Mart (BM)~\cite{bm_calc} (dot-dashed), Janssen {\it et al.} (JB)
\cite{jb_calc} (solid), and Guidal {\it et al.} (GLV) (dashed).  The 
parallel kinematics data point comes from Mohring {\it et al.}~\cite{mohring} 
(open square).}}
\label{l_sig}
\end{figure}

The structure functions $\sigma_T$ and $\sigma_L$ are plotted separately
in Figs.~\ref{t_sig} and \ref{l_sig} for $K^+\Sigma^0$ as a function of
$\cos \theta_K^*$.  $\sigma_T$ is seen to have a broad peaking at more
central angles for $K^+\Sigma^0$ compared to the $K^+\Lambda$ final state
and $\sigma_L$ is consistent with zero everywhere.  The data for $\sigma_T$ 
and $\sigma_L$ are consistent with the existing parallel kinematics 
measurement at $W$=1.85~GeV of Mohring {\it et al.}~\cite{mohring}
from Hall C at JLab.

The models compare poorly with the data for $\sigma_T$ for $K^+\Sigma^0$
underpredicting the strength of the data and missing the trends in
the angular dependence.  For $\sigma_L$, all models predict a small strength
in agreement with the data.  Given the size of the statistical and
systematic error bars on the data, not much more can be said with respect
to the model predictions.

\section{Conclusions}
\label{conclusions}

We have measured $K^+\Lambda$ and $K^+\Sigma^0$ electroproduction over 
a wide range of kinematics in the nucleon resonance region, including 
first-reported measurements over the full range of kaon angle.  We have 
presented data for the separated structure functions $\sigma_T$, 
$\sigma_L$, and the interference structure functions $\sigma_{LT}$ and 
$\sigma_{TT}$.  We conclude that $K^+\Lambda$ and $K^+\Sigma^0$ 
electroproduction dynamics are markedly different.  We find that 
$\sigma_U$ ($= \sigma_T + \epsilon \sigma_L$) and $\sigma_{TT}$ are 
forward-peaked (in kaon angle) for the $K^+\Lambda$ final state 
and peaked at more central angles for $K^+\Sigma^0$, with $\sigma_{TT}$ a
significant fraction of $\sigma_U$ for both hyperon final states.  For the 
$K^+\Lambda$ channel, $\sigma_{LT}$ is a significant fraction of 
$\sigma_U$ and negative, while it is consistent with zero for the 
$K^+\Sigma^0$ channel.  

The $W$ dependence of $\sigma_U$ for $K^+\Sigma^0$ shows a broad 
enhancement at $W \sim$ 1.85~GeV for $\sigma_U$ and $\sigma_{TT}$, 
presumably due to the various $\Delta^*$ resonances in this mass 
region.  The $W$ dependence for $K^+\Lambda$ production is more 
complicated, evolving from a single-peaked structure at forward angles 
to a double-peaked structure at backward angles at low $Q^2$, with the 
double-peaking not obvious above $Q^2$=1.0~GeV$^2$.  The longitudinal 
structure function $\sigma_L$ is consistent with zero for $K^+\Lambda$ 
across our full kinematic space except for our highest $W$ point at 
1.95~GeV, and is consistent with zero everywhere for our $K^+\Sigma^0$
data.  The transverse structure function $\sigma_T$ is forward-peaked
for $K^+\Lambda$ and peaked at more central angles for $K^+\Sigma^0$.

The $Q^2$ dependence of the structure functions is unremarkable; the
relatively slow fall-off, presumably reflecting the form factors of
the various exchanged kaons in the case of $t$-channel processes, and 
the analogous baryonic form-factors in $s$ and $u$-channel processes.  
Of interest is our observation that an extrapolation of $\sigma_U$ to 
$Q^2 = 0$ overshoots the photoproduction value for $\sigma_T$.  The 
obvious conjecture that this reveals the presence of a large value of 
$\sigma_L$ is not consistent with our direct measurements of $\sigma_L$, 
at least for lower values of $W$. 

Detailed calculations are needed in order to investigate whether our 
data indicate the presence of any new $N^*$ resonances in $K^+\Lambda$ 
production.  At the same time, the calculations must be able to fit 
the strong peak seen for $K^+\Sigma^0$ production and the shape of 
$\sigma_{TT}$ and $\sigma_{LT}$ in both channels.  The question of the 
presence of any new resonances must wait for further work with the 
existing hadrodynamic models and partial wave analyses applied to the 
full range of our data.  Fortunately, the new information we present
here, especially the interference terms, will impose stringent 
constraints on the amplitudes used to model electroproduction of 
$K^+\Lambda$ and $K^+\Sigma^0$ final states, making these models more 
reliable for future interpretation and prediction.

\vskip 0.3cm

We would like to acknowledge the outstanding efforts of the staff of the 
Accelerator and the Physics Divisions at JLab that made this experiment
possible. This work was supported in part by the U.S. Department of Energy,
the National Science Foundation, the Istituto Nazionale di Fisica Nucleare, 
the French Centre National de la Recherche Scientifique, the French 
Commissariat \`{a} l'Energie Atomique, and the Korean Science and 
Engineering Foundation. The Southeastern Universities Research
Association (SURA) operated Jefferson Lab under United States DOE
contract DE-AC05-84ER40150 during this work.

\vfil
\eject


\newpage

\begin{thebibliography}{99}

\bibitem{isgur}
R. Koniuk and N. Isgur, Phys. Rev. D {\bf 21}, 1868 (1980).

\bibitem{capstick} 
S. Capstick and W. Roberts, Phys. Rev. D {\bf 58}, 1 (1998).

\bibitem{klempt}
E. Klempt, nucl-ex/0203002, and references therein.

\bibitem{lee} 
T.-S.H. Lee and T. Sato, Proceedings of the NSTAR 2000 Conference, eds. 
V.D. Burkert, L. Elouadrhiri, J.J. Kelly, and R. Minehart, (World Scientific, 
Singapore, 2001), p. 215. 

\bibitem{database}
CLAS data base, URL: http://clasweb.jlab.org/physicsdb.

\bibitem{convention} 
Some authors use a pre-factor for the $\sigma_L$ ($\sigma_{LT}$) term of 
$\epsilon_L$ ($\sqrt{2\epsilon_L(\epsilon+1)}$) instead, where 
$\epsilon_L = \epsilon {Q^2}/{\nu_{cm}^2}$ parameterizes the longitudinal
polarization of the virtual photon.  Some also take a $\sin\theta_K^*$
($\sin^2\theta_K^*$) term out of the definition of $\sigma_{LT}$ 
($\sigma_{TT}$). 

\bibitem{knochlein} 
G. Kn\"ochlein, D. Drechsel, and L. Tiator, Z. Phys. A {\bf 352}, 327 (1995).

\bibitem{brown} 
C.N. Brown {\it et al.}, Phys. Rev. Lett. {\bf 28}, 1086 (1972).

\bibitem{bebek1} 
C.J. Bebek {\it et al.}, Phys. Rev. Lett. {\bf 32}, 21 (1974).

\bibitem{azemoon} 
T. Azemoon {\it et al.}, Nucl. Phys. B {\bf 95}, 77 (1975).

\bibitem{bebek2} 
C.J. Bebek {\it et al.}, Phys. Rev. D {\bf 15}, 594 (1977); 
C.J. Bebek {\it et al.}, Phys. Rev. D {\bf 15}, 3082 (1977).

\bibitem{close} 
F.E. Close, Nucl. Phys. B {\bf 73}, 410 (1974);
O. Nachtmann, Nucl. Phys. {\bf B74}, 422 (1974); 
J. Cleymans and F.E. Close, Nucl. Phys. B {\bf 85}, 429 (1975).

\bibitem{brauel}
P. Brauel {\it et al.}, Z. Phys. C {\bf 3}, 101 (1979).

\bibitem{mohring} 
R.M. Mohring {\it et al.}, Phys. Rev. C {\bf 67}, 055205 (2003); reanalysis
of G. Niculescu {\it et al.}, Phys. Rev. Lett. {\bf 81}, 1805 (1998).

\bibitem{coman}
Marius Coman, Ph.D. thesis, Florida International University (unpublished), 
(2005).

\bibitem{carman} 
D.S. Carman {\it et al.} {\it (CLAS Collaboration)}, Phys. Rev. Lett. 
{\bf 90}, 131804 (2003).

\bibitem{raue04} 
Brian A. Raue and Daniel S. Carman, Phys. Rev. C {\bf 71}, 065209 (2005).

\bibitem{saphir1} 
M.Q. Tran {\it et al.},  Phys. Lett. B {\bf 445}, 20 (1998).

\bibitem{saphir2} 
K.H. Glander {\it et al.}, Eur. Phys. J. A {\bf 19}, 251 (2004).

\bibitem{mcnabb} 
J. W. C. McNabb {\it et al.} {\it (CLAS Collaboration)}, Phys. Rev. C 
{\bf 69}, 042201(R) (2004).

\bibitem{bradford1}
R.K. Bradford {\it et al.} {\it (CLAS Collaboration)}, Phys. Rev. C
{\bf 73}, 035202 (2006).

\bibitem{bradford2}
R.K. Bradford, Ph.D. thesis, Carnegie Mellon University (2005) (unpublished).
Available at www.jlab.org/Hall-B/general/clas\_thesis.html.

\bibitem{leps1}
M. Sumihama {\it et al.} {\it (LEPS Collaboration)}, Phys. Rev. C {\bf 73},
035214 (2006); R.T.G. Zegers {\it et al.} {\it (LEPS Collaboration)}, Phys. 
Rev. Lett. {\bf 91}, 092001 (2003).

\bibitem{leps2}
H. Kohri {\it et al.} {\it (LEPS Collaboration)}, Preprint hep-ex/0602015 
(2006). 

\bibitem{janssen1}
S. Janssen {\it et al.}, Phys. Rev. C {\bf 67}, 052201 (2003).

\bibitem{bennhold} 
T. Mart and C. Bennhold, Phys. Rev. C {\bf 61}, 012201 (2000).  

\bibitem{haber} 
H. Haberzettl {\it et al.}, Phys. Rev. C {\bf 58}, R40 (1998).  

\bibitem{janssen2}
S. Janssen {\it et al.}, Eur. Phys. J. A {\bf 11}, 105 (2001); 
S. Janssen {\it et al.}, Phys. Rev. C {\bf 65}, 015201 (2001).

\bibitem{janssen3}
S. Janssen, Ph.D. Thesis, University of Gent, (2002).

\bibitem{gross}
F. Gross and D. Riska, Phys. Rev. C {\bf 36}, 1928 (1987).

\bibitem{saghai} 
B. Saghai, AIP Conf. Proc. {\bf 594}, 57 (2001).

\bibitem{ireland}
D.G. Ireland, S. Janssen, and J. Ryckebusch, Nucl. Phys. A {\bf 740},
147 (2004).

\bibitem{sarantsev}
A.V. Sarantsev {\it et al.}, Eur. Phys. J. A {\bf 25}, 427 (2005).

\bibitem{guidal}
M. Guidal, J.M. Laget, and M. Vanderhaeghen, Phys. Rev. C {\bf 61}, 025204 
(2000).

\bibitem{guidal1}
M. Guidal, J.M. Laget, and M. Vanderhaeghen, Nucl. Phys. A {\bf 627}, 645
(1997).

\bibitem{clas} 
B.A. Mecking {\it et al.}, Nucl. Inst. and Meth. A {\bf 503}, 513 (2003).

\bibitem{dcnim}
M.D. Mestayer {\it et al.}, Nucl. Inst. and Meth. A {\bf 449}, 
81 (2000).

\bibitem{ccnim}
G. Adams {\it et al.}, Nucl. Inst. and Meth. A {\bf 465}, 414 (2001).

\bibitem{scnim}
E.S. Smith {\it et al.}, Nucl. Inst. and Meth. A {\bf 432}, 265 (1999).

\bibitem{ecnim}
M. Amarian {\it et al.}, Nucl. Inst. and Meth. A {\bf 460}, 460 (2001).

\bibitem{geant}
R. Brun {\it et al.}, CERN-DD-78-2-REV, (1978).

\bibitem{motsai} 
L.W. Mo and Y. Tsai, Rev. Mod. Phys. {\bf 41}, 205 (1969); 
Y.S. Tsai, Preprint SLAC, PUB-848, 1971.

\bibitem{radcorr_ent} 
R. Ent {\it et al.}, Phys. Rev. C {\bf 64}, 054610 (2001).

\bibitem{afanasev} 
{\tt EXCLURAD} code is based upon A. Afanasev {\it et al.}, Phys. Rev. D 
{\bf 66}, 074004 (2002).

\bibitem{bm_calc} 
Code from T. Mart, private communication, (2000), based upon the model of 
Ref. \cite{bennhold}.

\bibitem{jb_calc} 
Code from model of S. Janssen, private communication, (2002), corresponding 
to Ref.~\cite{janssen1,janssen2}.  We show Model ``B''.

\bibitem{glv_calc} 
Curves from model code of M. Guidal, private communication,(2000), 
corresponding to Ref.~\cite{guidal}.

\bibitem{adel_saghai} 
R.A. Adelseck and B. Saghai, Phys. Rev. C {\bf 42}, 108 (1990).

\bibitem{deswart} 
J.J. deSwart, Rev. Mod. Phys. {\bf 35}, 916 (1963).

\bibitem{saghai1}
B. Saghai, nucl-th/0105001, (2001). 

\bibitem{pdg}
E.J. Weinberg and D.L. Nordstrom, Phys. Rev. D {\bf 66}, 01001 (2002).

\end{thebibliography}
\end{document}